
\message{......................}
\message{***************[Message from the authors]*****************}
\message{This uses two standard macros: harvmac.tex, epsf.tex.}
\message{macros in harvmac.tex is used only for title pages.}
\message{epsf.tex puts figures in the text. }
\message{epsf.tex works with dvips on UNIX and VAX,}
\message{but epsf.tex may not work with TeX on Mac.}
\message{****************[end of the message]*******************}

\input harvmac
\input epsf

\overfullrule=0pt

\def\Null#1{}

\def\cA{{\cal A}}

\def\bZ{{\bf Z}}

\def\End{{\rm End}}

\def\Rc{\check R}

\def\pn{\par\noindent}

\def\lra{\longrightarrow}

\def\Id{{\rm Id}}
\def\subskip{\vskip 0.3cm}
\def\sect#1{\vskip 1cm \centerline{\bf #1} }
\def\subsect#1{\vskip 0.6cm {\bf\noindent #1}}
\def\til{\tilde}
\def\qd{\Delta(u)}
\def\tensor{\otimes}

\def\YX{Y(X_r)}
\def\am{^{(a)}_m}
\def\om#1{\Lambda_{#1}}
\def\pr#1{P_{\Lambda_{#1}}}
\def\V#1{V_{\Lambda_{#1}}}
\def\W#1{W^{(#1)}_1(u)}
\def\Wnu#1{W^{(#1)}_1}
\def\tensor{\otimes}
\def\Rc{\check{R}}

\def\ba#1{\overline{#1}}
\def\End{{\rm End}}
\def\Id{{\rm Id}}
\def\emb{\hookrightarrow}
\def\rBax{1}
\def\rDr{2}
\def\rJim{3}
\def\rKS{4}
\def\rKNSii{5}
\def\rBP{6}
\def\rKRjp{7}
\def\rPea{8}
\def\rBRi{9}
\def\rKSZ{10}
\def\rKLPi{11}
\def\rKLPiii{12}
\def\rBRii{13}
\def\rResii{14}
\def\rKRS{15}
\def\rYY{16}
\def\rKu{17}
\def\rAlbZam{18}
\def\rKM{19}
\def\rAlbZamii{20}
\def\rKMii{21}
\def\rKN{22}
\def\rRav{23}
\def\rKil{24}
\def\rKZ{25}
\def\rKRii{26}
\def\rBD{27}
\def\rFR{28}
\def\rDJO{29}
\def\rMcG{30}
\def\rYa{31}
\def\rTar{32}
\def\rCPi{33}
\def\rSk{34}
\def\rJKMO{35}
\def\rChe{36}
\def\rCheii{37}
\def\rCPii{38}
\def\rDJKMONP{39}
\def\rPas{40}
\def\rCheiii{41}
\def\rMac{42}
\def\rCPiii{43}
\def\rKNii{44}
\def\rDrii{45}
\def\rJimi{46}
\def\rLus{47}
\def\rRA{48}
\def\rPS{49}
\def\rKel{50}
\def\rKac{51}
\def\rFVD{52}
\def\rFGP{53}
\def\rWal{54}
\def\rGN{55}
\def\rOW{56}
\def\rOgi{57}
\def\rResi{58}
\def\rORW{59}
\def\rFH{60}
\def\rZel{61}
\def\rAki{62}
\def\rRes{63}
\def\rABP{64}
\def\rBBP{65}
\def\rBaxii{66}
\def\rAMP{67}
\def\rKili{68}
\def\rDas{69}

\def\reference{
\vfill\eject
\pn
{\bf References}
\item{[\rBax]}{R.J.\ Baxter, {\sl Exactly Solved Models in Statistical
Mechanics},
(Academic Press, London, 1982)}
\item{[\rDr]}{V.G.\ Drinfel'd, Sov.\ Math.\ Dokl.\  {\bf 32} (1985) 254;
in {\it Proceedings of the International Congress of Mathematicians, Berkeley},
(American Mathematical Society, Providence, 1987)}
\item{[\rJim]}{M.\ Jimbo, Lett.\ Math.\ Phys.\  {\bf 10} (1985) 63}
\item{[\rKS]}{P.P.\ Kulish and E.K.\ Sklyanin, J.\ Sov.\ Math.\  {\bf
19} (1982) 1596; in
{\it Lecture Notes in Physics} {\bf 151} (Springer, Berlin, 1982) }
\item{[\rKNSii]}{A.\ Kuniba, T.\ Nakanishi and J.\ Suzuki, ``Functional
relations  in solvable lattice models II:
Applications", preprint, HUTP-93/A023}
\item{[\rBP]}{R.J.\ Baxter and P.A.\ Pearce,
J.\ Phys.\ A: Math.\ Gen.\ {\bf 15} (1982) 897; {\bf 16} (1983) 2239}
\item{[\rKRjp]}{A.N.\ Kirillov and N.Yu.\ Reshetikhin,
J.\ Phys.\ A: Math.\ Gen.\ {\bf 20} (1987) 1587}
\item{[\rPea]}{P.A.\ Pearce, Phys.\ Rev.\ Lett.\ {\bf 58} (1987) 1502}
\item{[\rBRi]}{V.V.\ Bazhanov and Yu.N.\ Reshetikhin, Int.\ J.\ Mod.\ Phys.\
{\bf A4}
(1989) 115}
\item{[\rKSZ]}{A.\ Kl\"umper, A.\ Schadschneider and J.\ Zittartz,
Z.\ Phys.\  {\bf B76} (1989) 247}
\item{[\rKLPi]}{A.\ Kl\"umper and P.A.\ Pearce, Phys.\ Rev.\ Lett.\  {\bf 66}
(1991) 974; J.\ Stat.\ Phys.\  {\bf 64} (1991) 13}
\item{[\rKLPiii]}{A.\ Kl\"umper and P.A.\ Pearce, Physica {\bf A183} (1992)
304}
\item{[\rBRii]}{V.V.\ Bazhanov and Yu.N.\ Reshetikhin,
 J.\ Phys.\ A: Math.\ Gen.\ {\bf 23} (1990) 1477}
\item{[\rResii]}{Yu.N.\ Reshetikhin, Lett.\ Math.\ Phys.\ {\bf 7}
(1983) 205}
\item{[\rKRS]}{P.P.\ Kulish, N.Yu.\ Reshetikhin and E.\ K.\ Sklyanin,
Lett.\ Math.\ Phys.\  {\bf 5} (1981) 393}
\item{[\rYY]}{C.N.\ Yang and C.P.\ Yang, J.\ Math.\ Phys.\  {\bf 10} (1969)
1115}
\item{[\rKu]}{A.\ Kuniba, Nucl.\ Phys.\ {\bf B389} (1993) 209}
\item{[\rAlbZam]}{Al.B.\ Zamolodchikov, Nucl.\ Phys.\  {\bf B342}
(1990) 695,
{\bf B358} (1991) 497}
\item{[\rKM]}{T.R.\ Klassen and E.\ Melzer,
 Nucl.\ Phys.\  {\bf B338} (1990) 485;
{\bf B350} (1991) 635}
\item{[\rAlbZamii]}{Al.B.\ Zamolodchikov,
  Phys.\ Lett.\  {\bf B253} (1991) 391}
\item{[\rKMii]}{T.R.\ Klassen and E.\ Melzer,
Nucl.\ Phys.\  {\bf B370} (1992) 511}
\item{[\rKN]}{A.\ Kuniba and T.\ Nakanishi, Mod.\ Phys.\ Lett.\  {\bf A7}
(1992) 3487}
\item{[\rRav]}{F.\ Ravanini, R.\ Tateo and A.\ Valleriani,
 Phys.\ Lett.\  {\bf B293} (1992) 361;
Int.\ J.\ Mod.\ Phys.\ {\bf A8} (1993) 1707}
\item{[\rKil]}{A.N.\ Kirillov, Zap.\ Nauch.\ Semin.\ LOMI {\bf 164}
(1987) 121 [J.\ Sov.\ Math.\  {\bf 47} (1989) 2450]}
\item{[\rKZ]}{V.G.\ Knizhnik  and A.B.\ Zamolodchikov, Nucl.\ Phys.\
{\bf B247} (1984) 83 }
\item{[\rKRii]}{A.N.\ Kirillov and N.Yu.\ Reshetikhin,
Zap.\ Nauch.\ Semin.\ LOMI {\bf 160} (1987) 211
[J.\ Sov.\ Math.\ {\bf 52} (1990) 3156]}
\item{[\rBD]}{A.A.\ Belavin and V.G.\ Drinfel'd, Funct.\ Anal.\ Appl.\  {\bf
16}
(1983) 159}
\item{[\rFR]}{I.B,\ Frenkel and N.Yu.\ Reshetkihin,
Commun.\ Math.\ Phys.\ {\bf 146} (1992) 1}
\item{[\rDJO]}{E.\ Date, M.\ Jimbo and M.\ Okado, Commun.\ Math.\ Phys.\
 {\bf 155} (1993) 47}
\item{[\rMcG]}{J.B.\ McGuire, J.\ Math.\ Phys.\  {\bf 5 } (1964) 622}
\item{[\rYa]}{C.N.\ Yang, Phys.\ Rev.\ Lett.\  {\bf 19} (1967) 1312}
\item{[\rTar]}{V.O.\ Tarasov, Theore.\ Math.\ Phys.\  {\bf 61}
 (1984) 163; {\bf 63} (1985) 175}
\item{[\rCPi]}{V.\  Chari and A.\ Pressley, L'Enseignement Math.\
{\bf 36} (1990) 267}
\item{[\rSk]}{E.K.\ Sklyanin, ``Quantum inverse scattering method.\
Selected topics'', preprint, hep-th/9211111}
\item{[\rJKMO]}{M.\ Jimbo, A.\ Kuniba, T.\ Miwa and M.\ Okado,
Commun.\  Math.\  Phys.\ {\bf 119} (1988) 543}
\item{[\rChe]}{I.\ Cherednik, Funct.\ Anal.\ Appl.\  {\bf 20} (1986) 76}
\item{[\rCheii]}{I.\ Cherednik, Soviet Math.\ Dokl.\  {\bf 33} (1986) 507
}
\item{[\rCPii]}{V.\  Chari and A.\ Pressley, Commun.\ Math.\ Phys.\  {\bf 142}
(1991) 261}
\item{[\rDJKMONP]}{E.\ Date, M.\ Jimbo, A.\ Kuniba, T.\ Miwa and M.\ Okado,
Nucl.\ Phys.\  {\bf B290} [FS20] (1987) 231}
\item{[\rPas]}{V.\ Pasquier, Commun.\ Math.\ Phys.\  {\bf 118} (1988) 335}
\item{[\rCheiii]}{I.\ Cherednik, in Proc.\ of the XVII International
Conference
on Differential Geometric Methods in Theoretical Physics, Chester,
ed.\ A.I.\ Solomon, (World Scientific, Singapore, 1989)
}
\item{[\rMac]}{I.G.\ Macdonald, {\sl Symmetric functions and Hall polynomials},
(Clarendon Press, Oxford, 1979)}
\item{[\rCPiii]}{V.\  Chari and A.\ Pressley, J.\ reine angew.\ Math.\
{\bf 417}
(1991)  87}
\item{[\rKNii]}{A.\ Kuniba and T.\ Nakanishi, ``Rogers dilogarithm
in integrable systems", hep-th/9210025, to appear in Proceedings of
the XXI Differential Geometry Methods in Theoretical Physics, Tianjin,
}
\item{[\rDrii]}{V.G.\ Drinfel'd, Soviet Math.\ Dokl.\  {\bf 36} (1988) 212}
\item{[\rJimi]}{M.\ Jimbo, {\sl Yang-Baxter equation in integrable
systems}, Ed.\ M.\ Jimbo, (World scientific, Singapore, 1990)}
\item{[\rLus]}{G.\ Lusztig, Contemporary Math.\  {\bf 82} (1989) 59}
\item{[\rRA]}{P.\ Roche and D.\ Arnaudon, Lett.\ Math.\ Phys.\  {\bf 17}
 (1989) 295}
\item{[\rPS]}{V.\ Pasquier and H.\ Saleur, Nucl.\ Phys.\  {\bf B330} (1990)
330}
\item{[\rKel]}{G.\ Keller, Lett.\ Math.\ Phys.\  {\bf 21} (1991) 273}
\item{[\rKac]}{V.G.\ Kac, {\sl Infinite dimensional Lie algebras,
3rd ed.\ },
(Cambridge University Press, Cambridge, 1990)}
\item{[\rFVD]}{J.\ Fuchs and P.\ van Driel,
Nucl.\ Phys.\ {\bf B 346} (1990) 632}
\item{[\rFGP]}{P.\ Furlan, A.Ch.\ Ganchev and V.B.\ Petkova,
Nucl.\ Phys.\ {\bf B 343} (1990) 205}
\item{[\rWal]}{M.\ Walton,
Phys.\ Lett.\ {\bf B 241} (1990) 365}
\item{[\rGN]}{F.\ Goodman and T.\ Nakanishi,
Phys.\ Lett.\ {\bf B 262} (1991) 259}
\item{[\rOW]}{E.I.\ Ogievetsky and P.B.\ Wiegmann,
Phys.\ Lett.\ {\bf B168} (1986) 360}
\item{[\rOgi]}{E.I.\ Ogievetsky, J.\ Phys.\ G {\bf 12} (1986) L105}
\item{[\rResi]}{N.Yu.\ Reshetikhin,  Zap.\ Nauch.\ Semin.\ LOMI {\bf 169}
(1988) 122}
\item{[\rORW]}{E.I.\ Ogievetsky, N.Yu.\  Reshetikhin and P.B.\ Wiegmann,
Nucl.\ Phys.\ {\bf B280} [FS 18] (1987) 45}
\item{[\rFH]}{W.\ Fulton and J.\ Harris,
 {\sl Representation theory: A first course},
(Springer, Berlin, 1990)}
\item{[\rZel]}{A.V.\ Zelevinskii, Funct.\ Anal.\ Appl.\  {\bf 21}
(1987) 152}
\item{[\rAki]}{K.\ Akin, J.\ Algebra {\bf 117}
(1988) 494}
\item{[\rRes]}{N.Yu.\ Reshetikhin,
 Lett.\ Math.\ Phys.\  {\bf 14} (1987) 235}
\item{[\rABP]}{
G.\ Albertini, B.M.\ McCoy and J.H.H.\ Perk, Phys.\ Lett.\ {\bf A135}
    (1989) 159}
\item{[\rBBP]}{R.J.\ Baxter, V.V.\ Bazhanov and J.H.H.\ Perk,
Int.\ J.\ Mod.\ Phys.\ {\bf B4} (1990) 803}
\item{[\rBaxii]}{R.J.\ Baxter, Phys. Lett. {\bf A146} (1990) 110.}
\item{[\rAMP]}{H.\ Au-Yang, B.M.\ McCoy, J.H.H.\ Perk,
S.\ Tang and M.\ Yan, Phys.\ Lett.\ {\bf A123} (1987) 219}
\item{}{ R.J.\ Baxter, J.H.H.\ Perk and H.\ Au-Yang, Phys. Lett. {\bf
 A128} (1988) 138
}
\item{[\rKili]}{A.N.\ Kirillov, Zap.\ Nauch.\ Semin.\ LOMI {\bf 134}
(1984) 169}
\item{[\rDas]}{S.\ Dasmahapatra, ``String hypothesis and characters of
coset
CFTs'', preprint, hep-th/9305024}
}

\Title{HUTP-93/A022\ \ hep-th/9309137}
 {\vbox{\centerline{Functional Relations
	in Solvable Lattice Models I:}
	\vskip3pt
	\centerline{Functional Relations and Representation Theory}}}
 \centerline{Atsuo Kuniba,$^{1}$
 \footnote{\null}
 {$^1$e-mail: kuniba@math.sci.kyushu-u.ac.jp}
 Tomoki Nakanishi$^{2}$
 \footnote{\null}
 {$^2$e-mail: nakanisi@string.harvard.edu}
 \footnote{\null}{Permanent address:
 Department of Mathematics, Nagoya University,
 Nagoya 464 Japan} and Junji
 Suzuki$^{3}$
 \footnote{\null}
 {$^3$e-mail: jsuzuki@tansei.cc.u-tokyo.ac.jp}}
 \vskip0.5cm
 \bigskip\centerline{
 $^1$Department of Mathematics, Kyushu University,
 Fukuoka 812 JAPAN}
 \bigskip\centerline{
 $^2$Lyman Laboratory of Physics, Harvard University,
 Cambridge, MA 02138 USA}
 \bigskip\centerline{
 $^3$Institute of Physics, University of Tokyo, Komaba,
 Meguro-ku, Tokyo 153 JAPAN}

 \vskip .3in
 Abstract. We study a system of functional relations among
  a commuting family of row-to-row transfer matrices in solvable
 lattice models. The role of exact sequences of the finite dimensional
  quantum group modules is
 clarified. We find a curious phenomenon that the solutions of those functional
 relations also solve the so-called thermodynamic Bethe ansatz
 equations in the high temperature limit for $sl(r+1)$ models.
 Based on this observation, we propose possible functional relations for models
 associated with all the simple Lie algebras.
 We show that these functional relations certainly fulfill strong
 constraints coming from the fusion procedure analysis.
 The application to the calculations of physical quantities will be
 presented in the subsequent publication.

 \Date{9/93}

 \sect{\bf{1. Introduction}}
 \subskip

 Studies of statistical systems on lattice greatly
 concern the spectra of transfer matrices.
 Solvable lattice models in two dimensions
 are especially interesting examples where
 one can actually compute them
 exactly [\rBax].
 At the heart of such a solvability
 resides, of course, the Yang-Baxter equation (YBE), which
 implies the local Boltzmann weights
 (or microscopic interactions) are governed by
 a certain symmetry algebra.
 The relevant algebras are
 quantum groups such as the
 quantized affine algebras $U_q({\hat g})$ [\rDr,\rJim] or their
 rational degenerations, Yangian $Y(g)$ [\rDr,\rKS]
 and their central extensions.
 Then a fundamental question would be:
 how does the symmetry algebra at such a microscopic level
 control the transfer matrix spectra which is macroscopic ?
 \par
 The purpose of the present paper (Part I) and the subsequent one
 (Part II) [\rKNSii] is to develop an approach for obtaining
 the spectra that resolves
 the above issue and possesses the applicability no less wide than
 the Bethe ansatz.
 It is based on the
 {\it functional relations} (FRs) among the row-to-row
 transfer matrices.
 We shall propose a family of
 FRs for the solvable lattice models associated to
 any simple Lie algebra $X_r$.
 It is a systematic extension of the earlier examples for
 $X_r = sl(2)$
 [\rBax, \rBP-\rKLPiii], $sl(n)$ [\rBRii] and $o(n)$ [\rResii] cases.
 The main idea in these works was
 to exploit the FRs
 with a few additional information on the analyticity to
 determine the spectra itself.
 We shall show that this is indeed possible also with our FRs and
 derive correlation lengths of vertex models and
 central charges of critical restricted solid-on-solid (RSOS) models.
 (Part II).
 The calculations generalize the earlier results [\rKSZ--\rKLPiii],
 demonstrating the efficiency of our FRs.
 \par
 The origin of those FRs lies directly in the symmetry algebra and
 its finite dimensional representations (FDRs).
 To sketch it roughly, consider the family of
 solvable vertex models generated from the fundamental rational $R$-matrices
 through the fusion procedure [\rKRS].
 The symmetry algebra in this case is a central extension
 $\til{Y}(X_r)$ of
 the Yangian $Y(X_r)$, which is generated by the $L$-operators [\rKS].
 Take any irreducible FDR (IFDR) $W$ and fix one FDR $H$
 of the $\til{Y}(X_r)$.
 Then one builds, by definition, a row-to-row transfer matrix
 $T_W \in \hbox{End} H$
 as the trace of some operator (monodromy matrix) over the auxiliary space $W$
 corresponding to the periodic boundary condition.
 The $T_W$ itself acts on the quantum space $H$,
 and carries the fusion type labeled by $W$.
 The transfer matrices $T_{W_0}, T_{W_1},
 \ldots \in \hbox{End} H$ so obtained for various IFDRs
 are commutative
 due to the fusion $R$-matrix
 $\in \hbox{Hom} (W_i \otimes W_j,W_j \otimes W_i)$.
 Now consider the product $T_{W_0} T_{W_1} \in \hbox{End} H$,
 which by construction is again a
 trace of a certain operator over $W_0 \otimes W_1$.
 Thus, if there exists an exact sequence of the $\til{Y}(X_r)$-modules, say,
 $$
 0 \rightarrow W_0 \otimes W_1 \rightarrow W_2 \otimes W_3
 \rightarrow W_4 \otimes W_5 \rightarrow 0,\eqno(1.1)
 $$
 connected by $\til{Y}(X_r)$-homomorphisms, it follows that the FR
 $$0 = T_{W_0} T_{W_1} - T_{W_2} T_{W_3} + T_{W_4} T_{W_5}
 \in \hbox{End} H \eqno(1.2)$$
 must hold for any choice of $H$.
 By regarding $T_{W_i}$'s as the eigenvalues,
 (1.2) exhibits how the ``microscopic data" (1.1)
 controls the spectra.
 As an example, for $X_r = sl(2)$, the
  FR in [\rKLPiii]
 $$
 T_m(u-{1\over 2})T_m(u+{1\over 2}) = T_{m+1}(u)T_{m-1}(u)
 + g_m(u) {\rm Id}.
 \eqno(1.3)
 $$
 can be reproduced in this manner and
 its representation theoretical background is thereby
 clarified.
 In (1.3),  $u$ denotes the spectral parameter that naturally
 enters IFDRs $W_i$ of $\til{Y}(X_r)$, $T_m(u)$
 stands for the transfer matrix for
 $m$-fold fusion model, and
 $g_m(u)$ is a scalar function
 satisfying $g_m(u+{1\over 2})g_m(u-{1\over 2}) = g_{m+1}(u)g_{m-1}(u)$
 and specified unambiguously from $H$.
 See (2.14) for the explicit formula.
 We shall call the FR among the commuting family of the transfer
 matrices like (1.3) as the $T$-{\it system}
 in this paper.
 To fully work out such $T$-system however requires the knowledge as (1.1)
 on a family of IFDRs of the symmetry algebra and
 we have only been able to derive it rigorously for
 $X_r = A_r$ so far.
 Thus, it appeared a challenge to
 proceed further to postulate a general form of
 the $T$-system for all the classical simple Lie algebras $X_r$.
 Nevertheless, we seem to manage it
 by a blend of ideas from
 another ingredient, namely,
 the thermodynamic Bethe ansatz (TBA) as we shall see below.
 \par
 The subject has a long history going back to [\rYY] but there was
 a renewed interest recently
 both in the lattice model context [\rKRjp,\rBRi,\rBRii,\rKu]
 and the perturbed conformal field theory (CFT) context
 [\rAlbZam--\rAlbZamii]
  as
 some fascinating structures were emerging.
 Let us recall two of them here
 exclusively for the case $X_r = sl(2)$ for simplicity.
 The first structure is the universal form of the
 so called TBA equation in the high temperature limit [\rAlbZamii--\rRav]:
 $$Y_m(u-{1\over 2})Y_m(u+{1\over 2}) = (1 + Y_{m+1}(u))(1 + Y_{m-1}(u)),
 \eqno(1.4)
 $$
 where the function $Y_m(u)$
 corresponds to the ratio of $m$-string and hole density functions
 in the context of lattice models [\rKu,\rKN].
 We call systems such as (1.4) the $Y$-{\it system}
 borrowing the naming in [\rRav].
 To see the second structure, recall the
 dilogarithm identity [\rKRjp,\rKil]
 $$
 {6 \over \pi^2} \sum_{m=1}^\ell L({1 \over Q_m^2}) =
 {3\ell \over \ell + 2}, \eqno(1.5)
 $$
 where $L(\cdot)$ is the Rogers dilogarithm and
 $\ell$ is any positive integer.
 The rhs is the well known central charge
 of the level $\ell$ ${\widehat{sl}}(2)$ WZW model [\rKZ].
 The quantity $Q_m$ is given by
 $$Q_m = {\sin{(m+1)\pi\over \ell + 2}\over \sin{\pi\over \ell + 2}},
 \eqno(1.6)$$
 which is the character of the
 $m+1$-dimensional representation
 of $sl(2)$  specialized at a rational point.
 The second structure relevant for us is the following
 character identity among them [\rKil,\rKRii]:
 $$Q_m^2 = Q_{m+1}Q_{m-1} + 1.\eqno(1.7)
 $$
 We call such relations among the characters as the $Q$-{\it system}.
 Having these structures at hand, we make the crucial
 observation;
 the $T$-system (1.3) is the Yang-Baxterization (i.e., spectral
 parameter dependent version)
 of the $Q$-system (1.7) such that
 the combination
 $y_m(u) = T_{m+1}(u)T_{m-1}(u)/g_m(u)$
 solves the $Y$-system (1.4) as follows:
$$\eqalignno{y_m(u-{1\over 2})y_m(u+{1\over 2})&=
 {T_{m+1}(u-{1\over2})T_{m+1}(u+{1\over2})T_{m-1}(u-{1\over2})
  T_{m-1}(u+{1\over2})\over g_m(u-{1\over2})g_m(u+{1\over2})}\cr
&={(T_{m+2}(u)T_m(u) + g_{m+1}(u))
  (T_{m-2}(u)T_m(u) + g_{m-1}(u))\over
  g_m(u-{1\over 2})g_m(u+{1\over2})}\cr
& = (y_{m+1}(u) + 1)(y_{m-1}(u) + 1).&(1.8)\cr}
$$
 %
 %
Though the meaning of this phenomenon is yet to be
clarified, it opens a  route
 to guess a general form of the $T$-system.
 Namely, ``Yang-Baxterize the $Q$-system
 so as to solve the $Y$-system", which can be tested because
 these systems are now explicitly known
 for all  $X_r$'s in
 [\rKRii] and [\rKN], respectively.
 See section 3 and Appendix B for details.
 We have found that there certainly exist such a
 Yang-Baxterization as listed in (3.20).
 Moreover, it is essentially unique
 for each $X_r$ up to some freedom responsible for
 the arbitrariness of choosing $H$ and  elementary
 redefinitions of the transfer matrices.
 \par
 We must then ask whether the $T$-system so obtained
  fits the representation theoretical scheme as
(1.1-3).  It turns out that the existence of  exact sequences  like (1.1)
 imposes strong constraints for a possible form
 of the $T$-system. Based on reasonable assumptions on the fusion
 procedure, we will show that our solution (3.20)
  indeed satisfies
 those constraints.
 \par
 Supported by these backgrounds we propose them
 as the $T$-system for  $X_r$ (main result of Part I).
 Its unrestricted and restricted versions (see section 2)
 are to hold for the vertex and the
 RSOS type solvable models, respectively.
 Part II will be devoted to further
 studies where we apply the $T$-system
 to those models and recover
 the known results or even generalize them.
 It is our hope that further studies
 unveil a deep structure
 connecting $T$, $Q$ and $Y$-systems
 and lead to a greater understanding of
 the related subjects,
 representation theories of quantum groups,
 dilogarithm identities, TBA, analytic Bethe ansatz and so forth.
 \par
 The outline of Part I is as follows.
 In section 2, we review the simplest example of
 the transfer matrix FRs for $sl(2)$ related models,
 and explain the connection to the exact sequences of
 the quantum group.
 We also include a proof of the $T$-system for $X_r = A_r$
 and thereby establish the same relation among  $Q$, $T$ and $Y$-systems
 as the $sl(2)$ case discussed above.
 This turns out to be a simple exercise using
 the FRs in [\rBRii].
 In section 3,
 we propose the $T$-system for every $X_r$ from the condition
that it solves the  $Y$-system as explained above.
 In section 4, we describe how our $T$-system
 is consistent with the representation theoretical
 viewpoint. Due to the lack of the knowledge of the
 fusion procedure, however, we leave the justification
 of the cases $E_7$, $E_8$ as
 a future problem.
 Section 5  presents interesting conjectures
 on certain determinant formulas
 in our $T$-system for $X_r = B_r$, $C_r$ and $D_r$.
 %
 %
 In section 6, we give a summary and discussion.
 Appendices A and B provide additional information
 on the $Q$ and $Y$-systems, respectively.
 The former contains a few new results on the special values of the
 quantity $Q^{(a)}_m$.
 \par
 Before closing the introduction,
 let us include one remark.
 Most of our arguments will be given
 upon the Yangian symmetry $Y(X_r)$ hence
 they are relevant to the models associated with rational
 $R$ matrices.
 However for $X_r = A_r$, they are also valid in
 trigonometric and even elliptic cases.
 We suppose this to be true for all the other $X_r$'s as well.
 Despite the absence of quasi-classical elliptic $R$-matrices
 for $X_r \neq A_r$ [\rBD],
 the elliptic SOS model will still exist [\rFR,\rDJO]
 in general, on which we expect our $T$-system.

 \sect{\bf{2. Functional Relations of Transfer Matrices}}\pn
 \subskip
 This is a warm-up section for
 getting familiar with the
 FRs and some background ideas.
 We will exclusively take examples from solvable models
 related to $A_r$.

 \subsect{2.1. A simplest example}
 \subskip

 We begin by reviewing an extremely simple case of
 the McGuire-Yang $R$-matrix [\rMcG,\rYa]
 $${\check R}(u) = 1 + uP \in {\rm End}(W\otimes W),
 \quad W = {\bf C}^2,\eqno(2.1)$$
 where $u$ is the spectral parameter and $P$ is the transposition
 operator $P(x\otimes y) = y \otimes x$.
 It satisfies the YBE
 $$\bigl(1\otimes {\check R}(u)\bigr)
 \bigl({\check R}(u+v)\otimes 1 \bigr)
 \bigl(1\otimes {\check R}(v)\bigr)  =
 \bigl({\check R}(v)\otimes 1 \bigr)
 \bigl(1\otimes {\check R}(u+v)\bigr)
 \bigl({\check R}(u)\otimes 1 \bigr),\eqno(2.2)
 $$
 and defines a solvable 2-state model (a rational limit of the 6-vertex
 model) on a 2D square lattice in the usual way.
 We depict the ${\check R}(u)$
 graphically as
 %
 $$
\epsfxsize=33.8pt\epsfysize=30.1pt \epsfbox{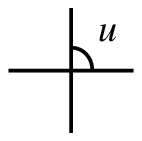}
 \quad,
 $$
 where the assigned little arc is also
 specifying that it acts from the
 SW to the NE direction.
 The $R$-matrix (2.1) has the spectral decomposition
 $${\check R}(u) = (1+u)P_2 \oplus (1-u)P_0,\eqno(2.3)
 $$
 where $P_{1\pm 1}$ are the projectors
 ${1\over 2}{\check R}(\pm 1)$ onto the spaces of
 symmetric and antisymmetric tensors.
 Regarding $W$ as the simplest non-trivial IFDR $W_1$ of $sl(2)$,
 we put $W_{1 \pm 1} = {\rm Im} P_{1 \pm 1}$.
 Thus
 $W_1 \otimes W_1 = W_2 \oplus W_0$ as the $sl(2)$-module
 (${\rm dim}W_m = m+1$) and ${\check R}(u)$ reduces to the projectors
 on the irreducible components at its singularities $u = \pm 1$.
 \par
 Consider the
 (homogeneous) transfer matrix $T_1(u) \in {\rm End}(W^{\otimes N})$
 defined graphically as
 $$
\epsfxsize=183.6pt\epsfysize=63.9pt \epsfbox{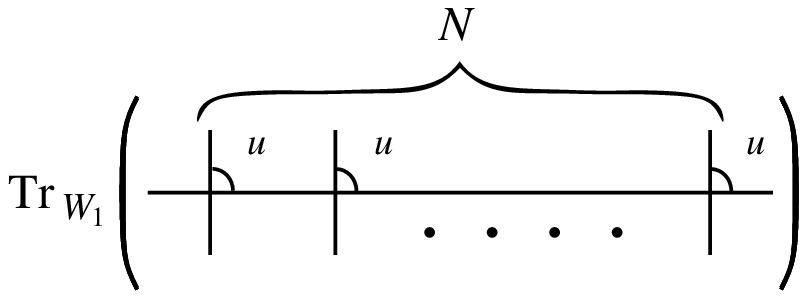}
 \quad,
 $$
 %
 %
where the sum is implied for each internal horizontal
edge as usual. The object inside the trace
 is the so called monodromy matrix:
 $W_1 \otimes W^{\otimes N} \rightarrow
 W^{\otimes N} \otimes W_1$ and
 the trace is over the auxiliary space
 $W_1 = W$ to account for
 the periodic boundary condition.
 The suffix 1 in $T_1(u)$ is to signify this fact,
 namely, its {\it fusion degree} is 1.
 The $R$-matrix ${\check R}(u-v)$ assures the commutativity
 $[T_1(u), T_1(v)]=0$ as is well known [\rBax].
 Take the product $T_1(u-{1\over 2})T_1(u+{1\over 2})$
 corresponding to the picture
 $$
\epsfxsize=203pt \epsfysize=58.7pt \epsfbox{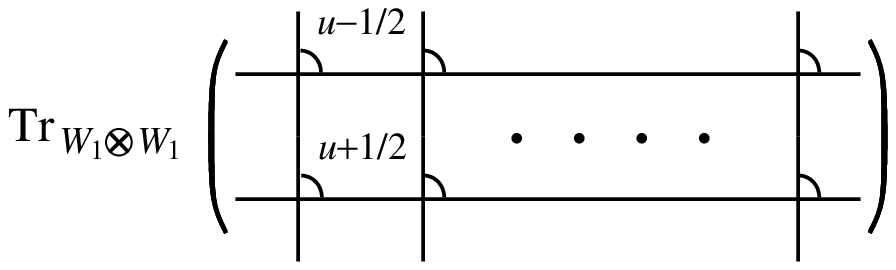}
 \quad,
 $$
 %
 %
 which is a trace over $W_1 \otimes W_1$.
 By inserting the operator ${\rm Id}_{W_1 \otimes W_1} =
 {1\over 2}{\check R}(1) + {1\over 2}{\check R}(-1)$
 at the left end of the monodromy matrix,
 this is equal to
 $$
\epsfxsize=258.9pt \epsfysize=126.4pt \epsfbox{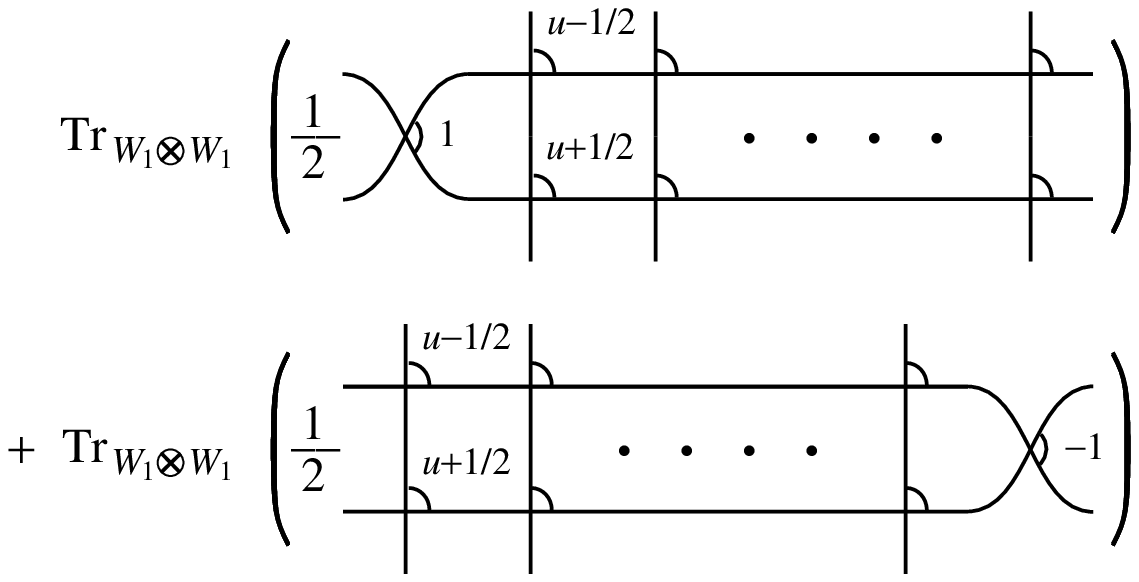}
 \quad,
 $$
 %
 %
 due to the cyclicity of the trace.
 With the aid of the YBE (2.1), the inserted pieces can now be
 slid to any position.
By  remembering that they are the projectors onto $W_{1 \pm 1}$,
 the rhs can be written as
 ${\rm Tr}_{W_2}(\cdots) + {\rm Tr}_{W_0}(\cdots)$.
 Obviously the two terms are the transfer matrices
 with fusion degrees 2 and 0, with the latter being
 a scalar matrix.
 Thus we get
 $$T_1(u+{1\over 2})T_1(u-{1\over 2}) = T_2(u)
 + (u-{1\over 2})^N(u+{3\over 2})^N {\rm Id},
 \eqno(2.4)
 $$
 where the last factor is
 an easy exercise.
 The commutativity $[T_m(u), T_{m^\prime}(v)] = 0$
 again holds owing to the fusion
 $R$-matrices [\rKRS]. See Fig.\ 1.
\par
\centerline{
\epsfxsize=295.5pt \epsfysize=83.55pt \epsfbox{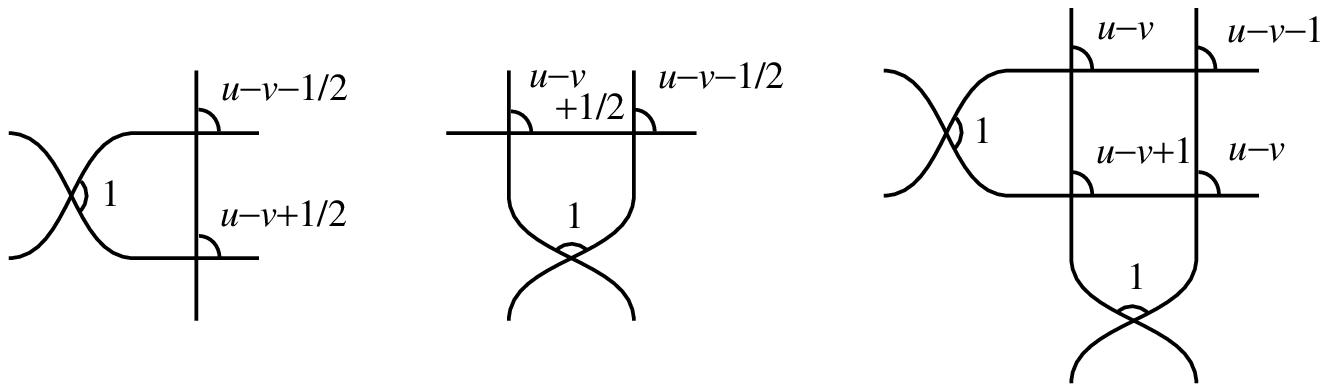}
}
\noindent
Figure 1. The fusion $R$-matrices
of the fusion degrees
 $(m,m')=(2,1),(1,2),(2,2)$  (from
 the left to the right)  along with the projectors.
\subskip
\par
 Eq.(2.4) is the simplest example of the (rational limit of) FR
 due to many authors [\rBP--\rBRi].
 There are already a lot of lessons gained from it.
 (i) The decomposition (2.4) looks
 analogous to the classical Clebsh-Gordan rule
 $W_1 \otimes W_1 = W_2 \oplus W_0$ but carries
 the spectral parameter $u$ non-trivially.
 This is a typical situation showing
 that the problem is in fact relevant to  FDRs of
 the Yangian $Y(sl(2))$
 and its central extension
  rather than $sl(2)$ itself.
 We will come to this point more later.
 (ii) The shift $u_0 = \pm 1$ in
 $T_1(u\mp{1\over 2})T_1(u\mp{1\over 2}+u_0)$ had to be
 a singularity of the $R$-matrix
 ${\rm det}{\check R}(u_0) = 0$ where it reduces to
 projectors onto invariant subspaces in $W_1 \otimes W_1$.
 Nothing interesting would have happened for generic shifts $u_0$.
 (iii) The essential structure of the FR is governed by
 the fusion procedure for the $R$-matrix.
 It uses the singularities at $u=\pm u_0$ and automatically guarantees
 $[T_m(u), T_{m^\prime}(v)] = 0$.
 (iv) The fusion in question concerns the $R$-matrix
 acting on the auxiliary spaces $W_1, W_2$, etc and
 {\it not} the quantum space $W^{\otimes N}$
 which enters (2.4) rather trivially only through the factor
 $(u-{1\over 2})^N(u+{3\over 2})^N$.
 \par
 One may proceed further to directly work out the
 FRs for the transfer matrices with higher fusion degrees.
 Such calculations are indeed possible [\rKRjp,\rBRi,\rKLPiii]
 leading to the forthcoming result (2.14) we want.
 However we will seek, in the next subsection, a more intrinsic
 understanding from the
 representation theoretical  viewpoint of our symmetry algebra,
 Yangian $Y(sl(2))$ and its central extension.
 As it turns out, it provides a natural framework
 in which all the observations
 (i)--(iv) can be understood most elegantly.
 (Parallel argument holds for $U_q(\widehat{sl}(2))$ hence for
 the trigonometric and even elliptic cases.)

 \subsect{2.2. Representation theoretical viewpoint}
 \subskip

\noindent {\it Yangian and its central extension}

 The Yangian $Y(X_r)$ is an important class of quantum groups [\rDr]
 such that to every its finite dimensional representation
 is associated a rational solution of the YBE (2.1).
 It is generated by the elements $\{x, J(x) \mid x \in X_r \}$
 under certain commutation relations and contains
 the universal enveloping algebra $U(X_r)$ as a subalgebra.
 Let us recall a few results [\rDr,\rTar,\rCPi]
 on its FDR theory for $X_r = sl(2)$, which are related to our FRs.
 \par\vskip0.4cm\noindent
 {\bf Fact 1} ([\rCPi] Proposition 2.5).
 There is an algebra homomorphism
 $\phi_u : Y(sl(2)) \rightarrow U(sl(2))$
 depending on a complex parameter $u$ such that
 $\phi_u(x) = x$, $\phi_u(J(x)) = ux$.
 \par\vskip0.4cm
 Suppose $\rho$ is a FDR of $sl(2)$, i.e.,
 $\rho: U(sl(2)) \rightarrow {\rm End} W$.
 {}From the Fact 1, one can then form a one parameter
 family of FDRs of $Y(sl(2))$ via the composition
 $$Y(sl(2)) \buildrel\phi_u\over\rightarrow U(sl(2))
 \buildrel\rho\over\rightarrow {\rm End} W.
 \eqno(2.5)$$
 The resulting $Y(sl(2))$-module will be denoted
 by $W(u)$ and called the evaluation representation.
 Let $W_m$ be the $(m+1)$-dimensional
 irreducible $sl(2)$-module and $W_m(u)$ be the
 associated evaluation representation which is also irreducible.
 Remarkably,
 every IFDR of $Y(sl(2))$ is known to be isomorphic to
 $W_{m_1}(u_1)\otimes \cdots \otimes W_{m_k}(u_k)$
 for some choice of
 $m_i \in {\bf Z}_{\ge 0}$ and $u_i \in {\bf C}$ [\rCPi].
 Note however that this statement is not asserting that
 such tensor products are always irreducible.
 In fact one has
 \par\vskip0.4cm\noindent
 {\bf Fact 2} ([\rCPi] Corollary 4.7, Proposition 4.2, Section 5).
 The tensor product $W_m(u)\otimes W_{m^\prime}(v)$ is
 reducible if and only if
 $$\vert u-v \vert = {m+m^\prime\over 2} - j + 1\eqno(2.6)$$
 for some $1 \le j \le {\rm min}(m,m^\prime)$.
 For all the other values of $u$ and $v$ than (2.6),
 the IFDRs $W_m(u)\otimes W_{m^\prime}(v)$ and
 $W_{m^\prime}(v)\otimes W_m(u)$ are isomorphic.
 The isomorphism is given by the unique intertwiner
 $\Rc_{m,m^\prime}(u-v)
 \in {\rm Hom}_{Y(sl(2))}(W_m(u)\otimes W_{m^\prime}(v),
 W_{m^\prime}(v)\otimes W_m(u))$ which
 agrees with the rational fusion $R$-matrices in [\rKRS].
 \par\vskip0.4cm
 \par
 There is a precise description of the
 reducibility content in the case (2.6)
 as follows.
 \par\vskip0.4cm\noindent
 {\bf Fact 3} ([\rCPi] Proposition 4.5).
 When (2.6),
 there are following exact sequences of the $Y(sl(2))$-modules:
 $$\eqalign{0 \rightarrow
 & W_{m-j}(u+{j\over 2})\otimes W_{m^\prime-j}(v-{j \over 2})
 \rightarrow
 W_m(u)\otimes W_{m^\prime}(v)\cr
 &\qquad\rightarrow W_{j-1}(u-{m-j+1\over 2})\otimes
 W_{m+m^\prime -j+1}(v+{m-j+1\over 2})
 \rightarrow 0\cr}\eqno(2.7{\rm a})
 $$
 for $u-v = {m+m^\prime\over 2} - j + 1$.
 $$\eqalign{0 \rightarrow
 &W_{j-1}(u+{m-j+1\over 2})\otimes
 W_{m+m^\prime -j+1}(v-{m-j+1\over 2})
 \rightarrow
 W_m(u)\otimes W_{m^\prime}(v)\cr
 &\qquad \rightarrow
 W_{m-j}(u-{j\over 2})\otimes W_{m^\prime-j}(v+{j\over 2})
 \rightarrow 0\cr}\eqno(2.7{\rm b})
 $$
 for $v-u = {m+m^\prime\over 2} - j + 1$.
 \par\vskip0.4cm
 The first spaces of these sequences are the
 proper subrepresentations of
 $W_m(u)\otimes W_{m^\prime}(v)$ and
 (2.7) may be viewed as a quantum version
 (or Yang-Baxterization) of the classical
 Clebsh-Gordan rule.
 Note however that
 the reducible space $W_m(u)\otimes W_{m^\prime}(v)$
 is no longer decomposable here in general, which is a
 distinct feature from the
 classical case.

To discuss the FRs, we actually need a central extension
       ${\tilde Y}$ of $Y(sl(2))$ [\rKS], which is most naturally
       seen in the alternative (and original) realization of
       the latter using $L$-operators.
 See the lecture note [\rSk] and the references therein for further
 information.
 Let  ${\cal L}_{ij}^{(k)}$ be  noncommutative elements,
 $(1\leq i,j\leq 2, k\geq 1)$, and let
 $$
 \eqalign{
 {\cal L}_{ij}(u)&=
\delta_{ij} +\sum_{k \geq 1} {\cal L}_{ij}^{(k)} u^{-k}, \cr
{\cal L}(u)&=\sum_{i,j=1}^2 {\cal L}_{ij}(u)E_{ij}, \cr
 }
 \eqno(2.8)
 $$
 where
 $E_{ij} \in \End W$ is $(E_{ij})_{k\ell}=\delta_{ik}\delta_{j\ell}$
 for a fixed basis of $W$.
 Consider  the associative noncommutative algebra  $\til{Y}$
 generated by the elements ${\cal L}_{ij}^{(k)}$ with the following
 relations
 $$
 \Rc(u)({\cal L}(u+v)\tensor 1)(1 \tensor {\cal L}(v))=
 ({\cal L}(v)\tensor 1)(1 \tensor {\cal L}(u+v))\Rc(u).
 \eqno(2.9)
 $$
 Then,
\par\vskip0.4cm\noindent
 {\bf Fact 4.}
 $\til{Y}$ is a Hopf algebra in a weak sense without the
 antipode.
There is a central element $\qd$ in $\til{Y}$ called
the {\it quantum determinant}.
Then,  $Y(sl(2))$ is isomorphic to the quotient of $\til{Y}$ by
 the ideal generated by $\qd - 1$  [\rDr].
  It is possible to construct all
 the IFDR of $\til{Y}$ using the elementary
 $R$-matrix (2.1) as a building block [\rTar].

 \par\vskip0.4cm

 Any IFDR of $\til{Y}$ is naturally viewed as
 an IFDR of $Y(sl(2))$. On the other hand, a lift of
 a given IFDR of $Y(sl(2))$ to  an IFDR of $\til{Y}$ is always
 possible, though not  unique.
 Thus, it will be  possible to lift the exact sequences (2.7) to
 exact sequences of $\til{Y}$-modules which are described
 completely
 in terms of the elementary $R$-matrix.
 The FRs are consequences of the latter exact sequences.
 Below we shall explain this lifting procedure in detail.

\subskip

\noindent {\it IFDRs of $\til{Y}$}

\par
To describe IFDRs of $\til{Y}$, it
 is convenient to assign a spectral parameter to
     each line of the $R$-matrix (2.1) which determines
     its argument by the following rule:
 $$
\epsfxsize=68.5pt \epsfysize=42.9pt \epsfbox{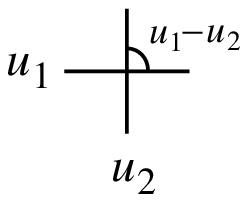}
 \quad.
 $$
 We introduce the IFDR of  $\til{Y}$ on the space
 $W_m \subset W^{\tensor m}$ [\rKRS]. Namely, the projection operator
 $ W^{\tensor m} \lra W_m $ is given by the product
 of the $R$-matrix as
 $$
\epsfxsize=258.9pt  \epsfysize=67.0pt \epsfbox{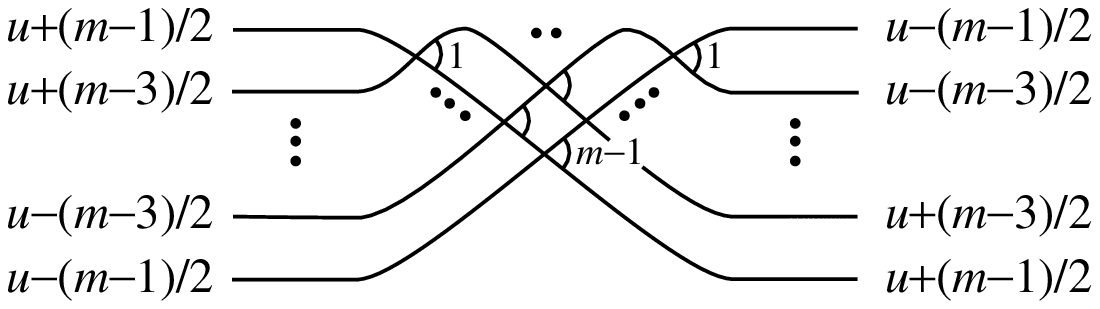}
 \quad.
 $$
 %
 %
 We remind the reader that operators here are considered
to act from the left
 to the right space.
The action (representation) of the generator ${\cal L}_{ij}(v)$
 (see (2.8)) is defined by the action of the operator
 $$
\epsfxsize=93.345pt \epsfysize=94.1pt \epsfbox{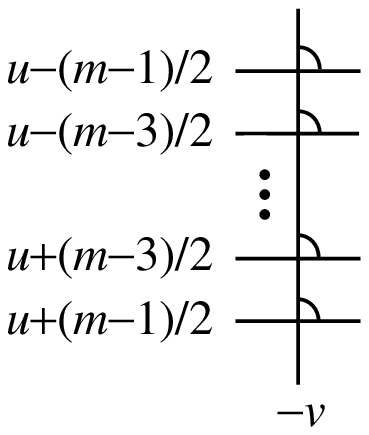}
 \quad
 $$
 on $W_m$,
 where the top and the bottom states  are fixed as  basis vectors of $W$
 corresponding to $i$ and $j$.\footnote{$^1$}{This
 defines only  the
 $\til{Y}$-action on the auxiliary space. The $\til{Y}$-action
 on the quantum space is defined similarly by reflecting
 the above diagrams along the SW-NE diagonal line.}
 Since this IFDR is isomorphic to $W_m(u)$ as a $Y(sl(2))$ representation,
we
 also write it  as $W_m(u)$.
 Keeping these definitions in mind, we also use the graphical abbreviation of
 the fusion $R$-matrix $\Rc_{m,m^\prime}(u-v)$ as
 $$
\epsfxsize=82.0pt \epsfysize=42.1pt \epsfbox{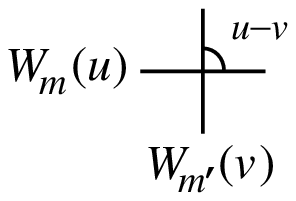}
 \quad.
 $$
 %
 %
 The tensor product  $W_m(u) \tensor W_{m'}(v)$ of IFDRs
 is naturally defined from (2.9) by aligning
 the spaces $W_m(u)$ and  $W_{m'}(v)$ from the top to the bottom
 (resp.\ from the right to the left) for the  $\til{Y}$-action
 on the auxiliary (resp.\ quantum) space.

\par
 Set the quantum space as
 $H = W_{m_1}(v_1)\otimes \cdots \otimes W_{m_N}(v_N)$,
 where the choice of $m_i$'s and $v_i$'s is arbitrary.
 It is natural to define the
 transfer matrix $T_m(u) \in {\rm End} H$
 with fusion degree $m$
 by $T_m(u) = {\rm Tr}_{W_m(u)} L$,
 where the monodromy matrix
 $L: W_m(u)\otimes H \rightarrow H\otimes W_m(u)$
 is defined by
 $$
\epsfxsize=293.5pt \epsfysize=45.1pt\epsfbox{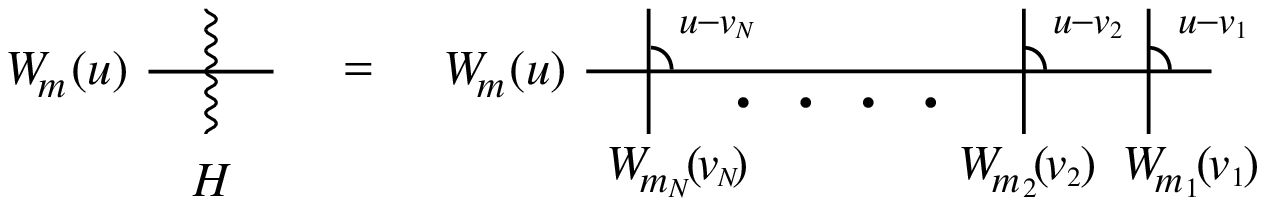}
\quad .
 $$
 %
 %
 We have abbreviated the quantum space $H$ by a single
 wavy line.
 The commutativity
 $[T_m(u), T_{m^\prime}(v)] = 0$ is valid
 owing to $\Rc_{m,m'}(u)$.
 The situation (2.6) corresponds to a
 singularity of the intertwiner $\Rc_{m,m^\prime}(u-v)$
 hence it is intimately connected to our FRs
 as noted in the lessons (ii) and (iii) previously.

\subskip

\noindent {\it Exact sequences of ${\tilde Y}$-modules}

\par
 We consider a lift of the exact sequence (2.7b) to
 an exact sequence of $\til{Y}$-modules  in the
  case $m^\prime = j = m$ in the following way:
 $$\eqalign{
 0 &\longrightarrow {\cal A}_1
 \buildrel\varepsilon_1\over\longrightarrow
 {\cal A}_0 \buildrel\varepsilon_0\over\longrightarrow
 {\cal A}_{-1} \longrightarrow 0,\cr
 {\cal A}_1 &=
 W_{m-1}(u)\otimes W_{m+1}(u),\cr
 {\cal A}_0 &=
 W_m(u-{1\over 2})\otimes W_m(u+{1\over 2}).\cr
 }\eqno(2.10)
 $$
 One must be careful for the third space ${\cal A}_{-1}$.
 As a $Y(sl(2))$ representation it is trivial, but
 as a $\til{Y}$ representation
  it is not necessary so because the central element
 $\qd$ may act non-trivially.
 In order to define ${\cal A}_{-1}$, it is necessary to introduce an additional
 IFDR $\til{W}_0(u)$
  of $\til{Y}$ on the antisymmetric subspace in $W^{\tensor 2}$.
As a $Y(sl(2))$ representation,  $\til{W}_0(u)$ is trivial.
 The  projection operator for $\til{W}_0(u)$,
 together with the specification of the spectral
 parameters on the lines,  is given by
 $$
\epsfxsize=78.2pt \epsfysize=39.8pt \epsfbox{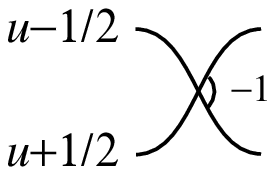}
\quad .
 $$
 Then, ${\cal A}_{-1}$ is defined by
 $$
 {\cal A}_{-1} =\til{W}_0(u-{m-1 \over 2})\otimes
 \til{W}_0(u-{m-3 \over 2})\otimes \cdots
 \tensor \til{W}_0(u+{m-1\over 2}).
 \eqno(2.11)
 $$
 The homomorphisms $\varepsilon_i$'s are given by
 successive multiplications of the $R$-matrices.
 The following example in the case $m=3$
$$
\epsfxsize=292.8pt \epsfysize=91.1pt \epsfbox{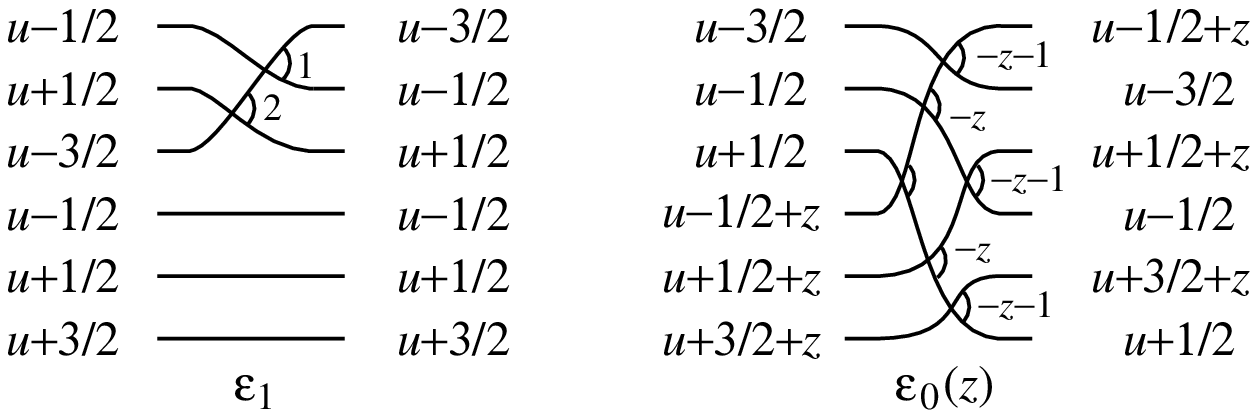}
 $$
 %
 clarifies the construction of $\varepsilon$'s for a general $m$ as well,
 where in the  figure for $\varepsilon_1$  straight lines means the
 identity operators on the corresponding tensor components, and $\varepsilon_0$
 is defined as the leading coefficient of the operator
 $\varepsilon_0(z)$ in
 the expansion $\varepsilon_0(z)=z^{m-1}\varepsilon_0 + o(z^{m-1})$.
 (See Lemma A.1 and Proposition A.2 of [\rJKMO], and  also [\rChe].)

 \par
 Let us check that the above constructions of $\varepsilon_i$'s
 indeed have desired properties.
 First, the fact that $\varepsilon_i$'s are $\til{Y}$-homomorphisms
(i.e., commute with ${\tilde Y}$-actions)
 is a consequence of the YBE. For example, for $\varepsilon_1$ the
 statement is equivalent to the following graphical equality:
$$
\epsfxsize=325.2pt \epsfysize=69.2pt \epsfbox{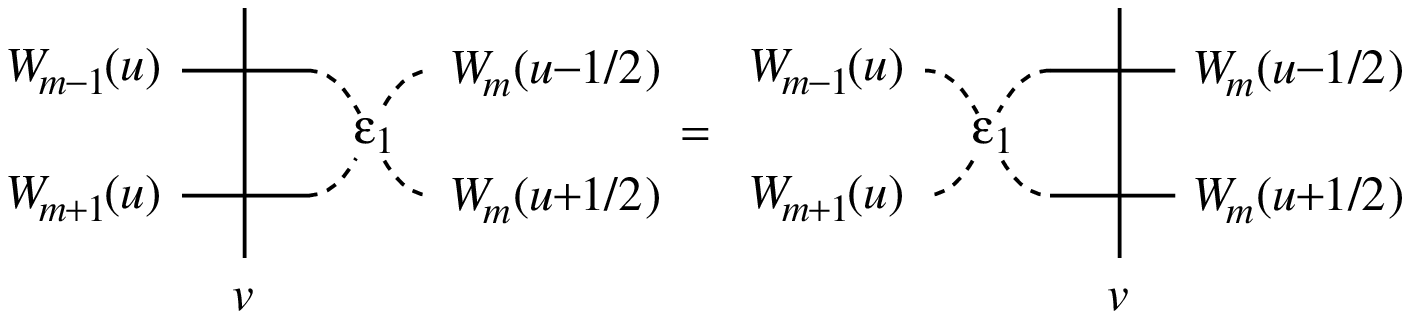}
 \quad.
 $$
 %
 %
 Second, to see the exactness we first note that
 ${\cal A}_1$ and  ${\cal A}_{-1}$ are irreducible by (2.6).
 Thus, by Schur's lemma, if  $\varepsilon_i$'s are nonvanishing,
 then  they should coincide with
 those in (2.7b) as  $Y(sl(2))$-homomorphisms,
 from which the exactness follows.
 To check the nonvanishment is straightforward.
 Thus, we conclude that (2.10--11) is a $\til{Y}$ lift of (2.7b).

\subskip

\noindent {\it Functional relations}

\par
 To derive the FR from (2.10),
 we introduce the composite monodromy matrices
 $L^{(i)}, i=0, \pm 1$ defined by the figures
$$
\epsfxsize=348.5pt \epsfysize=70.0pt \epsfbox{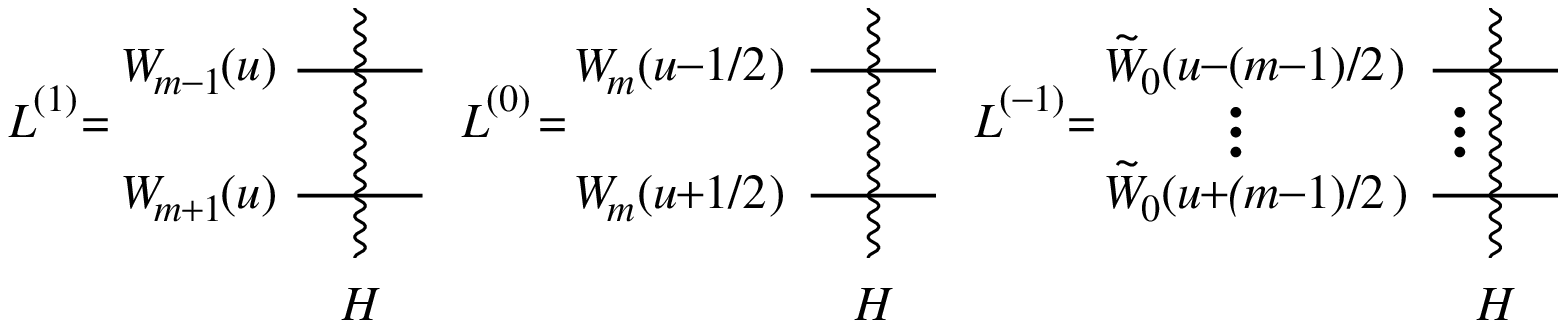}
 ,
$$
 %
 %
 where the arrangement of the horizontal lines
 for $L^{(i)}$
 corresponds to the auxiliary space
 ${\cal A}_i$ in (2.10--11) hence
 $L^{(i)}: {\cal A}_i \otimes H
 \rightarrow H \otimes{\cal A}_i$.
 Next we fix a basis $\{\alpha, \beta, \ldots\}$ of $H$
 and consider the operators
 $ L^{(i)}_{\alpha \beta} \in {\rm End} {\cal A}_i$.
 For fixed $\alpha$ and $\beta$,
 $L^{(i)}_{\alpha \beta}$ are the
 representations of a common element of $\til{Y}$
 on
 each space ${\cal A}_i$. Therefore,
 the exact sequence (2.10) of $\til{Y}$-modules
 implies the  commutative diagram
 $$
 \matrix{
 0 & \longrightarrow & {\cal A}_1\quad
   & \buildrel\varepsilon_1\over\longrightarrow & {\cal A}_0\quad
   & \buildrel\varepsilon_0\over\longrightarrow & {\cal A}_{-1}\quad
   & \longrightarrow  &0\cr
 & & \,\,\downarrow  L_{\alpha \beta}^{(1)} &
   & \,\,\downarrow  L_{\alpha \beta}^{(0)} &
   & \,\,\downarrow  L_{\alpha \beta}^{(-1)} & &\cr
 0 & \longrightarrow & {\cal A}_1\quad
   & \buildrel\varepsilon_1\over\longrightarrow & {\cal A}_0\quad
   & \buildrel\varepsilon_0\over\longrightarrow & {\cal A}_{-1}\quad
   & \longrightarrow  &0\cr
 }
 \eqno(2.12)
 $$
 {}From  the exactness of (2.12) it follows that
 $$0 = \sum_{i=0,\pm 1} (-)^i {\rm Tr}_{{\cal A}_i}
 \bigl(L^{(i)}_{\alpha \beta}\bigr).\eqno(2.13)
 $$
 This is nothing but the
 $(\alpha, \beta)$-matrix element of the FR $(m \geq 1)$:
 $$\eqalignno{
 0 &= T_{m-1}(u)T_{m+1}(u) - T_m(u-{1\over 2})T_m(u+{1\over 2})
 + g_m(u){\rm Id},&(2.14{\rm a})\cr
 g_m(u) &=
 {\rm Tr}_{{\cal A}_{-1}}
 (L^{(-1)}_{\alpha \alpha})
 \quad (\hbox{ for any } \alpha)&(2.14{\rm b})\cr
 &=\prod_{k=1}^N
 \Bigl(\prod_{i=0}^{m_k-1}\prod_{j=0}^{m-1}
 g(u_j-v_{k,i})\Bigr),
 &(2.14{\rm c})\cr
 g(u) &= (u-{1 \over 2})(u+{3\over 2}),&(2.14{\rm d})\cr
 u_j &= u - {m-1\over 2} + j,\quad
 v_{k,i} = v_k - {m_k-1\over 2} + i,&(2.14{\rm e})\cr}
 $$
 where we define $T_0(u) = 1$.
 This is the $T$-system we have mentioned in (1.3), which
 generalizes (2.4).
 Note that from (2.14b) and (2.11) one has
 $$g_m(u+{1\over 2})g_m(u-{1\over 2}) = g_{m+1}(u)g_{m-1}(u)
 \eqno(2.15)$$
 irrespective of the choice of the quantum space $H$.
 \par
 Although (2.14) is derived here for
 the rational vertex models, it is equally valid
 for the trigonometric and even
 the elliptic vertex models [\rCheii]
 if $g(u)$ in (2.14d) is replaced by its analogues in
 those functions.
 This is because
 the $R$-matrices in the trigonometric and elliptic
 cases still have the common features in
 their singularities and fusion procedures so that
 analogous construction of the sequences like (2.12) remains
 valid.
 For the trigonometric case,
 a parallel result to (2.7) for
 $U_q(\widehat{sl}(2))$ is also known as Proposition 4.9 in [\rCPii].
 \par
 The trigonometric and elliptic versions of the
 $T$-system (2.14) also apply
 to the critical and off-critical
 fusion RSOS models [\rDJKMONP], respectively.
 In the trigonometric case,
 this follows in principle from the relation
 of these models to the
 $U_q(\widehat{sl}(2))$
 $R$-matrix at $q = \exp({2\pi i\over \ell + 2})$ [\rPas].
 Here $\ell \in {\bf Z}_{\ge 1}$ denotes the level
 which is a characteristic label of the RSOS models.
 There is however one fundamental difference between the
 $T$-systems for the vertex and
 the level $\ell$ RSOS model.
 In the former, the index $m$ in (2.14) runs over
 the whole set of positive integers, while in the latter,
 the truncation
 $$T_{\ell+1}(u) \equiv 0\eqno(2.16)$$
 occurs hence (2.14) closes among finitely
 many $T_m(u)$'s with $1 \le m \le \ell$.
 We shall call these $T$-systems
 the unrestricted and (level $\ell$) restricted
 $T$-systems accordingly.
 Eq.(2.16) is due to the vanishment of
 corresponding symmetrizers observed in [\rBRii].
 Plainly, one can not fuse more than level times
 when $q = \exp({2\pi i\over \ell + 2})$.
 The transfer matrix $T_\ell(u)$ especially
 corresponds to a trivial model in which
 the configuration is frozen.
 See the remark at the bottom of pp235 in [\rDJKMONP]
 where $L-2$ and $N$ there are the level and the fusion degree,
 respectively.
 \vskip0.3cm\noindent
 {\it Remark 2.1.}
 The same argument using (2.7a) with
 $m^\prime = j = m$ also leads to (2.14).
 In fact, more  FRs follow
 from general cases of (2.7).
 However, we remark that (2.14) is a set of generating relations
 of the commutative algebra defined by the relations from
 (2.7).
 Namely, repeated applications of (2.14)
 with various $m$ allow one to express all the $T_m(u)$'s
 as polynomials in $T_1(u+{\rm shifts})$ and the scalar operators
 whereby the other $T$-systems
 will simply reduce to identities.
 See also [\rKRjp] and the Remark 2.2 below.
 Such recursive ``integrations" of the $T$-systems
 lead to curious determinant expressions
 which will be discussed in section 2.3 and section 5.
 \vskip0.3cm\noindent
 {\it Remark 2.2.}
 The $T$-system (2.14) was firstly
 obtained in [\rKLPiii] for the fusion RSOS models [\rDJKMONP]
 when $m_k$ and $v_k$ are $k$-independent.
 Their derivation is a direct computation based on the
 simpler FR eq.(3.19) in [\rBRi].
 The later FR corresponds, in the sense of Remark 2.1
 above, to the case $m^\prime = j = 1$ of (2.7a),
 which is also a set of generating relations of the
 commutative algebra mentioned in Remark 2.1.
 Our argument here clarifies the role of the exact sequence
 of the symmetry algebra
 and the idea is applicable
 in principle to all the other algebras $X_r$.

 \subsect{2.3.  $T$-system for $sl(r+1)$}
 \subskip

 Here we shall prove the following $X_r = A_r$ version
 of the $T$-system:
 $$
 T^{(a)}_m(u+{1\over 2})T^{(a)}_m(u-{1\over 2})
 = T^{(a)}_{m+1}(u)T^{(a)}_{m-1}(u) +
 T^{(a+1)}_m(u)T^{(a-1)}_m(u)\quad 1 \le a \le r,\eqno(2.17)$$
where $T^{(a)}_m(u) = 0$
 for $m < 0$,  $T^{(0)}_m(u) = T^{(a)}_0(u) = 1$.
 The symbol $T^{(a)}_m(u)$ denotes the
 row-to-row transfer matrix
 with fusion type $(a,m)$.
 By this we mean that
 the auxiliary space carries
 the $m$-fold symmetric tensor
 of the $a$-th fundamental representation of $sl(r+1)$.
 It is the irreducible $sl(r+1)$-module associated with the
 $a \times m$ rectangular Young diagram corresponding to
 the highest weight
 $m\Lambda_a$ ($\Lambda_a$ : $a$-th fundamental weight)
 and will be denoted by
 $W^{(a)}_m, 1 \le a \le r$.
 There is the associated evaluation representation
 $W^{(a)}_m(u)$ of $Y(sl(r+1))$
 analogous to the $sl(2)$ case [\rCPi].
 Let $W={\bf C}^{r+1}$ be the vector representation of $ sl(r+1)$.
 We consider the lift of $W^{(a)}_m(u)$
 to the $\til{Y}(sl(r+1))$ representation on the invariant subspace
 of $W^{\tensor am}$, and it will be also denoted by $W^{(a)}_m(u)$.
 The transfer matrix  $T^{(a)}_m(u)$ $(1\leq a \leq r)$
 is visualized in Fig.\ 2.

 \subskip
$$
\epsfxsize=194.9pt \epsfysize=54.2pt \epsfbox{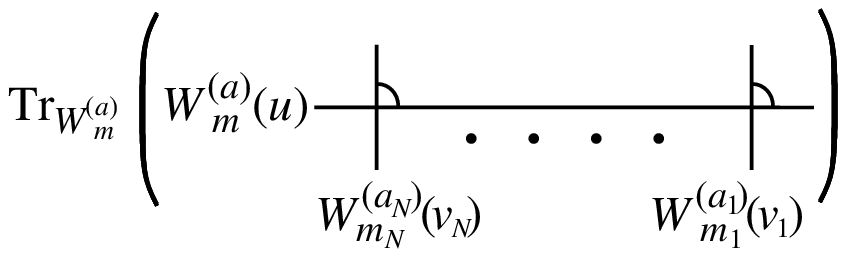}
$$
 %
 \centerline{Figure 2. The graphical representation of the transfer
 matrices $T\am(u)$.}
 \subskip

To define the extra matrix $T^{(r+1)}_m(u)$ in (2.17),
     we introduce the additional ${\tilde Y}(sl(r+1))$ representation
     ${\tilde W}^{(0)}_1(u)$ on the totally antisymmetric
     subspace of $W^{\otimes r+1}$. This is trivial
     as a $Y(sl(r+1))$-module  and analogous
     to  ${\tilde W}_0(u)$ defined for $sl(2)$ in section 2.2.
     See section 4.1.(or (4.5)) for a more description.
     The matrix $T^{(r+1)}_m(u)$ is then defined by
     Fig.\ 2 with the auxiliary space replaced by the following one:
 $$
 \til{W}^{(0)}_1(u-{m-1 \over 2})\tensor
 \til{W}^{(0)}_1(u-{m-3 \over 2})\tensor
 \cdots \tensor
 \til{W}^{(0)}_1(u+{m-1\over 2}),
 \eqno(2.18)
 $$
which is analogous to (2.11).
 As (2.14), $T^{(r+1)}_m(u)$ is a scalar operator.
 The explicit form is irrelevant here
 and we will later need only the property
 $$T^{(r+1)}_m(u+{1\over 2})T^{(r+1)}_m(u-{1\over 2})
 = T^{(r+1)}_{m+1}(u)T^{(r+1)}_{m-1}(u),\eqno(2.19)
 $$
 which is immediate from (2.18) and is a special case of
 (2.17) with ``$a=r+1$".
 The commutativity
 $[T^{(a)}_m(u), T^{(a^\prime)}_{m^\prime}(v)] = 0$
 is  valid due to the intertwining $R$-matrix
 as in the $sl(2)$ case.
 We shall henceforth treat the $T^{(a)}_m(u)$'s
 as commutative variables (or they may be regarded
 as eigenvalues).
 The underlying elliptic vertex and face models have actually been
 constructed in [\rCheii] and [\rJKMO], respectively.
 \par
 Our proof of (2.17) is based on
 the FR due to Bazhanov--Reshetikhin (eqs.(3.10,11) in [\rBRii]), which
 can be derived from the exact sequences (resolutions) by Cherednik [\rCheiii]
 in the same way as (2.14) was derived from (2.7) in the
 previous subsection.\footnote{$^2$}{We thank I.\ Cherednik for his
 communication about the relation between his result and the FR in [\rBRii].}
 Its relevant case to our problem here reads, in our notation, as follows:
 $$\eqalignno{
 T^{(a)}_m(u) &= \hbox{det}
 \bigl(T^{(1)}_{m-i+j}(u+{i+j-1-a\over 2})
 \bigr)_{1 \le i,j \le a}&(2.20{\rm a})\cr
 &= \hbox{det}
 \bigl(T^{(a-i+j)}_1(u+{m-i-j+1\over 2})
 \bigr)_{1 \le i,j \le m}.&(2.20{\rm b})\cr}
 $$
 The original FR of [\rBRii] provides
 determinant formulas for more general transfer matrices
 $T_Y(u)$ with fusion types
 parametrized by general Young diagrams $Y$ and
 (2.20) is the specialization to the case
 $Y = a \times m$ rectangular shape.\footnote{$^3$}{We thank V.V.\ Bazhanov
 for communicating the misprints contained in the published
 version, which have been corrected here in our convention of the
 definition of the transfer matrix.}
 The FR may be viewed as a quantum analogue of the second Weyl character
 formula (or the Jacobi-Trudi formula) [\rMac] as noted in [\rBRii].
 Put $t^a_k = T^{(a)}_1(u+{m+1-k\over 2})$
 and introduce the determinant $D$ of the $m+1$-
 dimensional matrix
 $$\pmatrix{
 t^a_1 & \,t^{a+1}_2 & \cdots & t^{a+m}_{m+1}\cr
 \,t^{a-1}_2 &t^a_3 & \cdots & \,t^{a-1+m}_{m+2}\cr
 \,\vdots &\vdots& \ddots &\vdots \cr
 t^{a-m}_{m+1} & t^{a-m+1}_{m+2} &     \cdots & t^a_{2m+1}\cr}.
 \eqno(2.21)
 $$
 Denoting by
 $D{i_1, i_2, \ldots  \atopwithdelims
 \lbrack \rbrack j_1, j_2, \ldots}$ its minor
 removing $i_k$'s rows and $j_k$'s columns,
 we find
 $$\eqalign{
 T^{(a)}_m(u+{1\over 2}) &=
 D{m+1 \atopwithdelims
 \lbrack \rbrack m+1},\quad
 T^{(a)}_m(u-{1\over 2}) =
 D{1 \atopwithdelims
 \lbrack \rbrack 1},\cr
 T^{(a)}_{m+1}(u) &= D, \quad\qquad\quad\quad
 T^{(a)}_{m-1}(u) =
 D{1, m+1 \atopwithdelims
 \lbrack \rbrack 1, m+1},\cr
 T^{(a+1)}_m(u) &=
 D{m+1 \atopwithdelims
 \lbrack \rbrack 1},
 \quad
 T^{(a-1)}_m(u)=
 D{1 \atopwithdelims
 \lbrack \rbrack m+1},\cr}
 \eqno(2.22)
 $$
 by virtue of (2.20b).
 Substituting these expressions into (2.17) one finds
 the result reduces to the so-called Jacobi identity among
 determinants (or an example of the Pl\"ucker relation)
  thereby completing the proof.
 A similar proof is also possible by using (2.20a).
 Although our proof relied on the FR (2.20),
 we expect there are the exact sequences of
 $Y(sl(r+1))$ and $U_q(\widehat{sl}(r+1))$-modules analogous to (2.10)
 that directly lead to (2.17) in the same sense as in the previous
 subsection.
 %
 %
 %
\par
 The restricted $T$-system
 relevant to the level $\ell$ $A^{(1)}_r$ RSOS models
 [\rJKMO] obeys a similar truncation to (2.16) as
 $$T^{(a)}_{\ell + 1}(u) \equiv 0\quad \hbox{ for all } a,
 \eqno(2.23)
 $$
 which may be deduced from eq.(3.14) in [\rBRii].
 \par\vskip0.3cm\par
 Having established the $T$-system (2.17), let us
 check if the curious phenomenon observed in the introduction is
 happening.
 Namely,
 {\it does (2.17) Yang-Baxterize the
 $Q$-system so as to satisfy the $Y$-system ?}
 {}From (3.8) and (B.30)
 the latter two are given by
 $$\eqalignno{
 &Q^{(a) 2}_m
 = Q^{(a)}_{m+1}Q^{(a)}_{m-1} +
 Q^{(a+1)}_mQ^{(a-1)}_m,\qquad Q^{(0)}_m=Q^{(r+1)}_m=1,
 &(2.24{\rm a})\cr
 &Y^{(a)}_m(u+{1\over 2})Y^{(a)}_m(u-{1\over 2})
 = {(1+Y^{(a)}_{m+1}(u))(1+Y^{(a)}_{m-1}(u))\over
 (1+Y^{(a+1)}_m(u)^{-1})(1+Y^{(a-1)}_m(u)^{-1})},&(2.24{\rm b})\cr
 }$$
 where the $Y$-system we have
 made a notational change
 ($Y^{(a)}_m(u+{k \over 2})$ here is
 $Y^{(a)}_m(u+ki)^{-1}$ in (B.30)) so as to suite
 (1.4).
 The $T$-system (2.17) is a Yang-Baxterization of
 (2.24a) in the obvious sense and it is elementary
 to see that the combination
 $$
 y^{(a)}_m(u) =
 {T^{(a)}_{m+1}(u)T^{(a)}_{m-1}(u)\over
 T^{(a+1)}_m(u)T^{(a-1)}_m(u)}
 \eqno(2.25)
 $$
 solves (2.24b) by means of (2.17) and (2.19).
 Thus we have established the same connection
 among the $T$, $Q$ and $Y$-systems also for
 $X_r = sl(r+1)$.

 \sect{\bf{3. $T$-system for a general $X_r$}}\pn
 \subskip
 Let us proceed to
 the $T$-systems for the other $X_r$'s.
 We introduce them through the curious connection to the
 $Q$ and $Y$-systems
 which has been already established  for $X_r = sl(r+1)$ in
 section 2.

 \subsect{3.1. Preliminaries}
 \subskip

 We firstly fix our notations.
 Let $X_r$ denote one of the classical
 simple Lie algebras
 $A_r (r \ge 1), B_r (r \ge 2), C_r (r \ge 2),
 D_r (r \ge 4), E_{6,7,8}, F_4$ and $G_2$.
 Let $\{ \alpha_a \mid 1 \le a \le r \}$,
 $\{ \Lambda_a \mid 1 \le a \le r \}$ and
 $(\cdot \mid \cdot)$ be
 the set of the simple roots, the fundamental weights and
 the invariant form on $X_r$, respectively.
We employ the convention $\Lambda_0 = 0$.
 We identify the Cartan subalgebra ${\cal H}$ and its dual
 ${\cal H}^\ast$ via the form $(\cdot \mid \cdot)$.
 We employ the normalization
 $\vert \hbox{ long root } \vert^2 = 2$ and
 put
 $$\eqalignno{
 t_a &= {2 \over (\alpha_a \vert \alpha_a)},\quad
 t_{a b} = \hbox{max}(t_a, t_b),&(3.1{\rm a})\cr
 C_{a b} &=
 {2(\alpha_a \vert \alpha_b) \over (\alpha_a \vert \alpha_a)},\quad
 B_{a b} = {t_b\over t_{a b}}C_{a b} = B_{b a},\quad
 I_{a b} = 2\delta_{a b} - B_{a b},&(3.1{\rm b})\cr
 \alpha^{\vee}_a &= t_a \alpha_a,&(3.1{\rm c})\cr
 g &= \hbox{dual Coxeter number of } X_r,&(3.1{\rm d})\cr
 \rho &= \Lambda_1 + \cdots + \Lambda_r.&(3.1{\rm e})\cr}
 $$
 Here $C_{a b}, B_{a b}$ and $I_{a b}$ are
 the Cartan, the symmetrized Cartan and the incidence matrix,
 respectively.
 By definition
 $(\alpha_a \vert \Lambda_b) = \delta_{a b}/t_a$ and $t_a = 1$ if
 $\alpha_a$ is a long root and otherwise
 $t_a = 2$ for $B_r, C_r, F_4$ and
 $t_a = 3$ for $G_2$.
 One may equivalently define
 $t_i$ by $t_i = a_i/a^{\vee}_i$
 with the Kac and dual Kac labels $a_i$ and $a^{\vee}_i$,
 respectively.
 In Table 1, we list the Dynkin diagram
 with the numeration of its nodes $1 \le a \le r$,
 $\hbox{ dim }X_r$ and $g$ for each $X_r$ along with
 those $t_a \neq 1$.
 In this paper and the subsequent Part II,
 the letters $a, b \ldots $ will mainly be used to denote
 the indices corresponding to the Dynkin diagram nodes.
 We shall also use the standard notations
 $$
 \eqalign{
 Q &= \bigoplus_{a=1}^r {\bf Z} \alpha_a,\quad
 Q^{\vee} = \bigoplus_{a=1}^r {\bf Z} \alpha^{\vee}_a,\quad \cr
 P &= \bigl( Q^{\vee} \bigr)^\ast
 = \bigoplus_{a=1}^r {\bf Z} \Lambda_a, \quad
 P _+
 = \bigoplus_{a=1}^r {\bf Z}_{\geq 0} \Lambda_a .\cr}
 \eqno(3.2)$$
 These are the root, coroot, weight lattices and the set of
 dominant weights, respectively.
 \par
 {In order to discuss the restricted $T$-systems,
 we fix an integer $\ell \in {\bf Z}_{\ge 1}$
 and set
 $$
 \eqalignno{
 \ell_a &= t_a \ell,&(3.3{\rm a})\cr
 G &= \{ (a,m) \mid 1 \le a \le r, \, 1 \le m \le \ell_a - 1 \} .
 &(3.3{\rm b})\cr}
 $$
 Let
 $$
 P_{\ell}=
 \{ \Lambda \in P_+ | (\Lambda| {\rm maximal\ root})
 \leq \ell \},
 \eqno(3.4)
 $$
 which is the classical projection of the set of the
 dominant integral
 weights of the affine Lie algebra $X^{(1)}_r$ at level
 $\ell$.

 \subsect{3.2. IFDR of Yangian and $Q$-system}
 \subskip

  Let us recall some basic properties of FDRs of the
  Yangian $Y(X_r)$ for a general $X_r$ [\rDr,\rKRii,\rCPiii].
  Firstly, $Y(X_r)$ contains $U(X_r)$ as a subalgebra
  hence any FDR of the former is decomposed into
  a direct sum of the IFDRs of the latter.
  A practical way of finding IFDRs of $Y(X_r)$
  would be to seek a surjection
  $Y(X_r) \rightarrow X_r$ which  pulls back
  any IFDR of $X_r$ to that of $Y(X_r)$.
  This is indeed known to be possible for $X_r= A_r$,
  yielding the so-called evaluation representations as
  noted for $sl(2)$ in section 2.2.
  However, for $X_r \neq A_r$, there is no obvious
  analogue of such an evaluation map that the
  restriction on $X_r$ is an identity map [\rDr,\rCPiii].
  In general, IFDRs of $Y(X_r)$ are no longer
  irreducible and possess nontrivial reducibility contents
  under the decomposition as the $X_r$-modules.
  Secondly, the Yangian $Y(X_r)$ is equipped with an automorphism
  depending on a complex parameter. Thus, any given IFDR $W$
  of $Y(X_r)$ can be pulled back by it to a one parameter
  family of the IFDRs, which will be denoted by
  $W(u)$, $u \in {\bf C}$.
  As an $X_r$-module, $W(u) \simeq W$ holds and
  one may regard $W = W(0)$ for example.
  The parameter $u$ is related to the spectral parameter
  of $R$-matrices as in section 2.2.

\par Hereafter we concentrate on  the family of IFDRs
 $W\am(u)$ ($1\leq a \leq r$, $1 \leq m$) of  $\YX$ introduced in
 [\rKRii] that are relevant to the  $T$-system
 (though there is no written proof of
 the irreducibility  in [\rKRii]).
$W^{(a)}_m(u)$ is a quantum analogue of the
$m$-fold symmetric tensor of $W^{(a)}_1(u)$.
It
is characterized by the Drinfel'd polynomials [\rCPiii,\rDrii]  as
$P^+_b(v) = 1$ for $b \neq a$ and $P^+_a(v) =
\prod_{j=1}^m(v-u + (m-2j)/ 2t_a + c)$ (see (4.14)).
Here $c$ is a constant independent of
$a$ and $m$, and irrelevant to our discussion.
  Let $W\am$ denote
 $W^{(a)}_m(u)$ as an $X_r$-module, and
 $V_{\Lambda}$  be the  IFDR of $X_r$ with highest weight
 $\Lambda$. Then, its reducibility content is given as [\rKRii]
 $$
 \eqalignno{
 W^{(a)}_m &= \sum_{\Lambda} Z(a,m,m\Lambda_a -\Lambda)
	     V_{\Lambda},\quad
 &(3.5{\rm a}) \cr
 Z(a, m, \sum_{b=1}^{r}{n_b\alpha_b})
  &= \sum_{\nu} \prod_{b=1}^r \prod_{k=1}^{\infty}
 \left ( \matrix{{\cal P}^{(b)}_k + \nu^{(b)}_k \cr
		 \nu^{(b)}_k \cr} \right ),
 &(3.5{\rm b})\cr
{\cal P}^{(b)}_k &= \delta_{a b}\hbox{min}(m,k)
  - \sum_{c=1}^r\sum_{j=1}^\infty (\alpha_b \vert \alpha_c)
  \hbox{min}(t_c k, t_b j) \nu^{(c)}_j.
 &(3.5{\rm c})\cr}
$$
 Here in (3.5a) $\Lambda$-sum extends over
 those $\Lambda$ such that
 $m\Lambda_a - \Lambda \in
 \sum_{a=1}^r {\bf Z}_{\ge 0}\alpha_a$
hence $\forall n_b \in {\bf Z}_{\ge 0}$ is implied
    in the lhs of (3.5b).
 In (3.5b) the symbol
 $\left( {\alpha  \atop \beta}\right)$ means the binomial
 coefficient and $\nu$-sum extends over all possible
 decompositions
 $\{ \nu^{(b)}_k | n_b=\sum_{k=1}^{\infty} k \nu^{(b)}_k,
 \nu^{(b)}_k \in \bZ_{\geq 0}, 1 \leq b\leq r, k\geq 1 \}
 $
 such that  ${\cal P}^{(b)}_k \geq 0$ for all $1\leq b \leq r$
 and $k \geq 1$.
 More explicit formulas of (3.5) are also known
 [\rKu, \rKRii] in many cases. See Appendix A.1 for a further
 information on (3.5).

 %
 \par
 The  case $X_r = A_r$ is simplest in that
 $W^{(a)}_m
  = V_{m\Lambda_a}$ holds,
which is consistent to the  notation in section 2.3.
 In general,  the decomposition of $W^{(a)}_m$ always contains
 the module $V_{m\Lambda_a}$  as the highest term
 w.r.t the root system  with multiplicity one.
  $W^{(a)}_1(u)$'s are identified with the {\it fundamental representations}
 of $Y(X_r)$ in the sense of [\rCPiii].
\par
There are remarkable polynomial relations among the characters of
$W\am$ as we describe below.
 Let $Q\am=Q\am(z)$ ($z \in {\cal H}^\ast$) be the character
 of $W\am$, namely,
 $$\eqalignno{
 Q^{(a)}_m(z) &= \sum_{\Lambda} Z(a,m,m\Lambda_a-\Lambda)
	     \chi_{\Lambda}(z), &(3.6{\rm a})\cr
 \chi_{\Lambda}(z) &=
 \hbox{Tr}_{V_{\Lambda}}
 \hbox{ exp}\bigl(-{2\pi i\over \ell + g}(z + \rho)\bigr),
 &(3.6{\rm b})\cr
 }$$
 where  $\ell\in {\bf Z}_{\ge 0}$ has been fixed in (3.3), and
  $Q^{(a)}_0= 1$, $Q^{(a)}_{-1}=0$ for all $a$.
 %
 The following  relation holds [\rKil,\rKRii],
which we call the unrestricted $Q$-system:
 \par
 \noindent
 {\it Unrestricted $Q$-system}
 $$
 Q^{(a)}_m{}^2 = Q^{(a)}_{m-1}Q^{(a)}_{m+1} +
 Q^{(a)}_m{}^2 \prod_{b=1}^r\prod_{k=0}^{\infty}
 Q^{(b)}_k{}^{-2J^{k\, m}_{b\, a}}\,\,\hbox{ for }
 1 \le a \le r, \, m \ge 1, \eqno(3.7)
 $$
 where $2J^{k m}_{b a}$ is
 given by (B.22b)
and all the powers
   of $Q^{(a)}_m$'s are nonnegative integers.
 The $k$-product in (3.7) actually extends over
 only finite support of $J^{k\, m}_{b\, a}$.
The second term on the rhs of (3.7) contains
    0 (i.e., equal to 1),1, 2 or 3 $Q$-factors
    depending on the choice of $a, m$ and $X_r$.
 %
 %
 %
 For simply-laced algebras $X_r = A_r, D_r$ and $E_{6,7,8}$,
 (3.7) takes an especially simple form due to
 (B.32b) as
 $$
 Q^{(a)}_m{}^2 = Q^{(a)}_{m-1}Q^{(a)}_{m+1} +
 \prod_{b=1}^r Q^{(b)}_m {}^{I_{a\, b}},\eqno(3.8)
 $$
 where $I_{a b}$ is the incidence matrix (3.1).

 An interesting feature emerges when one specializes $z$ to
 $0 \in P$. In this case, all  $Q\am(0)$ $(a,m) \in G$
 are real positive.
 Moreover, admitting (A.6{\rm a}) of [\rKu] and (A.7),
we  see that
$$
\eqalignno{
Q^{(a)}_{\ell_a +1}(0)&=0, &(3.9{\rm a})\cr
Q^{(a)}_{\ell_a}(0)& = 1. &(3.9{\rm b})\cr
}
$$
 Therefore (3.7)
 closes among $Q^{(a)}_m(0)$ for $(a,m) \in G$,
which we call the restricted $Q$-system,
 i.e.,
 \par
 \noindent
 {\it Restricted $Q$-system}
 $$
 Q^{(a)}_m(0)^2 = Q^{(a)}_{m-1}(0)Q^{(a)}_{m+1}(0) +
 Q^{(a)}_m(0)^2 \prod_{(b,k) \in G}
 Q^{(b)}_k(0)^{-2J^{k\, m}_{b\, a}}\,\hbox{ for }
 (a,m) \in G. \eqno(3.10)
 $$
In section 3.4 we shall consider the Yang-Baxterization of (3.7) and
(3.10).
See Appendix A for more general restricted systems.
Those restricted $Q$-systems are related to the dilogarithm identities
[\rKRjp,\rKN,\rKil],
on which further analyses will be given in Part II.

The reducibility content (3.5) and the $Q$-system (3.7)
among the characters are the basic features
in the IFDRs $W^{(a)}_m$ of the Yangian $Y(X_r)$.
However, those properties are also known to be common in
 the quantum affine algebra $U_q(X_r^{(1)})$ for
$X_r = A_1$ [\rCPii]. We assume this for all
$X_r$'s in this paper and  Part II. Namely, for each
$W^{(a)}_m(u)$, we postulate the existence of its natural $q$-analogue
having the properties (3.5) and (3.7).
We shall write the IFDR of $U_q(X_r^{(1)})$ so postulated
also as $W^{(a)}_m(u)$.

 \subsect{3.3. Models with fusion type $W^{(a)}_m$}
 \subskip

 Having identified the spaces $W^{(a)}_m$, we can now
 describe a set of models associated with them.
  Though our description here might seem formal,
  most known examples of the trigonometric
 vertex and RSOS weights obey this scheme (cf.\
[\rJimi]). See also remark 2.5 in [\rKu].

\par
Let $W^{(a)}_m(u)$ and $W^{(a^\prime)}_{m^\prime}(v)$
    be IFDRs of $U_q(X^{(1)}_r)$ as described in section 3.2.
   For generic $u-v$ and $q$, their tensor products are irreducible and
    isomorphic in both order, which implies the existence of the unique
    $U_q(X^{(1)}_r)$-isomorphism
    $\Rc(u-v) \in {\rm Hom}_{U_q(X^{(1)}_r)}(
W^{(a)}_m(u)\otimes W^{(a')}_{m'}(v),
	      W^{(a')}_{m'}(v)\otimes W^{(a)}_{m}(u)
)$
    solving the Yang-Baxter equation.
    Regarding its matrix elements as the Boltzmann weights,
    one has the vertex model with fusion type labeled by
    the pair $(W^{(a)}_m, W^{(a^\prime)}_{m^\prime})$.
    Each horizontal (vertical) edge of the vertex
    carries the IFDR $W^{(a)}_m$ ($W^{(a^\prime)}_{m^\prime}$)
    as in section 2.2
\par
 From the vertex model, one  can in principle construct the level $\ell$
 $U_q(X^{(1)}_r)$ RSOS models with the same fusion type
 when $q=e^{{2\pi i} \over {\ell+g}}$
 through the vertex-IRF correspondence [\rPas], which
 we shall briefly sketch below.
 Let $V_{{\lambda}}$  ($\lambda \in P_{\ell}$)
 and $W\am$ be the $U_q(X_r)$-modules which
 are uniquely lifted from the corresponding $X_r$-modules.
 When $q$ is not a root of unity, the tensor product
 decomposes as an $U_q(X_r)$-module as
 $$
 V_{\lambda} \otimes W\am \tensor W^{(a')}_{m'}=
 V_{\lambda} \otimes   W^{(a')}_{m'}   \tensor W\am
 =\bigoplus_{\mu\in P_+} \Omega({\lambda})_\mu \otimes V_{\mu},
 \eqno{(3.11)}
 $$
 where $\Omega(\lambda)_\mu$ is the space of highest weight
 vectors with weight $\mu$.
 Since $\Rc(u)$ commutes with $U_q(X_r)$, the space
 $\Omega({\lambda})_{\mu}$ is invariant under the action of
   $\hbox{id} \otimes \Rc(u)$:  $V_{\lambda} \otimes W\am
 \tensor W^{(a')}_{m'} \lra V_{\lambda} \otimes   W^{(a')}_{m'}
    \tensor W\am$.
 When $q=e^{{2\pi i} \over {\ell+g}}$, the decomposition
 (3.11) no longer holds [\rLus,\rRA].
 However,  in the case $A_1$ it is known [\rPas] that
 the action of   $\hbox{id} \otimes \Rc(u)$ remains well-defined on
 the quotient of the rhs
 of (3.11) divided by the Type I modules [\rPS,\rKel]
  (i.e., indecomposable modules with quantum dimension zero).
 It is natural to
assume the same prescription holds for any $X_r$. Then,
  the RSOS Boltzmann weights should be defined by
 the matrix elements of  the operator $\hbox{id} \tensor \Rc (u)$
 acting on the quotient  space of $\Omega({\lambda})_\mu$
  and they satisfy the YBE.
\par

 The transfer matrices $T\am(u)$ are defined as in Fig.\ 2
 by using the above
 vertex (or RSOS) weights.
 For vertex models, we consider  fusion type (of
 auxiliary space)  $W\am$ with $1 \le a \le r,
 m \ge 1 $, while for RSOS models with
$1 \le a \le r, 1 \le m \le \ell_a$ (cf.\ (3.9)).
 The commutativity of transfer matrices with the same quantum space
 is assured by the YBE.

\par
The RSOS model described above has the
    admissibility matrix which is remarkably consistent with
    the TBA analyses  [\rBRi,\rKu]
as we shall explain below.
 The fluctuation variables of the RSOS model are
attached to both the sites and the edges [\rJKMO]. In our model,
the site variable belongs to $P_\ell$. To describe the edge
variable, we consider the decomposition
$$
 V_{\lambda}\otimes W^{(a)}_m=
 \bigoplus_{\lambda' \in P_+} \overline{\Omega}_{\lambda' \lambda}
\otimes
V_{\lambda'}
 \eqno{(3.12)}
 $$
at generic $q$.
When
$q=e^{{2\pi i} \over {\ell+g}}$,
 we need to take the quotient of
the rhs by the Type I modules,  and this induces the quotient
$\Omega_{\lambda' \lambda}$ of $\overline{\Omega}_{\lambda' \lambda}$.
Then, the edge variable associated to the
$W\am$ fusion belongs to the space
$\Omega_{\lambda' \lambda}$. Let $M_{\lambda' \lambda}=
\dim\Omega_{\lambda' \lambda}$ and
$\overline{M}_{\lambda' \lambda}=\dim
\overline{\Omega}_{\lambda' \lambda}$. We call the matrix
$M$  the admissibility matrix of fusion type $W\am$.
 Following the standard argument of the fusion algebra in  conformal field
 theories [\rKac--\rGN],
the matrix $M$ is related to $\overline{M}$ as
 $$
 M_{\lambda' \lambda}
 =\sum_{w \in {\cal W}} \det w \cdot
\overline{M}_{{\dot w} ( \lambda')\,\lambda},
 \qquad \lambda',\, \lambda \in P_{\ell},
 \eqno{(3.13)}
 $$
 where ${\cal W}$ is the  Weyl group of  $X_r^{(1)}$ at level $\ell$
and ${\dot w}(\lambda)=w(\lambda+\rho)-\rho$.
 Then, from the pseudo-Weyl invariance of the
 specialized characters [\rWal],
it follows that
 $$
 M \cdot {\vec{v}} =Q^{(a)}_m (0) \, {\vec{v}},
 \eqno{(3.14)}
 $$
 where $(\vec{v})_{\lambda}=\chi_{\lambda}(0)$.
 Since $M$ and ${\vec{v}}$ are nonnegative matrix and vector,
 $Q^{(a)}_m (0)$ is nothing but the maximum eigenvalue of
 $M$.
Thus, for the quantum space
     $H_N$  with lattice size $N$, we have
 $$
 \lim_{N\rightarrow \infty} ({\hbox{dim}}{H_N})^{1/N} =
 \lim_{N\rightarrow \infty} ({\hbox{Tr}}\, M^N )^{1/N}
 =Q^{(a)}_m (0),
 \eqno{(3.15)}
 $$
 where we have assumed the periodic boundary condition for the
 first equality.
 This is quite consistent with the result of the TBA  analysis
on a corresponding quantum spin chain model
 that the high temperature limit of
 entropy per edge is equal to $\log Q^{(a)}_m (0)$ [\rBRi,\rKu].
 Note in particular from (3.9b) that
 the fusion type $W^{(a)}_{\ell_a}$ corresponds
 to a ``frozen model", which is consistent to the
 comments after (2.16).

 \subsect{3.4. $T$-system for
  arbitrary classical Lie algebras}
 \subskip

 Let us proceed to the main problem of this paper.  What
 are functional relations among the transfer matrices described in
 section 3.3?
Our strategy here is to apply the working hypothesis
   that the $T$-system Yang-Baxterizes the $Q$-system
in such a way that particular
   combinations of transfer matrices solve the $Y$-system.
   This has been established for $X_r=A_r$ in section 2.

 We start with the consideration on  the  $T$-systems
that correspond to the unrestricted $Q$-systems (3.7).
 A general relation in  (3.7) has the following form:
 $$
 Q^{(a)}_m{}^2 = Q_{m+1}^{(a)} Q_{m-1}^{(a)} +Q^{(b)}_k \cdots Q^{(c)}_j .
 \eqno{(3.16)}
 $$
 Accordingly, we take the following ansatz for the
 $T$-system:
 $$
 T^{(a)}_m(u-{1 \over 2{t_a}}) T^{(a)}_m(u+{1 \over 2{t_a}})
 = T_{m+1}^{(a)}(u) T_{m-1}^{(a)}(u)+
  g^{(a)}_m(u) T^{(b)}_k(u+\alpha)\cdots T^{(c)}_j(u+\beta),
 \eqno{(3.17)}
 $$
where all  $T^{(a)}_m$'s may be treated as
commutative variables.
The functions $g^{(a)}_m$ play a similar role to those in (2.14), (2.19),
and they are related to a central extension of $Y(X_r)$.
Below its explicit form is not needed but only the property
 $$
 g^{(a)}_{m}(u-{1 \over{2 t_a}}) g^{(a)}_{m}(u+{1 \over{2 t_a}})
 =g^{(a)}_{m-1}(u) g^{(a)}_{m+1}(u),
 \eqno{(3.18)}
 $$
 which will be justified in section 4 in many cases.
In (3.17) $\alpha, \beta$ etc.\ are parameters to be determined in the
   following way.  Let $y^{(a)}_m(2iu)$ be the
 combination:\footnote{$^4$}{
 Our normalization of the spectral parameter
 here differs from the one  in Appendix B by the factor
 $2i$. For a notational simplification,
the $y^{(a)}_m$'s in (1.8) and (2.25) have been taken
to be the inverse of (3.19).
}
 $$
 y^{(a)}_m(2iu)={{g^{(a)}_m(u) T^{(b)}_k(u+\alpha)
  \cdots T^{(c)}_j(u+\beta)}
	 \over{T_{m+1}^{(a)}(u) T_{m-1}^{(a)}(u)}}.
 \eqno{(3.19)}
 $$
 We then demand that this  $y^{(a)}_m(2iu)$ solves the unrestricted
 $Y$-system (cf.(B.6) and remarks following it).
 Remarkably, this fixes the parameters $\alpha ,\beta$ etc.\  uniquely
 up to trivial freedoms which can be absorbed into
 the  redefinition of the transfer matrices by
 $T^{(a)}_m(u) \rightarrow T^{(a)}_m(u+\gamma^{(a)}_m)$.
We note that the presence of $g^{(a)}_m(u)$ is
irrelevant to fixing $\alpha, \beta$, etc.\ as long as
(3.18) is satisfied.

 In this way, we reach the following conjecture of the functional
 equations for the transfer matrices on the vertex
 models associated to the simple Lie algebras. We call them
{\it the unrestricted $T$-systems}.
 \par\noindent $X_r=A_r,D_r, E_6,E_7,E_8:$
 $$
 T^{(a)}_m(u-{1 \over 2}) T^{(a)}_m(u+{1 \over 2})
 = T_{m+1}^{(a)}(u) T_{m-1}^{(a)}(u)+
 g^{(a)}_m(u)\prod_{b=1}^{r} T^{(b)}_m(u){}^{I_{ab}}.
 \eqno{(3.20{\rm a})}
 $$
 \noindent $X_r=B_r:$
 $$\eqalignno{
 T^{(a)}_m(u-{1 \over 2})T^{(a)}_m(u+{1 \over 2})
  &=T_{m-1}^{(a)}(u) T_{m+1}^{(a)}(u)+
			   g^{(a)}_m(u)T^{(a-1)}_m(u)T^{(a+1)}_m(u) ,\cr
 &\hskip130pt
			 (1 \le a \le r-2 ),\cr
 T^{(r-1)}_m(u-{1 \over 2})T^{(r-1)
 }_m(u+{1 \over 2}) &=T_{m-1}^{(r-1)}(u) T_{m+1}^{(r-1)}(u)\cr
 &\qquad +
			   g^{(r-1)}_m(u) T^{(r-2)}_m(u)T^{(r)}_{2m}(u) , \cr
 T^{(r)}_{2m}(u-{1 \over 4})T^{(r)}_{2m}(u+{1 \over 4})
      &=T_{2m-1}^{(r)}(u) T_{2m+1}^{(r)}(u)\cr
 &\qquad +
      g^{(r)}_{2m}(u) T^{(r-1)}_m(u-{1 \over 4})T^{(r-1)}_m(u+{1 \over 4}), \cr
 T^{(r)}_{2m+1}(u-{1 \over 4})T^{(r)}_{2m+1}(u+{1 \over 4})
      &=T_{2m}^{(r)}(u) T_{2m+2}^{(r)}(u) \cr
 & \qquad +
      g^{(r)}_{2m+1}(u) T^{(r-1)}_m(u)T^{(r-1)}_{m+1}(u) .
		      &(3.20{\rm b})}
 $$
 \noindent $X_r=C_r:$
 $$\eqalignno{
 T^{(a)}_m(u-{1 \over 4})T^{(a)}_m(u+{1 \over 4})
	&=T_{m-1}^{(a)}(u) T_{m+1}^{(a)}(u)+
	g^{(a)}_m(u)T^{(a-1)}_m(u)T^{(a+1)}_m(u), \cr
 &\hskip120pt
				      \qquad  (1 \le a \le r-2) ,\cr
 T^{(r-1)}_{2m}(u-{1 \over 4}) T^{(r-1)}_{2m}(u+{1 \over 4})
      &=T_{2m-1}^{(r-1)}(u) T_{2m+1}^{(r-1)}(u) \cr
 &\qquad +
      g^{(r-1)}_{2m}(u) T^{(r-2)}_{2m}(u)
      T^{(r)}_m(u-{1 \over 4})T^{(r)}_m(u+{1 \over 4}),\cr
 T^{(r-1)}_{2m+1}(u-{1 \over 4})T^{(r-1)}_{2m+1}(u+{1 \over 4})
      &=T_{2m}^{(r-1)}(u) T_{2m+2}^{(r-1)}(u) \cr
 &\qquad +     g^{(r-1)}_{2m+1}(u) T^{(r-2)}_{2m+1}(u)
      T^{(r)}_m(u)T^{(r)}_{m+1}(u), \cr
 T^{(r)}_m(u-{1 \over 2})T^{(r)}_m(u+{1 \over 2})
  &=T_{m-1}^{(r)}(u) T_{m+1}^{(r)}(u)+
	      g^{(r)}_m(u) T^{(r-1)}_{2m}(u).&
				  (3.20{\rm c})}
 $$
 \noindent $X_r=F_4:$
 $$\eqalignno{
 T^{(1)}_m(u-{1 \over 2})T^{(1)}_m(u+{1 \over 2})
  &=T_{m-1}^{(1)}(u) T_{m+1}^{(1)}(u)+
	      g^{(1)}_m(u) T^{(2)}_{m}(u), \cr
 T^{(2)}_m(u-{1 \over 2})T^{(2)}_m(u+{1 \over 2})
  &=T_{m-1}^{(2)}(u) T_{m+1}^{(2)}(u)+
	      g^{(2)}_m(u) T^{(1)}_{m}(u)T^{(3)}_{2m}(u),  \cr
 T^{(3)}_{2m}(u-{1 \over 4}) T^{(3)}_{2m}(u+{1 \over 4})
      &=T_{2m-1}^{(3)}(u) T_{2m+1}^{(3)}(u)\cr
 &\qquad +
      g^{(3)}_{2m}(u) T^{(2)}_m(u-{1 \over 4})T^{(2)}_m(u+{1 \over 4})
	    T^{(4)}_{2m}(u)                   ,              \cr
 T^{(3)}_{2m+1}(u-{1 \over 4}) T^{(3)}_{2m+1}(u+{1 \over 4})
      &=T_{2m}^{(3)}(u) T_{2m+2}^{(3)}(u) \cr
 &\qquad +
      g^{(3)}_{2m+1}(u) T^{(2)}_m(u)T^{(2)}_{m+1}(u)
	    T^{(4)}_{2m+1}(u)                    ,             \cr
 T^{(4)}_m(u-{1 \over 4})T^{(4)}_m(u+{1\over 4})
       &=T_{m-1}^{(4)}(u) T_{m+1}^{(4)}(u)+
	      g^{(4)}_m(u) T^{(3)}_{m}(u).
				  &(3.20{\rm d})
 }$$
 \noindent $X_r=G_2:$
 $$\eqalignno{
 T^{(1)}_m(u-{1 \over 2})T^{(1)}_m(u+{1 \over 2})
  &=T_{m-1}^{(1)}(u) T_{m+1}^{(1)}(u)+
	      g^{(1)}_m(u) T^{(2)}_{3m}(u) , \cr
 T^{(2)}_{3m}(u-{1\over 6})T^{(2)}_{3m}(u+{1\over 6})
	  &=T_{3m-1}^{(2)}(u) T_{3m+1}^{(2)}(u) \cr
 &\qquad +
	      g^{(2)}_{3m}(u)   T^{(1)}_m(u-{1\over 3})
	      T^{(1)}_m(u) T^{(1)}_m(u+{ 1 \over 3})  ,  \cr
 T^{(2)}_{3m+1}(u-{1\over 6})T^{(2)}_{3m+1}(u+{1\over 6})
	  &=T_{3m}^{(2)}(u) T_{3m+2}^{(2)}(u)\cr
 &\qquad +
	  g^{(2)}_{3m+1}(u)   T^{(1)}_m(u-{1\over 6})
	       T^{(1)}_m(u+{1\over 6}) T^{(1)}_{m+1}(u)  , \cr
 T^{(2)}_{3m+2}(u-{1\over 6})T^{(2)}_{3m+2}(u+{1\over 6})
	  &=T_{3m+1}^{(2)}(u) T_{3m+3}^{(2)}(u)\cr
 & \qquad +
	  g^{(2)}_{3m+2}(u) T^{(1)}_{m}(u) T^{(1)}_{m+1}(u-{1\over 6})
	       T^{(1)}_{m+1}(u+{1\over 6})  .  \cr
			     &     &(3.20{\rm e})
 }$$
 In the above the subscripts of the transfer matrices in the
lhs extend over positive integers, and
we adopt the conventions $T^{(0)}_m=T^{(a)}_0=1$.
\par
For the level $\ell$ RSOS models, we suppose the same
system (3.20) is valid except that
the truncation
 $$
 T^{(a)}_{\ell_a+1}(u)=0 \qquad {\hbox{for\ all\  }} a
 \eqno(3.21)
 $$
takes place based on the $X_r=A_r$ case (2.16), (2.23) and
the fact (3.9a).
 Then (3.20) reduces to a finite set of functional
 relations among $T^{(a)}_m$  for $1 \le a \le r, 1 \le m \le \ell_a$.
 We call it the (level $\ell$) {\it restricted $T$-system}.
Since (3.21) leads to $y^{(a)}_{\ell_a}{}^{-1} = 0$
in (3.19), the restricted $T$-system is a
Yang-Baxterization of the restricted $Q$-system (3.10)
solving the restricted $Y$-system (B.6) in the parallel sense
to the unrestricted case.
 Notice that $T^{(a)}_{\ell_a}$ $(1\leq a \leq r)$ satisfy
  closed relations among themselves:
 $$
 T^{(a)}_{\ell_a}(u-{1 \over 2{t_a}})T^{(a)}_{\ell_a}(u+{1 \over 2{t_a}})
 =g^{(a)}_{\ell_a}(u)
\prod_{{b=1 \atop  b \neq a}}^{r} \prod_{k=1}^{-C_{ab}}
   T^{(b)}_{\ell_b}(u+{{2k-1+C_{ab}}\over 2{t_a}}
 ).
 \eqno(3.22)
 $$
As remarked in the end of section 3.3,
the $T^{(a)}_{\ell_a}$'s are the transfer matrices
for the frozen RSOS models.
 The restricted $T$-system is invariant under the simultaneous
transformations
 $$
 T^{(a)}_m(u) \rightarrow Z_a T^{(a)}_{\ell_a-m} (u+{\hbox{constant}}),
 \eqno(3.23)
 $$
for all $a$
 if the operators $ Z_a $ satisfy
 $$
 \eqalign{
 [Z_a,Z_b]&=[Z_a, T^{(b)}_k(u)] =0 \qquad {\hbox{for }} \forall a, b, k,\cr
 \prod_{a=1}^r (Z_a)^{C_{ba}} &= \hbox{Id} .
 }
 \eqno(3.24)
 $$
 This may be viewed as a generalization of the property found in [\rBRi]
 for the $sl(2)$ case.

Finally we remark that
the level 1 restricted $T$-systems for non-simply laced
algebras $X_r = B_r, C_r, F_4, G_2$ reduce to
those for $A_1,A_{r-1},A_2,A_1$ of levels $2,2,2,3$,
respectively.
These reductions are consistent to the equivalences
of the associated RSOS models indicated from the TBA
analysis. See eq.(3.12) in [\rKu].

 \sect{4. Fusion contents  and  $T$-system}

In this section we examine our $T$-system (3.20) from the
the representation theoretical scheme as in section 2.2.
For definiteness,
 we  restrict our description to the unrestricted $T$-system of
 the rational vertex models throughout the section.

 \subsect{4.1.  Fusion contents}
 \subskip

In this subsection we reconsider the $T$-system (2.17) for
$A_r$.
We begin by recalling the fusion procedure to
  build the space $W^{(a)}_m(u)$ and the data encoding it.
 The rational $R$-matrix  (2.1) of the vector representation
 of $Y(A_r)$ has the
 following spectral decomposition form:
 $$
 \Rc(u)=\Rc_{\om{1},\om{1}}(u)=(1+u)P_{2\om{1}} \oplus (1-u)\pr{2},
 \eqno(4.1)
 $$
 where $P_\Lambda $ is the projection operator on the space $V_\Lambda$.
 From (4.1), ${\rm Im} \Rc(1)=W^{(1)}_2$ and
 ${\rm Im} \Rc(-1)=W^{(2)}_1$. Thus, one can define the IFDRs
$ W^{(1)}_2(u)$ and $ W^{(2)}_1(u)$ through the embeddings
 $$
 \eqalign{
 W^{(1)}_2(u) \emb
 & W^{(1)}_1(u-{1 \over 2}) \tensor W^{(1)}_1(u+{1\over 2}) , \cr
  W^{(2)}_1(u) \emb
& W^{(1)}_1(u+{1\over 2}) \tensor W^{(1)}_1(u-{1 \over 2}) ,\cr
 }
 \eqno(4.2)
 $$
which are the simplest cases of the fusion procedures [\rKRS].
The fusion procedures for  $ W^{(1)}_2(u)$
 and $ W^{(2)}_1(u)$ above are compactly summarized by the sequences
of the spectral parameter shifts
$(-{1\over 2}$,${1\over 2})$
  and $({1\over 2}$,$-{1\over 2})$, respectively.
For a sequence of complex numbers
$ (u_1,\dots,u_n) $, the permutation of the
adjacent elements, say, $u_i$ and $u_{i+1}$ will be called
nonsingular if $u_i-u_{i+1}\neq \pm 1$. The product of
two nonsigular permutations is also called nonsingular.
 Two sequences $ (u_1,\dots,u_n) $ and $ (v_1,\dots,v_n) $
   are    {\it equivalent}
if one is obtained from the other by applying a nonsingular
permutation.
 For a sequence
$(u_1,\dots,u_n)$, let  $[u_1,\dots,u_n]$ denote
its equivalence class by the above equivalence relation.
Then,
we call the equivalence class $[u_1,\dots,u_n]$
the {\it fusion content} of $W(u)$
 if there is an embedding
$W(u)\emb W^{(1)}_1(u+u_1)\tensor \cdots\tensor
W^{(1)}_1(u+u_n)$.
 The product of two fusion contents is naturally defined as
 $$
 [k_1,\dots,k_n] \cdot [j_1,\dots,j_m]=[k_1,\dots,k_n,j_1,\dots,j_m].
 \eqno(4.3)
 $$
 Thus, the fusion content of the tensor product of two FDRs
    is the product of their fusion contents with the same
    ordering.
 We say two fusion contents are {\it similar} if all the elements
 of both sequences coincide ignoring the ordering.
\par
It is known that
the IFDR $W\am(u)$ has the following fusion content [\rKRS,\rJKMO,\rChe]
 $$
 W\am(u+{a+m \over 2}-1): \
 [a-1,\cdots,1,0]\cdot [a,\cdots,2,1]\cdots[a+m-2,\cdots,m,m-1].
 \eqno(4.4)
 $$
 We also define the IFDR $\til{W}^{(0)}_1(u)$ to be the unique
 one dimensional  module having the fusion content
 $$
 [{r\over 2},{r \over 2}-1,\cdots,-{r \over 2}].
 \eqno(4.5)
 $$
\par
 Let us consider the implication of the fusion contents to
 the Yang-Baxterization of the $Q$-system (2.24a) such as
 $$
 \eqalign{
 T^{(a)}_m(u+c_1)T^{(a)}_m(u+c_2)&
 = T^{(a)}_{m+1}(u+c_3)T^{(a)}_{m-1}(u+c_4)  \cr
 &\qquad +
 T^{(a+1)}_m(u+c_5)T^{(a-1)}_m(u+c_6),  \cr}
 \eqno(4.6)
 $$
where the meaning of  $T^{(r+1)}_m(u)$ is undefined for a while.
\par
 We first consider the case $1\leq a \leq r-1$ of (4.6).
 Our basic assumption here is the existence of
 the exact sequences of $\til{Y}(A_r)$-modules
 $$
 0 \lra \cA_1(u) \lra \cA_2(u) \lra \cA_3 (u)\lra 0
 \eqno(4.7)
 $$
 such that
 $$
 \eqalignno{
 \cA_1(u)&=W^{(a)}_{m+1}(u+c_3)\tensor W^{(a)}_{m-1}(u+c_4)\  {\rm or}  \
 W^{(a)}_{m-1}(u+c_4)\tensor W^{(a)}_{m+1}(u+c_3), \cr
 \cA_2(u)&=W^{(a)}_m(u+c_1)\tensor W^{(a)}_m(u+c_2)\   {\rm or} \
 W^{(a)}_m(u+c_2)\tensor W^{(a)}_m(u+c_1), \cr
 \cA_3(u)&=W^{(a+1)}_m(u+c_5)\tensor W^{(a-1)}_m(u+c_6)\  \ \cr
 &\hskip70pt {\rm or} \
 W^{(a-1)}_m(u+c_6)\tensor W^{(a+1)}_m(u+c_5). &
 (4.8)\cr}
 $$
 Then, (4.6) is a consequence of (4.7--8) as explained in section 2.2.
 In section 2.2, we saw that a $\til{Y}(A_r)$-homomorphism
 only interchanges the
 ordering of the elements of the fusion contents. Thus,
 a necessary condition for (4.7) to exist is that
 the fusion contents
 of all the three representations $\cA_i(u)$ ($i=1,2,3$) are similar.
 This condition, in fact,  {\it almost}
 determines the possible form of the Yang-Baxterization. To
see it, we let $c_1=0$,
 $c_2 \geq 0$ without losing the generality, then the above condition
 is satisfied if and only if
 $$
 c_2=1,\quad c_3=c_4=c_5=c_6={1 \over 2},
 \eqno(4.9)
 $$
 or
 $$
 \eqalign{
 &c_2=a,\quad (c_3,c_4)=({1 \over 2},a+{1\over 2})\ {\rm or}\ (a-{1 \over 2},
 -{1 \over 2}), \quad\cr
 &(c_5,c_6)=({1 \over 2},a+{1 \over 2})\ {\rm or}\
 (a-{1 \over 2},-{1 \over 2}),\hskip50pt {\rm if}\ a=m\geq 2.}
 \eqno(4.10)
 $$
 The solution (4.9) reproduces our $T$-system (2.17) for $a\leq r-1$.
 On the other hand, the known formula (2.20) tells that there are no
 exact sequences of the form (4.7) corresponding to the solutions (4.10),
  thus, they are irrelevant to
 the $T$-system.

\par
In the case $a=r$, the numbers of the elements in
  the  fusion contents of ${\cal A}_2(u)$ and ${\cal A}_3(u)$
    can be matched by replacing ${\cal A}_3(u)$ in (4.8) with
 $$
\cA_3(u)= \til{W}^{(0)}_1(u+d_1)\tensor \cdots
 \tensor \til{W}^{(0)}_1(u+d_m)\tensor W^{(r-1)}_m(u+c_6).
 \eqno(4.11)
 $$
 The solution analogous to (4.9) is
 $$
 c_2=1,\quad c_3=c_4=c_6={1 \over 2},\quad d_k=-{m \over 2}+k,
 \eqno(4.12)
 $$
 which agrees with the definition of the scalar factor $T^{(r+1)}_m(u)$ defined
through (2.18).
 Again, there is an extra solution when $m=r$ as in (4.10), to
which we cannot
 associate an exact sequence.

 To summarize, the assumption of the existence of the exact sequences (4.7)
together with  the fusion contents data imposes
 a  strong constraint on the possible form of the $T$-system
 and the scalar functions $g\am (u)$.

 \subsect{4.2. General cases}
 \subskip

 Let us study (3.20) for other $X_r$'s by the method described in the previous
 subsection. We consider the algebra $\til{Y}(X_r)$ of the $L$-operators
analogously defined as (2.8-9). The algebra $\til{Y}(X_r)$ is a central
extension of $Y(X_r)$ [\rDr].
As in the case $A_r$, the IFDRs of $\til{Y}(X_r)$
are constructed though the fusion procedure using the
 $R$-matrix of the ``minimal" representation as building
 blocks [\rKRii,\rOW,\rCPiii].
 The list of the
 minimal representations  is
 $$
 \matrix{
 A_r: & W^{(1)}_1(u), & &
 E_6:&W^{(1)}_1(u), W^{(5)}_1(u),&&
 F_4:&W^{(4)}_1(u),\cr
 B_r: & W^{(r)}_1(u) ,&&E_7:&W^{(6)}_1 (u), &&
 G_2: &W^{(2)}_1 (u).\cr
 C_r:& W^{(1)}_1(u) , && E_8:&W^{(1)}_1(u) ,&&\cr
 D_r:& W^{(r-1)}_1(u),W^{(r)}_1(u) ,&&&&&&\cr
 }\eqno(4.13)
 $$
We consider the fusion contents  with respect to these
minimal representations. For that purpose, in the definition
of the nonsingular exchanges described
after (4.2), the condition $u_i-u_{i+1}\neq \pm 1$ should be replaced
by
$u_i-u_{i+1}\notin \{u|\det \Rc (u)=0 \}$
for the $R$-matrix $\Rc(u)$ of the minimal representation.
Then, we call the equivalence class $[u_1,\dots,
u_n]$ the fusion content of $W (u)$ if
there is an embedding
$ W(u) \emb W_{\rm min}(u+u_1)\tensor \cdots \tensor
 W_{\rm min}(u+u_n)$ with $W_{\rm min}(u)$ being
the minimal representation.
(In the cases $D_r,E_6$,
each factor $W_{\rm min}(u+u_k)$ in the above is
considered to be  one of two minimal representations,
and we distinguish the spectral shifts of these two representations.
See (4.31c) and (4.33c). The definition of the equivalence classes
also needs to be  modified adequately.)
We remark that
as $X_r$-modules,
the minimal representations
   in (4.13) are just $V_{\Lambda_a}$ with the only exception
   for $E_8$ where $W^{(1)}_1 = V_{\Lambda_1} \oplus V_0$.
 The analysis as in section 4.1
is straightforward once  the fusion contents of all
  $W\am (u)$'s are known.
 In our normalization of the
 spectral parameter here, the higher representations are assumed to be
 obtained by the fusion
 $$
W\am(u)\emb
 W^{(a)}_1(u-{m-1\over 2t_a})\tensor
 W^{(a)}_1(u-{m-3\over 2t_a})\tensor
 \cdots \tensor
 W^{(a)}_1(u+{m-1\over 2t_a}).
 \eqno(4.14)
 $$
 Though we lack a general proof of (4.14),
  in most cases the assumption (4.14) itself is a necessary
 condition for the exact sequences questioned to exist.
  From (4.14), it is enough to know
  the fusion contents
 of the fundamental representations.

 Let us illustrate the analysis in the case $G_2$ in
 some detail.
 By explicitly calculating the coefficients in (3.5a), we have [\rKu]
 $$
 W^{(1)}_1= \V{1}\oplus V_0, \
 W^{(2)}_1= \V{2}, \
 W^{(2)}_2= V_{2\om{2}} \oplus \V{2}.
 \eqno(4.15)
 $$
 The $R$-matrix for the minimal representation $\W{2}$ is  [\rOgi]
 $$
 \eqalign{
 \Rc(u)  &=
 (1+3u)(4+3u)(6+3u)P_{2\om{2}} \oplus (1-3u)(4+3u)(6-3u) P_{0} \cr
 &\quad  \oplus (1-3u)(4+3u)(6+3u)\pr{1} \oplus
 (1+3u)(4-3u)(6+3u)\pr{2}. \cr
 } \eqno(4.16)
 $$
 By comparing (4.15) and (4.16), $W^{(2)}_2(u)$ and
$W^{(1)}_1(u)$ can be defined through the
fusion procedures
 $$
 \eqalign{
  W^{(2)}_2(u) \emb
& W^{(2)}_1(u-{1 \over 6}) \tensor W^{(2)}_1(u+{1 \over 6})
 , \cr
W^{(1)}_1(u) \emb
 & W^{(2)}_1(u+{1 \over 6})
  \tensor W^{(2)}_1(u-{1 \over 6})  . \cr
 }
 \eqno(4.17)
 $$
 In addition, we define $\til{W}^{(2)}_1(u)$ by the fusion
 $$
\til{W}^{(2)}_1(u) \emb
  W^{(2)}_1(u+{2 \over 3}) \tensor W^{(2)}_1(u-{2\over 3})
   ,
 \eqno(4.18)
 $$
 where $\til{W}^{(2)}_1(u)$ and $W^{(2)}_1(u)$ are isomorphic as
 $Y(G_2)$-modules, but  distinguished as $\til{Y}(G_2)$-
 modules.
 \Null{ This is a common feature for all the exceptional
 algebras in contrast to the situation for the non-exceptional algebras,
 where
 such a non-trivial isomorphism
 only occurs through the tensoring of the factor
 $\til{W}^{(0)}_1(u+u_i)$.}
 By (4.16--17),  we have the fusion contents
 $$
 W^{(1)}_1(u):  [{1\over6}, -{1\over6}],\quad
 W^{(2)}_1(u):  [0], \quad
 \tilde{W}^{(2)}_1(u):  [ {2\over3}, -{2\over3}].
 \eqno(4.19)
 $$
 \par
 Now let us consider the necessary condition for the existence of
 (4.7) corresponding to  the  relation
 $$
 Q^{(1)}_1{}^2 =
 Q^{(1)}_{2}+  Q^{(2)}_{3}
 \eqno(4.20)
 $$
in (3.7).
 We first assume the forms of $\cA_1(u)$ and $\cA_2(u)$ as
 (we shall
ignore the precise ordering of the tensor products in this argument),
 $$
 \eqalign{
 \cA_1(u)&=W^{(1)}_2(u), \cr
 \cA_2(u)&=W^{(1)}_1(u+c_1)\tensor W^{(1)}_1(u+c_2),\quad (c_1 \leq
c_2) .\cr
 }
 \eqno(4.21)
 $$
In order for (4.7) to exist, their fusion contents
   must be similar (denoted by $\sim$), therefore we have
 $$
   [-{1\over3},-{2\over3},{2\over3},{1\over3}] \sim
   [c_1+{1\over6},c_1 -{1\over6}]  \cdot
 [c_2+{1\over6}, c_2-{1\over6}],
  \eqno(4.22)
 $$
by (4.14) and (4.19). It is easy to see
 that the unique solution of (4.22) is
 $$
 c_1=-{1\over 2},\quad c_2={1\over 2}.
 \eqno(4.23)
 $$
 The third space $\cA_3(u)$ requires  more  consideration. If we simply assume
 $$
 \cA_3(u)=W^{(2)}_3(u+c_3),
 \eqno(4.24)
 $$
 then, its fusion content becomes
 $[c_3-{1\over3},c_3,c_3+{1\over3}]$ and there is no way
 to match it with (4.22). A trick is to consider the fusion
 $$
\til{W}^{(2)}_3(u) \emb
  W^{(2)}_1(u-{1 \over 3}) \tensor \til{W}^{(2)}_1(u)
 \tensor W^{(2)}_1(u+{1 \over 3})
 \eqno(4.25)
 $$
 instead of (4.14).
 The fusion content of the rhs is then $[-{1\over3},
 {2\over3},-{2\over3},{1\over3}]$,  which is similar to (4.22). So
 we may identify $\cA_3(u)=\til{W}^{(2)}_3(u)$.
 It follows that the only possible $T$-system for (4.20) based on
 the assumption of the exact sequence of type (4.7) is
 $$
 T^{(1)}_1(u-{1\over 2}) T^{(1)}_1(u+{1\over2})=
 T^{(1)}_{2}(u)
  +g(u) \, T^{(2)}_{3}(u),
 \eqno(4.26)
 $$
 and the function $g(u)$ is defined through the factorization
 $$
 \tilde{T}^{(2)}_1(u)=g(u)\, T^{(2)}_1(u),\eqno(4.27)
 $$
 where $\tilde{T}^{(2)}_1(u)$ is the transfer matrix associated to
 $\tilde{W}^{(2)}_1(u)$. Repeating these arguments, one finds
 that in the case $G_2$, our $T$-system (3.20e) is the only possible solution
 under the assumptions  (4.7) and (4.14). The scalar functions
 in (3.20e) are fixed as
 $$
 g^{(1)}_m(u)=
 g(u-{m-1\over 2})
 g(u-{m-3\over 2})
 \cdots
 g(u+{m-1\over 2}), \quad g^{(2)}_m(u)=1,
 \eqno(4.28)
 $$
 where $g(u)$ is given in (4.27).
 In particular,  (3.18) is automatically satisfied.

 Similar analyses have been
 done for all  $X_r$'s except $E_7$ and $E_8$.
 In all the cases,
 (3.20)  has certainly a structure consistent with the fusion procedure,
 and furthermore the only possible solution in most
 cases.

 Below we only present the necessary data of  (a) the  $R$-matrix,
 (b) the explicit form of (3.5a) for most fundamental representations
 [\rKRii,\rOW],
(c) the fusion contents of fundamental representations
 and (d) the nontrivial scalar functions $g\am (u)
\neq 1$.
All the spectral decomposition formulas of the
rational $R$-matrices in data (a)
are found in the references for
$B_r$, $D_r$ [\rResi], $C_r$ [\rORW], $E_6$, $F_4$ [\rOW].
Additional comments are included  when necessary.
For $D_r$ and  $E_6$,
 there is an additional type of
isomorphisms by the conjugate transformation. Those are
indicated in data (c)  by the symbol $\simeq$.
In data (d),
  $\til{T}^{(a)}_1(u)$ is the transfer
matrix associated to $\til{W}^{(a)}_1(u)$
which is found in data (c).

\subskip
 $X_r=B_r$:  
$$
\eqalignno{
&((2r-1+2u)(2r-3+2u)\cdots (1+2u))^{-1} \Rc(u) \cr
&\qquad=P_{2\om{r}}  \oplus
\bigoplus_{i=2 \atop i: {\rm even}}^r
{(2i-1-2u)(2i-5-2u)\cdots(3-2u) \over
 (2i-1+2u)(2i-5+2u)\cdots(3+2u)} \pr{r-i} \cr
&\qquad\qquad \oplus
\bigoplus_{i=1 \atop i: {\rm odd}}^r
{(2i-1-2u)(2i-5-2u)\cdots(1-2u) \over
 (2i-1+2u)(2i-5+2u)\cdots(1+2u)} \pr{r-i}. & (4.29{\rm a}) \cr
}
$$
$$
\Wnu{a}=
 \V{a}\oplus\V{a-2}\oplus \cdots \ , \  (1 \leq a \leq r-1) ,
 \quad  \Wnu{r}=  \V{r}.  \eqno(4.29{\rm b})
$$
%
$$
\eqalignno{
&W^{(a)}_1(u): [{2r-2a-1\over 4},-{2r-2a-1\over 4}],\  (1\leq a\leq r-1), \cr
&W^{(r)}_1(u): [0], \quad
\tilde{W}^{(0)}_1(u): [{2r-1 \over 4},-{2r-1 \over 4}]. &
(4.29{\rm c}) \cr}
$$
%
$$
\eqalignno{
&g^{(1)}_m(u)=\prod_{k=1}^m g(u-{m-2k+1 \over 2}),\quad
\tilde{T}^{(0)}_1(u)=g(u)\, \Id .  & (4.29{\rm d}) \cr}
$$

\subskip
$X_r=C_r$:
$$
\eqalignno{
\Rc(u)&=(1+2u)(r+1+2u)P_{2\om{1}}\oplus (1+2u)(r+1-2u)P_{0}\cr
&\qquad \oplus
 (1-2u)(r+1+2u)\pr{2}. &
(4.30{\rm a})\cr}
$$
$$
\Wnu{a} =\V{a}, \quad (1\leq a \leq r) .
\eqno(4.30{\rm b})
$$
$$
W^{(a)}_1(u): [{a-1 \over 4},{a-3 \over 4},\dots,
-{a-1 \over 4}],\  (1\leq a\leq r), \quad
\tilde{W}^{(0)}_1(u): [{r+1 \over 4},-{r+1 \over 4}].
\eqno(4.30{\rm c})
$$
$$
\eqalignno{
&g^{(r)}_m(u)=\prod_{k=1}^m g(u-{m-2k+1 \over 2}),\quad
\tilde{T}^{(0)}_1(u)=g(u)\, \Id .  & (4.30{\rm d}) \cr}
$$

\subskip
 $X_r=D_r$: We have two minimal representations $W^{(r-1)}_1(u)$ (conjugate
spin reps.) and $W^{(r)}_1(u)$ (spin reps.). Thus, we need two $R$-matrices
$\Rc_{\om{r-1},\om{r}}(u)$, $\Rc_{\om{r},\om{r}}(u)$
below as
 building
blocks of the $\tilde{Y}(D_r)$ representation. We distinguish the fusion
contents
 relevant to
$W^{(r-1)}_1(u+j)$ and $W^{(r)}_1(u+j)$
 as $[\ba{j}]$ and $[j]$, respectively. There are
isomorphisms of IFDRs for the ``vector'' type representations $W\am(u)$
 $(1\leq a \leq r-2)$
as in
(4.31c). The spectral decomposition of $\Rc_{\om{r},\om{r-1}}(u)$
coincides with that of $\Rc_{\om{r-1},\om{r}}(u)$. The spectral
decomposition of $\Rc_{\om{r-1},\om{r-1}}(u)$ is obtained by that of
$\Rc_{\om{r},\om{r}}(u)$ with $P_{2\om{r}}$ replaced with
$P_{2\om{r-1}}$.
$$
\eqalignno{
&((2[(r-1)/2]+u)\cdots (4+u)(2+u))^{-1} \Rc_{\om{r-1},\om{r}}(u) \cr
&\qquad=P_{\om{r-1}+\om{r}}  \oplus
\bigoplus_{i=1}^{[(r-1)/2]}
{(2i-u)\cdots(4-u)(2-u) \over
 (2i+u)\cdots(4+u)(2+u)} \pr{r-2i-1}, \cr
&((2[r/2]-1+u)\cdots(3+u) (1+u))^{-1} \Rc_{\om{r},\om{r}}(u) \cr
&\qquad=P_{2\om{r}}  \oplus
\bigoplus_{i=1}^{[r/2]}
{(2i-1-u)\cdots(3-u)(1-u) \over
 (2i-1+u)\cdots(3+u)(1+u)} \pr{r-2i}. & (4.31{\rm a})\cr
}
$$
$$
\Wnu{a}=
 \V{a}\oplus\V{a-2}\oplus \cdots \ , \  (1 \leq a \leq r-2) ,
\ \Wnu{r-1}=  \V{r-1},
 \   \Wnu{r}=  \V{r}.  \eqno(4.31{\rm b})
$$
$$
\eqalignno{
&W^{(a)}_1(u):
\cases{ [{r-a-1 \over 2}, -{r-a-1 \over 2} ]
\simeq [\ba{{r-a-1 \over 2}},-\ba{{r-a-1 \over 2}}], &
 $r-a$: even, \cr
 [{{r-a-1 \over 2}},-\ba{r-a-1 \over 2}]
\simeq [\ba{{r-a-1 \over 2}},-{r-a-1 \over 2}], &
  $r-a$: odd, \cr} \cr
&\hskip150pt (1 \leq a\leq r-2),\cr
&W^{(r-1)}_1(u):  [\ba{0}], \quad
W^{(r)}_1(u): [0], \cr
&\tilde{W}^{(0)}_1(u):  \cases{
[{r-1 \over 2},-{r-1 \over 2}] \simeq [\ba{{r-1 \over 2}},-\ba{{r-1 \over 2}}],
&
$r$: even, \cr
[{r-1 \over 2},-\ba{{r-1 \over 2}}] \simeq [\ba{{r-1 \over 2}},-{r-1 \over 2}],
&
$ r$: odd. \cr} & (4.31{\rm c})\cr
}
$$
$$
\eqalignno{
&g^{(1)}_m(u)=\prod_{k=1}^m g(u-{m-2k+1 \over 2}),\quad
\tilde{T}^{(0)}_1(u)=g(u)\, \Id .  & (4.31{\rm d}) \cr}
$$
\subskip

$X_r=F_4$: The fusion content of $W^{(2)}_1(u)$
can be determined from the fusion
$W^{(2)}_1(u)\emb W^{(3)}_1(u+{1\over4})
\tensor W^{(4)}_1(u-{1\over2})$.
$$
\eqalignno{
\Rc(u)  &=
(1+2u)(4+2u)(6+2u)(9+2u)P_{2\om{4}} \oplus (1+2u)(4-2u)(6+2u)(9-2u) P_{0} \cr
&\quad  \oplus (1+2u)(4-2u)(6+2u)(9+2u)\pr{1} \oplus
(1-2u)(4+2u)(6+2u)(9+2u)\pr{3} \cr
& \quad \oplus (1-2u)(4+2u)(6-2u)(9+2u) \pr{4}. &(4.32{\rm a})\cr
}
$$
$$
\W{1}=  \V{1}\oplus V_0, \
\W{3}=  \V{3}\oplus \V{4}, \
\W{4}=  \V{4}.
\eqno(4.32{\rm b})
$$
$$
\eqalignno{
&W^{(1)}_1(u): [1,-1],\quad
W^{(2)}_1(u): [{1\over 2},0,-{1\over2}], \quad
W^{(3)}_1(u): [{1\over 4},-{1\over 4}], \quad \cr
&W^{(4)}_1(u): [0], \quad
\tilde{W}^{(4)}_1(u): [{3\over 2},-{3\over 2}]. &(4.32{\rm c}) \cr
}
$$
$$
\eqalignno{
&g^{(1)}_m(u)=\prod_{k=1}^m g(u-{m-2k+1 \over 2}),\quad
\tilde{T}^{(4)}_1(u)=g(u)\, T^{(4)}_1(u).  & (4.32{\rm d}) \cr}
$$

\subskip

$X_r=E_6$: Like the $D_r$ case,
 we use the conjugate pair $W^{(1)}_1(u)$ and
 $W^{(5)}_1(u)$ as building blocks.
 The spectral decomposition of $\Rc_{\om{5},\om{1}}(u)$
coincides with that of $\Rc_{\om{1},\om{5}}(u)$. The spectral
decomposition of $\Rc_{\om{5},\om{5}}(u)$ is obtained from that of
$\Rc_{\om{1},\om{1}}(u)$ by the simultaneous replacements
$\Lambda_a \leftrightarrow \Lambda_{6-a}, 1 \le a \le 5$.
(There is a discrepancy between the expressions of the $R$-matrix
$\Rc_{\om{5},\om{1}}(u)$ in  refs.\ [\rOW] and [\rCPiii] and the
latter  seems erroneous.)
The fusion content of $W^{(3)}_1(u)$ is obtained from the fusions
$W^{(3)}_1(u) \emb W^{(2)}_1(u+{1\over2})\tensor W^{(1)}_1(u-1)$ and
$W^{(3)}_1(u) \emb W^{(4)}_1(u+{1\over2})\tensor W^{(5)}_1(u-1)$
$$
\eqalignno{
\Rc_{\om{1},\om{1}}(u) & =(1+u)(4+u)P_{2\om{1}} \oplus (1-u)(4+u)\pr{2}
\oplus (1-u)(4-u)\pr{5}, \cr
\Rc_{\om{5},\om{1}}(u) & =(3+u)(6+u)P_{\om{1}+\om{5}} \oplus (3-u)(6-u)P_{0}
\oplus (3-u)(6+u)\pr{6}.\cr
&& (4.33{\rm a})\cr
}
$$
$$
\Wnu{1}=  \V{1}, \
\Wnu{2}=  \V{2}\oplus \V{5}, \
\Wnu{5}= \V{5}, \
\Wnu{6}=  \V{6}\oplus V_{0}.
\eqno(4.33{\rm b})
$$
$$
\eqalignno{
&W^{(1)}_1(u): [0], \
W^{(2)}_1(u): [{1\over 2},-{1\over 2}], \
W^{(3)}_1(u): [1,0,-1]\simeq [\ba{1},\ba{0},\ba{-1}], \cr
&W^{(4)}_1(u): [\ba{{1\over 2}},-\ba{{1\over 2}}], \
W^{(5)}_1(u): [\ba{0}], \
W^{(6)}_1(u): [{3\over 2},-\ba{{3\over 2}}]\simeq
[\ba{{3\over 2}},-{3\over 2}], \cr
&\tilde{W}^{(5)}_1(u): [2,-2].&(4.33{\rm c}) \cr
}
$$
$$
\eqalignno{
&g^{(6)}_m(u)=\prod_{k=1}^m g(u-{m-2k+1 \over 2}),\quad
\tilde{T}^{(5)}_1(u)=g(u)\, T^{(5)}_1(u).  & (4.33{\rm d}) \cr}
$$

\sect{5. Determinant formulas for $B_r, C_r$ and $D_r$}
\subskip

In this section we consider the classical series
$X_r = B_r, C_r, D_r$ and observe that (3.20)
leads to remarkable conjectures of determinant formulas analogous to (2.20).
We regard these as  further supports of our $T$-system from an
algebraic viewpoint.

As is well known, by the tensor product the FDRs of $X_r$ generate a ring
$R(X_r)$ called the representation ring [\rFH]. The character (3.6b) is a
one-dimensional representation of $R(X_r)$. From the definition
of $Q\am$
in (3.6a), the $Q$-system is a set of  relations of
the ring $RQ(X_r)\subset R(X_r)$ generated by $Q\am$'s.
In the same way, the restricted $Q$-system is that of
the ring $RQ_{\ell}(X_r) \subset R_{\ell}(X_r)=R(X_r)/I_{\ell}$
generated by the images of $Q\am$'s.
Here $I_{\ell}$ is the ideal generated by the FDRs of $X_r$-modules
whose quantum dimensions at $q=\exp(2\pi i/(\ell+g))$ are zero.
The  matrices $\overline{M}$ and $M$
in (3.13) are representations of $RQ(X_r)$ and $RQ_{\ell}(X_r)$,
respectively.
It is now natural to regard the rings $RT(X_r)$ and
   $RT_\ell(X_r)$ of the transfer matrices for the vertex and
   level $\ell$ RSOS models as quantum analogues of
   the $RQ(X_r)$ and $RQ_\ell(X_r)$, respectively.
 Though
the tensor products of quantum group modules here are indecomposable,
the transfer matrices have a decomposability property
and play a role of the ``character''
in $RT(X_r)$.
\par
For our $T$-system (3.20),
one can in principle solve it successively for $T\am(u)$ in terms
 of the $T^{(a)}_1(u)$'s. Based on  studies of several examples,
we
conjecture
that $T\am(u)$ is expressed as a polynomial in $T^{(a)}_1(u)$'s,
i.e.,
 the ring $RT(X_r)$ is generated by
$T^{(a)}_1(u)$'s for any $X_r$. This is a natural quantum analogue of
the fact that
$R(X_r)$ is generated by the fundamental representations [\rFH].

\par
Furthermore, in the $A_r$ case, all the $T\am(u)$ are
expressed  through a  beautiful
determinant formula (2.20b).  This was
a consequence of
the resolutions of  quantum group modules [\rCheiii], whose
 classical counterpart has been also
 studied in [\rZel, \rAki].
Below we present
the  conjectures on
determinant formulas for $B_r,C_r,D_r$ analogous to (2.20b).
They seem novel even in the classical context to authors'
knowledge.
%
%
The following notations are adopted below.
\item{1.} $T^{(a)}_1(u+k)=x^{a}_k$ for $0 \le a \le r$.
\item{2.} For a semi-infinite dimensional matrix
$M = (M_{i j})_{1 \le i,j < \infty}$,
we denote by $M(a,b,m)$
the $m \times m$ submatrix
$(M_{i j})_{a \le i \le a+m-1, b \le j \le b+m-1}$.\pn
\subskip\pn
$X_r=B_r$:\par
We find it convenient to deal with the
following $T$-system which is equivalent to
(3.20b) through some elementary redefinitions.
$$\eqalign{
T^{(a)}_m(u)T^{(a)}_m(u+2)
&= T^{(a)}_{m+1}(u)T^{(a)}_{m-1}(u+2) +
T^{(a+1)}_m(u)T^{(a-1)}_m(u+2),\cr
&\hskip130pt  (1\le a \le r-2),\cr
T^{(r-1)}_{m}(u)T^{(r-1)}_{m}(u+2)
&= T^{(r-1)}_{m+1}(u)T^{(r-1)}_{m-1}(u+2) +
T^{(r-2)}_m(u+2)T^{(r)}_{2m}(u+1),\cr
T^{(r)}_{2m}(u)T^{(r)}_{2m}(u+1)
&= T^{(r)}_{2m+1}(u)T^{(r)}_{2m-1}(u+1) +
T^{(r-1)}_m(u)T^{(r-1)}_{m}(u+1),\cr
T^{(r)}_{2m+1}(u)T^{(r)}_{2m+1}(u+1)
&= T^{(r)}_{2m+2}(u)T^{(r)}_{2m}(u+1) +
T^{(r-1)}_m(u+1)T^{(r-1)}_{m+1}(u).\cr
}\eqno(5.1)
$$
In the above, $T^{(0)}_m(u)$
is to be understood as the scalar function  obeying
$T^{(0)}_m(u) = T^{(0)}_1(u)T^{(0)}_1(u+2) \cdots T^{(0)}_1(u+2m-2)$
corresponding to  (4.29d).
Solving (5.1) recursively, one finds, for example
    in $B_3$ case, that
$$\eqalignno{
T^{(1)}_2(u)
   =&x^{1}_0 x^{1}_2-x^{0}_2 x^{2}_0,
                          \cr
T^{(2)}_2(u)
   =&x^{2}_0 x^{2}_2-
     x^{1}_2 x^{3}_1 x^{3}_2+ x^{1}_2 x^{2}_1,
                          \cr
T^{(3)}_2(u)
   =&x^{3}_0 x^{3}_1 -x^{2}_0  ,
                          \cr
T^{(1)}_3(u)
   =&x^{1}_0 x^{1}_2 x^{1}_4 -
     x^{0}_4 x^{1}_0 x^{2}_2 -
     x^{0}_2 x^{1}_4 x^{2}_0
    -x^{0}_2 x^{0}_4 x^{2}_1
    -x^{0}_2 x^{0}_4 x^{3}_1 x^{3}_2,
                           \cr
T^{(2)}_3(u)
   =&-x^{1}_2 x^{1}_3 x^{1}_4 +
      x^{1}_4 x^{2}_0 x^{2}_3
     +x^{1}_2 x^{2}_1 x^{2}_4 +
      x^{2}_0 x^{2}_2 x^{2}_4
      -x^{1}_2 x^{2}_4 x^{3}_1 x^{3}_2
    + x^{1}_2 x^{1}_4 x^{3}_1 x^{3}_4\cr
   &    -x^{1}_4 x^{2}_0 x^{3}_3 x^{3}_4
      + x^{0}_4 x^{1}_3 x^{2}_2
      + x^{0}_4 x^{2}_1 x^{2}_3
      - x^{0}_4 x^{2}_3 x^{3}_1 x^{3}_2
     - x^{0}_4  x^{2}_2  x^{3}_1  x^{3}_4  \cr
     &- x^{0}_4  x^{2}_1  x^{3}_3  x^{3}_4
     + x^{0}_4    x^{3}_1  x^{3}_2   x^{3}_3  x^{3}_4,\cr
T^{(3)}_3(u)
    =&-x^{2}_1 x^{3}_0 - x^{2}_0 x^{3}_2 +
      x^{3}_0 x^{3}_1 x^{3}_2 .
          &(5.2)             \cr
}$$
Then, we find the following determinant expressions.
$$\eqalignno{
T^{(1)}_2(u) &=             \hbox{det} \Bigl(
\pmatrix{x^{1}_0&          0&  x^{2}_0\cr
                 0&          0&          0\cr
         x^{0}_2&          0&  x^{1}_2\cr} +
\pmatrix{        0&        x^{3}_1  &          0\cr
                 0&          1&   0\cr
                 0&          0&          0 \cr}
         \Bigr),
                          &(5.3{\rm a}) \cr
T^{(2)}_2(u)             &=\hbox{det} \Bigl(
\pmatrix{x^{2}_0& 0 & -x^{2}_1\cr
                 0&          0&  -x^{3}_2       \cr
         x^{1}_2&          0&  x^{2}_2\cr} +
\pmatrix{        0&  x^{3}_1      &          0\cr
                 0&           1&  0\cr
                 0&          0&          0 \cr}
         \Bigr),
                          &(5.3{\rm b}) \cr
T^{(1)}_3(u)           &=\hbox{det} \Bigl(
\pmatrix{x^{1}_0&          0&  x^{2}_0&0& -x^{2}_1\cr
                 0&          0&          0&          0&
        -x^{3}_2\cr
         x^{0}_2&          0&  x^{1}_2&          0&  x^{2}_2\cr
                 0&          0&          0&          0&          0\cr
                 0&          0&  x^{0}_4&          0&  x^{1}_4\cr}
         +
\pmatrix{        0&      x^{3}_1    &          0&
       0&          0\cr
                 0&          1&0 &         -1&          0\cr
                 0&          0&          0&    x^{3}_3     &          0\cr
                 0&          0&          0&          1&  0\cr
                 0&          0&          0&          0&          0\cr}
           \Bigr),\cr
                   &   &(5.3{\rm c})\cr
T^{(2)}_3(u)           &=\hbox{det} \Bigl(
\pmatrix{x^{2}_0& 0& -x^{2}_1&          0& -x^{1}_3\cr
                 0&          0&          -x^{3}_2&          0&          0\cr
         x^{1}_2&          0&  x^{2}_2& 0 & -x^{2}_3\cr
                 0&          0&          0&          0&         - x^{3}_4\cr
         x^{0}_4&          0&  x^{1}_4&          0&  x^{2}_4\cr}
         +
\pmatrix{        0&      x^{3}_1    &          0&
         0& 0\cr
                 0&          1&0 &         -1&          0\cr
                 0&          0&          0&    x^{3}_3     &          0\cr
                 0&          0&          0&          1&  0\cr
                 0&          0&          0&          0&          0\cr}
           \Bigr). \cr                     &      &(5.3{\rm d})
}$$
In (5.3) we separate the matrices into two parts by intention.
As seen immediately, two matrices in (5.3a,b)  are submatrices of
the corresponding ones in (5.3c,d).
Furthermore, the second matrices in (5.3c) and (5.3d)
    are identical while the first ones seem to be submatrices
    of a common bigger size matrix.
These observations are extended to a general conjecture for $B_r$
 as follows.
Let us
define the semi-infinite
dimensional matrices ${\cal T}^{B_r}$ and $\varepsilon^{B_r}$ by
$$\eqalign{
{\cal T}^{B_r}_{i j} &= \cases{
x^{{j-i \over 2}+1}_{i-1} &
if $i \in 2{\bf Z}+1$
and ${i-j \over 2} \in \{1,0,\ldots,2-r\}$,\cr
-x^{{i-j \over 2}+2r-2}_{j-2r+2} &
if $i \in 2{\bf Z}+1$ and
${i-j \over 2} \in
\{1-r,-r,\ldots,2-2r\}$,\cr
-x^r_i & if $i \in 2{\bf Z}$ and $j=i+2r-3$,\cr
0 &otherwise,\cr}\cr
\varepsilon^{B_r}_{i j} &= \cases{
\pm 1 & if $i = j - 1 \pm 1$ and $i \in 2{\bf Z}$,\cr
x^r_i & if $i = j - 1$ and $i \in 2{\bf Z} + 1$,\cr
0 & otherwise.\cr}\cr}\eqno(5.4)$$
For example, for $B_4$, they look as
$$
{\cal T}^{B_4}=
\pmatrix{     x^1_0&     0&       x^2_0&      0&
          x^{3}_0&     0&  -x^{3}_1&      0& -x^{2}_3&     0&
             -x^1_{5}&      0&         -x^0_{7}&     \cr
                  0&     0&           0&      0&
                  0&     0&     -x^{4}_2&     0&          0&     0&
                      0&      0&         0 &      \cr
             x^0_2& 0& x^1_2& 0& x^{2}_2& 0&  x^{3}_2&
         0& -x^{3}_3& 0& -x^2_{5}& 0& -x^1_{7}& \cdots \cr
             0& 0&     0& 0&         0& 0&        0&
         0&   -x^{4}_4& 0&  0& 0& 0 &  \cr
        &  &     &   &  &  &  & \vdots & &
          & & &   & \ddots \cr
}, \eqno(5.5{\rm a})
$$
$$
\varepsilon^{B_4}=
\pmatrix{ 0& x^4_1&  0&     0&     0&     0&     0&     0&     0&  \cr
          0&     1&  0&    -1&     0&     0&     0&     0&     0&  \cr
          0&     0&  0& x^4_3&     0&     0&     0&     0&     0&  \cr
          0&     0&  0&     1&     0&    -1&     0&     0&     0& \ldots \cr
          0&     0&  0&     0&     0& x^4_5&     0&     0&     0&  \cr
          0&     0&  0&     0&     0&     1&     0&    -1&     0&  \cr
          0&     0&  0&     0&     0&     0&     0& x^4_7&     0&  \cr
      &&    &       & \vdots&       &       &       &       & \ddots \cr
         }.
\eqno(5.5{\rm b})
$$
Then we have a conjecture
$$
T^{(a)}_{m}(u) = \hbox{det}({\cal T}^{B_r}(1,2a-1,2m-1)+
\varepsilon^{B_r}(1,1,2m-1) ), \eqno(5.6)
$$
for $1\le a\le r-1, m \ge 1$.
This has been verified up to $r= 5, m=4$.
\subskip\pn
$X_r=C_r$:\par
We consider the
following $T$-system which is equivalent to
(3.20c) through some elementary redefinitions.
$$\eqalign{
T^{(a)}_m(u)T^{(a)}_m(u+1)
&= T^{(a)}_{m+1}(u)T^{(a)}_{m-1}(u+1) + T^{(a+1)}_m(u)T^{(a-1)}_m(u+1),
\cr
&\hskip130pt
(1 \le a \le r-2),\cr
T^{(r-1)}_{2m}(u)T^{(r-1)}_{2m}(u+1)
&= T^{(r-1)}_{2m+1}(u)T^{(r-1)}_{2m-1}(u+1) \cr
&\qquad +
T^{(r-2)}_{2m}(u+1)T^{(r)}_{m}(u)T^{(r)}_m(u+1),\cr
T^{(r-1)}_{2m+1}(u)T^{(r-1)}_{2m+1}(u+1)
&= T^{(r-1)}_{2m+2}(u)T^{(r-1)}_{2m}(u+1) \cr
&\qquad +
T^{(r-2)}_{2m+1}(u+1)T^{(r)}_{m}(u+1)T^{(r)}_{m+1}(u),\cr
T^{(r)}_{m}(u)T^{(r)}_{m}(u+2)
&= T^{(r)}_{m+1}(u)T^{(r)}_{m-1}(u+2) +
g_m(u)T^{(r-1)}_{2m}(u+1).\cr
}\eqno(5.7)
$$
Here we take $T^{(0)}_m(u) \equiv 1$ in the first equation
and the scalar in the last equation
satisfies $g_m(u) = g_1(u)g_1(u+2) \cdots g_1(u+2m-2)$
corresponding to (4.30d).
Define the semi-infinite
dimensional matrix ${\cal T}^{C_r}$ by
$$\eqalign{
{\cal T}^{C_r}_{i j} &= \cases{
x^{j-i+1}_{i-1} & if $i - j \in \{1,0, \ldots, 1-r\}$,\cr
-y_{i j} x^{i-j+2r+1}_{j-r-1} & if $i - j \in \{-1-r,-2-r, \ldots,
-1-2r\}$,\cr
0 & otherwise, \cr}\cr
y_{i j} &= g_1(u+i-1)g_1(u+i) \cdots g_1(u+j-r-2),\cr}
\eqno(5.8)
$$
where we assume $\forall x^0_k = 1$. For example, for $C_4$, it looks as
$$
{\cal{T}}^{C_4}=\pmatrix{
x^1_0&  x^2_0&  x^3_0&    x^4_0&     0&   -y_{16}x^4_1& -y_{17}x^{3}_2&
       -y_{18}x^{2}_3&   -y_{19}x^1_4&
 \cr
    1&  x^1_1&  x^2_1& x^{3}_1&  x^4_1&        0&     -y_{27}x^4_2&
       -y_{28}x^{3}_3&    -y_{29}x^2_4&  \cdots \cr
    0&      1&  x^1_2&  x^{2}_2&x^{3}_2&  x^4_2&        0&
        -y_{38}x^{4}_3&      -y_{39}x^3_4 &  \cr
&        &        &
&         & \vdots &    &                   &      & \ddots
\cr }.
\eqno(5.9)
$$
Then we have a conjecture
$$
T^{(a)}_{m}(u) = \hbox{det}{\cal T}^{C_r}(1,a,m),
\eqno(5.10)
$$
for $1 \le a \le r-1, m \ge 1$, which can be shown
to actually satisfy the first equation in (5.7)
as in section 2.3.
{}From (5.10), $T^{(r)}_m(u)$ can be solved recursively as
$$
T^{(r)}_m(u)=T^{(r)}_{m-1}(u)
{{\hbox{det}{\cal T}^{C_r}(1,r,2m-1)}\over
{\hbox{det}{\cal T}^{C_r}(1,r,2m-2)}}.
\eqno(5.11)
$$
We have verified that (5.11) actually yields a polynomial in
$x^a_k$'s
up to $r= 5, m=4$.
\subskip
\pn
$X_r=D_r$:\par
We consider the
following $T$-system which is equivalent to
(3.20a) through some elementary redefinitions.
$$\eqalign{
T^{(a)}_m(u)T^{(a)}_m(u+1)
=& T^{(a)}_{m+1}(u)T^{(a)}_{m-1}(u+1) + T^{(a+1)}_m(u)T^{(a-1)}_m(u+1),
\cr
& \hskip100pt (1 \le a \le r-3),\cr
T^{(r-2)}_m(u)T^{(r-2)}_m(u+1)
=& T^{(r-2)}_{m+1}(u)T^{(r-2)}_{m-1}(u+1) \cr
&\qquad
+ T^{(r-3)}_m(u+1)T^{(r-1)}_m(u+1)T^{(r)}_m(u+1),\cr
T^{(b)}_m(u)T^{(b)}_m(u+1) =& T^{(b)}_{m+1}(u)T^{(b)}_{m-1}(u+1)
+ T^{(r-2)}_m(u)\quad b = r-1,r.\cr}
\eqno(5.12)
$$
As in $B_r$ case, we here interpret $T^{(0)}_m(u)$ as the
scalar function obeying
$T^{(0)}_m(u) = T^{(0)}_1(u)T^{(0)}_1(u+1) \cdots T^{(0)}_1(u+m-1)$
corresponding to (4.31d).
Define the semi-infinite
dimensional matrices ${\cal T}^{D_r}$ and $\varepsilon^{D_r}$ by
$$\eqalignno{
{\cal T}^{D_r}_{i j} &= \cases{
x^{{j-i \over 2}+1}_{i-1\over 2} &
if $i \in 2{\bf Z}+1$ and
${i-j\over 2} \in \{1,0,\ldots,3-r\}$,\cr
-x^{r-1}_{i+1\over 2}&if $i \in 2{\bf Z}+1$ and
${i-j\over 2}={5\over 2}-r$,\cr
-x^{r}_{i+1\over 2}&if $i \in 2{\bf Z}+1$ and
${i-j\over 2}={3\over 2}-r$,\cr
-x^{{i-j\over 2}+2r-3}_{j-2r+3 \over 2}
&if $i \in 2{\bf Z}+1$ and
${i-j\over 2} \in \{1-r,-r,\ldots,3-2r\}$,\cr
0&otherwise,\cr}
&(5.13{\rm a})\cr
\varepsilon^{D_r}_{i j} &= \cases{
\pm 1 &if $i=j-2 \pm 2$ and $i \in 2{\bf Z}$,\cr
x^{r-1}_{j-2r+5 \over 2}&if $i=j-3$ and $i \in 2{\bf Z}$,\cr
x^{r}_{j-2r+5 \over 2}&if $i=j-1$ and $i \in 2{\bf Z}$,\cr
0&otherwise.\cr}&(5.13{\rm b})\cr}$$
For example, for $D_4$, they look as
$$
{\cal T}^{D_4} =\pmatrix{
x^1_0&      0&    x^2_0&
   -x^{3}_1&        0&    -x^{4}_1& -x^{2}_1&
        0& -x^{1}_2&             0&    -x^0_{3}&
                       \cr
    0&         0&            0&
            0&        0&           0&
         0&         0&          0&             0&     0&
                          \cdots \cr
    x^0_1&      0&    x^1_1&                0&x^{2}_1&  -x^{3}_2&
        0&  -x^{4}_2& -x^{2}_2&         0&-x^1_{3}&
                     \cr
    0&      0&            0&            0&
            0&        0&           0&
      0&         0&          0&          0&
                   \cr
&      &           &              &
      &         &  \vdots          &          &
         &         &      &
                      \ddots \cr
},\eqno(5.14{\rm a})
$$
$$
\varepsilon^{D_4}=
\pmatrix{
         0&        0&        0&          0&
 0&         0&          0&
         0&        0&   &   \cr
         0&        1& x^4_{0}&         0& x^{3}_{1}&       -1&
          0&
         0&        0&   \ldots&   \cr
         0&        0&        0&          0&            0&
       0&          0&
         0&        0&   &   \cr
         0&        0&        0&          1&    x^4_{1}&
        0& x^{3}_{2}&
       -1&          0&  &  \cr
   &        &         &     &             &     \vdots
    &           &
         &    & \ddots   \cr
}. \eqno(5.14{\rm b})
$$
Then we have a conjecture
$$
T^{(a)}_m(u) =\hbox{det}({\cal T}^{D_r}(1,2a-1,2m-1) +
            \varepsilon^{D_r}(2a-1,2a-1,2m-1) ),
\eqno(5.15)
$$
for $1\le a \le r-2,~~m \ge 1$.
%
\subskip

We have not found simple expressions for
$T^{(r)}_m(u)$ for $B_r$ and
$T^{(r-1)}_m(u)$ and $T^{(r)}_m(u)$ for
$D_r$
but checked that they are expressed as polynomials in
$T^{(a)}_1(u)$'s at least for small $r$ and $m$.

\sect{6. Summary}
\subskip

In this paper we have proposed the $T$-system (3.20),
a new set of functional relations among the commuting family
of the row-to-row transfer matrices, for a class of solvable
lattice models.
They are the vertex and
RSOS type models associated with the  simple Lie
algebras $X_r$ as sketched in section 3.3.
For $X_r = sl(2)$, we have shown
in section 2.2 that the $T$-system is
governed by exact sequences of the quantum group modules,
which  provide a representation theoretical background
for a general $X_r$ case as well.
An intriguing connection has thereby been proved
in section 2.3 among the $T$, $Q$ and
$Y$-systems for $X_r = A_r$, indicating deep relations to
the TBA and dilogarithm identities, etc.
Though the meaning of the connection is yet to be
understood, postulating it for the other  $X_r$'s
has led to our main proposal (3.20).
In fact, (3.20) is almost the unique possible
system of the FRs once one admits the underlying
exact sequences as argued in section 4.
We have also observed that our $T$-system leads
to remarkable determinant formulas for $B_r, C_r$ and $D_r$
in section 5, which may be viewed as a further
support for our proposal.
The $T$-system is to hold either for
  critical or off-critical models
  and for finite or infinite system sizes.
  Its application to the calculation of various
  thermodynamic quantities will be presented in Part II.
\par
An important open problem is to understand the
meaning of the $Y$-system (B.6) and its connection
to the $T$-system (3.19).
While the $T$-system involves the scalar functions (3.18),
the $Y$-system does not. This indicates that the latter is
an object related to $Y(X_r)$ or $U_q(X^{(1)}_r)$
and not their central extensions.
We believe that the connection (3.19) is deep
and its explanation will shed a new light to the
theories of solvable lattice models and related subjects.
\par
There are other kinds of FRs known
for many models described in section 3.3.
For instance, those in
 eqs.(9.8.40) and (10.5.32) of [\rBax] are typical such examples, which
may be viewed more broadly as the FRs
from the analytic Bethe ansatz [\rRes].
It would be interesting
to investigate the implication of
our approach in the light of the analytic Bethe ansatz.
Finally, we remark that the FRs of the transfer matrices
have also been studied [\rABP--\rBaxii]
 for the chiral Potts model [\rAMP].
The relevant representations there are so-called
cyclic representations, which only exist at
$q$ a root of unity. To widen our scope to such
``genuine" quantum representations would be an
important subject.

\sect{Acknowledgements}
\subskip

The authors would like to thank V.V.\ Bazhanov, I.\ Cherednik,
G.\ Delius, E.\ Frenkel, T.\ Hayashi,
M.\ Hashimoto, S.\ Hosono, M.\ Jimbo, A.\ Matsuo, T.\ Miwa, Y-H.\ Quano and
V.O.\ Tarasov for their helpful discussions.
This work is supported in part by JSPS fellowship,
NSF grant PHY-92-18167 and Packard fellowship.

\sect{Appendix A. Remarks on $Q$-system}
\subsect{A.1. Origin of eq.(3.5)}
\subskip

Let us sketch a derivation of (3.5) (eq.(21) in [\rKRii]).
Let $W_{m_j}^{(a_j)}(u_j)$,
$j=1,2,\ldots, N$ be the IFDRs of $Y(X_r)$ as
described in section 3.2
for some integers $m_j \ge 0$, $1 \le a_j \le r$ and
complex parameters $u_j$.
For those $u_j$'s in a general position,
the $N$-fold tensor product
$\bigotimes_{j=1}^N W_{m_j}^{(a_j)}(u_j)$
is known to be independent of the order and
irreducible as a $Y(X_r)$-module.
Decomposing it into IFDRs of $X_r \subset Y(X_r)$,
one has
$$
\bigotimes_{j=1}^N W_{m_j}^{(a_j)}(u_j) = \bigoplus_{\Lambda}
Z(\{a_j \} , \{m_j \}, \sum_{j=1}^N m_j \Lambda_{a_j}
 - \Lambda) V_\Lambda,
\eqno({\rm A}.1)
$$
where
$V_\Lambda$ is the IFDR of $X_r$ with highest weight $\Lambda$
and $Z(\cdots)$
denotes its multiplicity which is
independent of $u_j$'s in a general position.
By the definition, the sum in (A.1) extends over
those $\Lambda$ having the form
$$
\Lambda=\sum_{j=1}^N m_j \Lambda_{a_j} -\sum_{b=1}^r n_b \alpha_b
\eqno({\rm A}.2)
$$
for some non-negative integers $n_b$.
The $Z$ in (3.5) is the
special case of the above multiplicity
corresponding to the simplest situation
$N=1, a_1=a, m_1 = m$.
The function
$Z(\{a_j \} , \{m_j \},\sum_{b=1}^r n_b \alpha_b)$
for general $\{a_j \}$, $\{m_j \}$
has been computed in [\rKRii,\rKili] from
rational Bethe ansatz equations.
Their calculation essentially reduces to counting available
ranges for ``branch cut integers" of the Bethe ansatz equation
(BAE) (B.2).
\par
Here we provide an alternative
prescription to recover their formula at least formally based on the
observation in [\rDas].
The argument is essentially equivalent to [26]
   but closer to the TBA context in appendix B.
Consider the ``inhomogeneous version" of the BAE (B.2)
that corresponds to
the quantum space choice (A.1)
in the rational limit.
The equation (B.3) may then formally be replaced by
$${1 \over N}\sum_{j=1}^N\delta_{a a_j}{\cal A}^{m_j m}_{a_j a}
= \sigma^{(a)}_m + \sum_{(b,k) \in G}
{\cal K}^{m k}_{a b}\ast
\rho^{(b)}_k
\eqno({\rm A}.3)
$$
for $N$ tending to infinity.
Taking the $x=0$-th Fourier component (B.8) of this, we have
$$\sum_{j=1}^N\delta_{a a_j}\hat{{\cal A}}^{m_j m}_{a_j a}(0)
= M^{(a)}_m + \sum_{(b,k) \in G}
K^{m k}_{a b} N^{(b)}_k
\eqno({\rm A}.4)
$$
by using (B.21a). Here, $N^{(a)}_m$, $M^{(a)}_m$ are the total
number (infinite actually)
of color $a$ $m$-strings and holes by which
we have put $\hat{\rho}^{(a)}_m(0) = N^{(a)}_m/N$,
$\hat{\sigma}^{(a)}_m(0) = M^{(a)}_m/N$.
By considering the rational limit $\ell \rightarrow \infty$ further,
(A.4) can be solved for $M^{(a)}_m$ as
$$M^{(a)}_m = \sum_{j=1}^N \delta_{a a_j}
\hbox{min}(m, m_j) - \sum_{b=1}^r\sum_{k=1}^\infty
(\alpha_a \mid \alpha_b) \hbox{min}(t_bm, t_ak) N^{(b)}_k
\eqno({\rm A}.5)
$$
by using (B.11) and (B.21b).
Now regard (A.5) as a formal relation among the
infinitely many integers $M^{(a)}_m$, $N^{(a)}_m$
$(1 \le a \le r, m \in {\bf Z}_{\ge 1})$ for
any {\it finite} $N$.
This makes sense since $N$ now only appears as a parameter.
Then the multiplicity formula in [\rKRii] is given as follows:
$$
Z(\{a_j \} , \{m_j \},\sum_{b=1}^r n_b \alpha_b)
= \sum_{\{N^{(b)}_k\}}\prod_{a=1}^r\prod_{m=1}^\infty
\left( {N^{(a)}_m + M^{(a)}_m \atop
                N^{(a)}_m } \right),
\eqno({\rm A}.6{\rm a})
$$
where the sum extends over all the non-negative
integers $\{N^{(b)}_k\}$ such that
$$\eqalignno{
n_b &= \sum_{k=1}^\infty kN^{(b)}_k\quad \hbox{for all }
1 \le b \le r,&({\rm A}.6{\rm b})\cr
M^{(a)}_m& \hbox{ in (A.5) } \ge 0 \quad \hbox{for all }
1 \le a \le r, m \ge 1.&({\rm A}.6{\rm c})\cr}$$
The formula is just counting the total number of those
string-hole arrangements that obey the BAE constraint (A.5).

\subsect{A.2. General specializations, charge function and congruence}
\subskip

It is possible to generalize the restricted $Q$-system (3.10)
by considering the specializations $z=\Lambda \in P_\ell$.
Supported by numerical experiments we assume that
$$Q^{(a)}_m(\Lambda) = Q^{(a)}_{\ell_a}(\Lambda)
Q^{(a)}_{\ell_a-m}(\Lambda)^\ast \quad
\hbox{for any }\,\,
-1 \le m \le \ell_a+1 \,\,\hbox{ and }\,\,\Lambda \in P_\ell,
\eqno({\rm A}.7)$$
where $\ast$ denotes complex conjugation.
Note in particular that it implies
$$Q^{(a)}_{\ell_a+1}(\Lambda) = 0, \eqno({\rm A}.8)$$
from which one finds that the $Q$-system (3.7)
closes within those $Q^{(a)}_m(\Lambda)$
with $0 \le m \le \ell_a$.
Due to (A.7) and (B.23b), we have
\pn
{\it $\Lambda$-restricted $Q$-system}
$$
Q^{(a)}_m(\Lambda)^2 = Q^{(a)}_{m-1}(\Lambda)Q^{(a)}_{m+1}(\Lambda) +
Q^{(a)}_m(\Lambda)^2 \prod_{b=1}^r \prod_{k=1}^{\ell_b}
Q^{(b)}_k(\Lambda)^{-2J^{k\, m}_{b\, a}}\,\hbox{ for }
(a,m) \in G, \eqno({\rm A}.9{\rm a})
$$
$$
\prod_{b=1}^r Q^{(b)}_{\ell_b}(\Lambda) {}^{C_{a b}} = 1
\quad \hbox{ for } 1 \le a \le r.
\eqno({\rm A}.9{\rm b})
$$
It is possible to determine the values $Q^{(a)}_{\ell_a}(\Lambda)$ as
we shall describe below. Thus,
(A.9a) can be also viewed as relations among
$\{Q^{(a)}_m(\Lambda) \mid \, (a,m) \in G \}$
with $Q^{(a)}_{\ell_a}(\Lambda)$
as an external input.

The eq.(A.9b) says that
$Q^{(a)}_{\ell_a}(\Lambda)$'s are $\kappa$-th roots of unity with
$\kappa=\hbox{det}\bigl(C_{a b}\bigr)_{1 \le a, b \le r}$ =
number of the $X^{(1)}_r$ Kac labels $a_i (0 \le i \le r)$ equal to 1.
Explicitly, $\kappa$ is given by
$$\vbox{
\settabs 10 \columns
\+ $X_r$ & $A_r$ & $B_r$ & $C_r$& $D_r$ & $E_6$ & $E_7$ & $E_8$
& $F_4$ & $G_2$ \cr
\+ $\kappa$   & $r+1$ & 2 & 2 & 4 & 3 & 2 & 1 & 1 & 1 \cr}.
\eqno({\rm A}.10)$$
Being $\kappa$-th roots of unity, their actual values can easily be fixed
from a numerical analysis.
The following is the so obtained result and was announced
earlier in [\rKNii].
(Eqs.(6) and (7) there contain misprints and should read as
(A.11,12) below.)
$$\eqalign{
&Q^{(a)}_{t_a\ell}(\Lambda)\cr
=&
\hbox{exp}(-2\pi i \Gamma(\Lambda){\overline \gamma}_a/\kappa)
\,\, \hbox { for } X_r \neq D_r,\cr
=& \hbox{exp}(2\pi i ((r\Gamma_1(\Lambda)+\Gamma_2(\Lambda))\gamma^{(1)}_a
                    + \Gamma_1(\Lambda)\gamma^{(2)}_a)/\kappa)
\,\, \hbox { for } X_r = D_r,\, r: \hbox{even},\cr
=& \hbox{exp}(2\pi i \Gamma_2(\Lambda)((r+1)\gamma^{(1)}_a
                    + \gamma^{(2)}_a)/\kappa)
\,\, \hbox { for } X_r = D_r,\, r: \hbox{odd}.\cr
}\eqno({\rm A}.11)$$
Here, for $\mu = \sum_{a=1}^r \mu_a \Lambda_a \in {\cal H}^\ast$,
we have set
$$\eqalign{
\Gamma(\mu)
&= \sum_{a=1}^r \gamma_a \mu_a \, \quad
\hbox{ for } X_r \neq D_r, \cr
\Gamma_i(\Lambda)
&= \sum_{a=1}^r \gamma_a^{(i)} \mu_a \,\, (i=1,2)
\, \hbox{ for } X_r = D_r,\cr
}\eqno({\rm A}.12{\rm a})$$
and $\gamma = (\gamma_1, \ldots, \gamma_r)$ is
the rank-dimensional integer vector given by
$$
\eqalign{
&A_r, C_r, E_6: \gamma_a = a,\cr
&B_r: \gamma = (0,\ldots,0,1),\cr
&D_r: \gamma^{(1)}=(0,\ldots,0,1,1), \quad
\gamma^{(2)} = (2,4,6,\ldots,2(r-2),r-2,r),\cr
&E_7: \gamma = (0,0,0,1,0,1,1),\cr
&E_8,F_4,G_2: \gamma = (0,\ldots,0).\cr
}\eqno({\rm A}.12{\rm b})$$
The ${\overline \gamma}$ in the first line of (A.11) is specified from
the above $\gamma$ by the rule
${\overline \gamma} = -\gamma$ if $X_r = A_r$ and
${\overline \gamma} = \gamma$ of the dual algebra of  $X_r$ if
$X_r \neq A_r, D_r$.
We note that when $X_r = D_r$, there is an obvious constraint
${1-(-)^r \over 2}\Gamma_1(\Lambda) + \Gamma_2(\Lambda) \in 2{\bf Z}$
for $\Lambda \in P$.
It is zero for $\Lambda = 0$ hence (3.9b) is just the
special case of (A.11).
We call $\Gamma(\mu)$ the charge function.
Below we summarize its properties that will be of use
    in Part II.

The charge function dictates the
congruence property of the weight lattice element
$\mu = \sum_{a=1}^r \mu_a \Lambda_a \in P$
mod root lattice $Q$ as follows,
$$\eqalign{
\mu \equiv 0 \,\, \hbox{ mod } Q
& \leftrightarrow \Gamma(\mu) \equiv 0 \,\, \hbox{ mod } \kappa
\,\hbox{ for } \, X_r \neq D_r, \cr
& \leftrightarrow \Gamma_i(\mu) \equiv 0 \,\, \hbox{ mod } 2i
\,\, (i = 1,2)\,\hbox{ for } \,  X_r = D_r. \cr
}\eqno({\rm A}.13)$$
This originates in the relation
$\alpha_a = \sum_{b=1}^r C_{b a} \Lambda_b$ and the property
of the $\gamma-$vectors in (A.12) as follows:
$$\eqalign{
&\sum_{a=1}^r \gamma_a C_{a b}\equiv
\sum_{b=1}^r C_{a b}{\overline  \gamma}_b \equiv 0 \,\, \hbox{ mod } \kappa
\,\, \hbox{ for } \, X_r \neq D_r,\cr
&\sum_{a=1}^r \gamma^{(i)}_a C_{a b} \equiv 0 \,\, \hbox{ mod } 2i
\,\,(i=1,2)\,\,\hbox{ for } \, X_r = D_r.\cr
}\eqno({\rm A}.14)$$
As is well known, the weight lattice is divided into $\kappa$
distinct classes each of which consists of the elements
congruent under the root lattice $Q$.
In particular for $X_r = E_8, F_4$ and $G_2$,
one has $\kappa = 1$ hence every integral weight is
congruent under $Q$.
The following formula will also be useful as well as (A.13),
$$\eqalignno{
&{1 \over \kappa}\sum_{1 \le a, b \le r}
\gamma_a C_{a b} {\overline \gamma}_b \equiv 1 \,\,
\hbox{ mod } \kappa \,\,\hbox{ for } X_r \neq D_r,
&({\rm A}.15{\rm a})\cr
&{1 \over \kappa}\sum_{1 \le a, b \le r}
\gamma^{(i)}_a C_{a b} \gamma^{(j)}_b
= \cases{ r& if $i=j=2$\cr
          1& otherwise\cr}
\,\,\hbox{ for } X_r = D_r. &({\rm A}.15{\rm b})\cr}
$$

\sect{\bf Appendix B. $Y$-system}

\subsect{B.1. $Y$-system from TBA}
\subskip

Let us recapitulate the $U_q(X^{(1)}_r)$ functional
relation introduced in [\rKN] based on the
TBA analyses [\rBRi,\rBRii,\rKu].
Formally the same relation has been
noticed for some $X_r$ in the context of the TBA for
the perturbed CFTs [\rAlbZam--\rRav].
Here we call the functional relation simply the $Y$-system borrowing
the naming in [\rRav].
See section 3.1 for the notations.
\par
Take any  simple Lie algebra $X_r$ and fix
the integers $\ell \ge 1$, $p$ and $s$
   so that $(p,s) \in G$.
Choose a positive integer $N$ such that
$$N_a \buildrel\rm def\over =
Ns\bigl(C^{-1}\bigr)_{ap} \in {\bf Z}\quad\hbox{ for all }\,\,
1 \le a \le r.
\eqno({\rm B}.1)
$$
Then the following Bethe ansatz equation (BAE)
for $\{u^{(a)}_j \vert 1 \le a \le r, 1 \le j \le N_a \}$
was considered in [\rBRi,\rBRii,\rKu]:
$$
\Biggl({\sinh\Bigl({\pi\over 2L}\bigl(
u^{(a)}_j + i
{s \over t_p}\delta_{a p}
\bigr)\Bigr) \over
\sinh\Bigl({\pi\over 2L}\bigl(
u^{(a)}_j - i
{s \over t_p}\delta_{a p}
\bigr)\Bigr)}
\Biggr)^N = \Omega^{(a)}_j \prod_{b=1}^r\prod_{k=1}^{N_b}
{\sinh\Bigl({\pi\over 2L}\bigl(
u^{(a)}_j - u^{(b)}_k + i(\alpha_a \vert \alpha_b)\bigr)\Bigr) \over
\sinh\Bigl({\pi\over 2L}\bigl(
u^{(a)}_j - u^{(b)}_k - i(\alpha_a \vert \alpha_b)\bigr)\Bigr)},
\eqno({\rm B}.2)
$$
where $L = \ell + g$ and
$\Omega^{(a)}_j$ is some phase factor without which
(B.2) is essentially the BAE proposed in [\rRes].
The above BAE is indeed valid [\rBRi,\rBRii] for the
$A^{(1)}_r$ RSOS model [\rJKMO] and is a candidate
describing the transfer matrix eigenvalues of the
level $\ell$ critical $X^{(1)}_r$ RSOS model with fusion type
$W^{(p)}_s$
in general.
In Part II, we will investigate such
RSOS models with the aid of the $T$-system (3.20).
However there is yet an alternative approach, namely,
the TBA analysis of (B.2) upon the
special {\sl string hypothesis} employed in [\rBRi,\rBRii,\rKu].
The analysis was the source of our
$Y$-system and
the rest of this section is a quick digest of it
following section 2 in [\rKu].
\par
Denote by ${\cal N}^{(a)}_m$ the number of
$u^{(a)}_j$'s that form the color $a$ $m$-string
$\{u + it_a^{-1}(m+1-2n) \vert 1 \le n \le m \}$
$(u \in {\bf R})$ in the
thermodynamic limit $N \rightarrow \infty$.
Then the hypothesis is to assume
$\lim_{N \rightarrow \infty}
\sum_{m=1}^{\ell_a}{\cal N}^{(a)}_m/N_a = 1$ for all
$1 \le a \le r$.
It means that for color $a$, only those strings with length $\le \ell_a$
contribute to the thermodynamic quantities.
Then one introduces the string and hole densities
$\rho^{(a)}_m(u)$ and $\sigma^{(a)}_m(u)$, respectively
($1 \le a \le r, 1 \le m \le \ell_a)$ and
transforms (B.2) into an integral equation.
This is a fairly standard calculation going back to [\rYY].
A peculiar feature here is that
$\sigma^{(a)}_{\ell_a}(u) \equiv 0$ follows automatically
from the resulting equation, string hypothesis and (B.1).
After some reduction owing to this, the integral equation becomes
$$\delta_{p a} {\cal A}^{s m}_{p a} =
\sigma^{(a)}_m + \sum_{(b,k) \in G}
{\cal K}^{m k}_{a b}\ast
\rho^{(b)}_k \quad\hbox{ for }\,\,
(a,m) \in G. \eqno({\rm B}.3)
$$
See section B.2 for the definitions of
${\cal A}^{m k}_{a b}, {\cal K}^{m k}_{a b}$
and the symbol $\ast$.
Under the constraint (B.3), one demands the free energy be extremum
with respect to $\rho^{(a)}_m$ and
deduces the equilibrium condition
$$\eqalignno{
{\epsilon t_a\delta_{p a}\delta_{s m} \over 4 T \cosh(t_a\pi u/2)}
&=  \sum_{n=1}^{\ell_a - 1}\int_{-\infty}^\infty dv
{\cal K}^{m n}_a(u-v) \log
\bigl(1 + \hbox{exp}(\epsilon^{(a)}_n(v))\bigr) \cr
- &\sum_{(b,k) \in G}\int_{-\infty}^\infty dv
{\cal J}^{m k}_{a b}(u-v)
\log\bigl(1 + \hbox{exp}(-\epsilon^{(b)}_k(v))\bigr),
&({\rm B}.4{\rm a})\cr
\sigma^{(a)}_m(u) / \rho^{(a)}_m(u) &=
\hbox{exp}(\epsilon^{(a)}_m(u)),&({\rm B}.4{\rm b})\cr}
$$
for $(a,m) \in G$.
Here $T$ denotes the temperature and
$\epsilon = \pm 1$ on the lhs, which specifies the two
regimes as in [\rBRii,\rKu].
${\cal K}^{m n}_a$ and ${\cal J}^{m k}_{a b}$
are available in (B.16) and (B.19).
Eq.(B.4a) is called the TBA equation and played
a central role in the study of
thermodynamics of the RSOS models [\rBRi,\rBRii,\rKu].
The rhs is universal in that it is only
governed by the data $X_r$, $\ell$ and reflects the
structure of the rhs in BAE
(B.2) under the string hypothesis.
On the other hand, the lhs of (B.4a) depends on the
model details like fusion type $W^{(p)}_s$ as well as regimes.
Now consider the high temperature limit
$T \rightarrow \infty$, where one formally
drops the lhs and is left with
the equation
$$\eqalignno{
&\sum_{n=1}^{\ell_a - 1}
{\cal K}^{m n}_a \ast\log
\bigl(1 + Y^{(a)}_n{}^{-1}\bigr) =
\sum_{(b,k) \in G}
{\cal J}^{m k}_{a b}\ast
\log\bigl(1 + Y^{(b)}_k\bigr),
&({\rm B}.5{\rm a})\cr
&Y^{(a)}_m(u) =
\hbox{exp}(-\epsilon^{(a)}_m(u)),&({\rm B}.5{\rm b})\cr}
$$
which is universal in the above sense.
Formally passing to the Fourier components
by using (B.16)--(B.19) and transforming back after some
rearrangements, one can also rewrite (B.5) in the form
$$\eqalignno{
&Y^{(a)}_m(u+{i\over t_a})Y^{(a)}_m(u-{i\over t_a}) =
{\prod_{b=1}^r \prod_{k=1}^3
F_k(a,m,b;u)^{I_{ab} \delta_{t_ak, t_{ab}}} \over
\bigl(1+Y^{(a)}_{m-1}(u)^{-1}\bigr)
\bigl(1+Y^{(a)}_{m+1}(u)^{-1}\bigr)},
&({\rm B}.6{\rm a})\cr
&F_k(a,m,b;u) = \prod_{j=-k+1}^{k-1}\prod_{n=0}^{k-1-\vert j \vert}
\Bigl(1 + Y^{(b)}_{t_bm/t_a + j}
\bigl(u + i(k-1-\vert j \vert - 2n)/t_b\bigr)\Bigr),
&\cr
&&({\rm B}.6{\rm b})\cr}
$$
where by convention
$Y^{(a)}_0(u)^{-1} = Y^{(a)}_{\ell_a}(u)^{-1} = 0$
in (B.6a) and
$Y^{(a)}_m(u) = 0$ if $m \not\in {\bf Z}$ in (B.6b).
Eq.(B.6) closes among $(a,m) \in G$.
However, it extends to an infinite system of simultaneous equations
for all $m \ge 1$ if we disregard
$Y^{(a)}_{\ell_a}(u)^{-1} = 0$.
We call them the
(level $\ell$)
restricted and unrestricted $Y$-system, respectively.
The former refers to the data $X_r$ and $\ell$
while the latter is specified only by $X_r$.
They were firstly introduced in the above
generality in [\rKN].
The level $\ell$ restricted $Y$-system is a
closed and finite set of functional relations.
Repeated use of it seems to yield the following
periodicity in general:
$$
Y^{(a)}_m(u) = Y^{(a)}_m(u + 2i(\ell + g))\quad
\hbox{ for }\,\,(a,m) \in G.\eqno({\rm B}.7)
$$
In view of the identification
$Y^{(a)}_m(u) = \rho^{(a)}_m(u)/\sigma^{(a)}_m(u)$,
this is consistent with the invariance
of the BAE (B.2) under
$u^{(a)}_j \rightarrow u^{(a)}_j + 2i(\ell + g)$.
Note however that the period can be
a divisor of the above.
For example in the simplest $X_r = A_1$, $\ell = 2$
case, one has $Y^{(1)}_1(u-i)Y^{(1)}_1(u+i) = 1$ hence
the period is $4i$, which is the half of (B.7).

\subsect{B.2. Summary of relevant functions}
\subskip

Let us summarize the definitions of
the functions
${\cal A}^{m k}_{a b}$ and ${\cal K}^{m k}_{a b}$ etc and their properties
relevant to the TBA analysis.
They will be also of use in many places in Part II.
Given a function $f(u)$ we shall frequently
work with its Fourier component ${\hat f}(x)$ defined as
$$f(u) = \,\check{ }\, ({\hat f})(u) = {1 \over 2 \pi}
\int_{-\infty}^\infty {\hat f}(x) e^{i u x} dx,\quad
{\hat f}(x) = \int_{-\infty}^\infty f(u) e^{-i u x} du.
\eqno({\rm B}.8)$$
The convolution of the
two functions will be denoted by
$$
(f_1 \ast f_2)(u) = \int_{-\infty}^\infty dv f_1(u-v)f_2(v).
\eqno({\rm B}.9)
$$
We shall often suppress the arguments as
$f = f(u)$, ${\hat f} = {\hat f}(x)$.
Below we list the definitions and the useful properties.
The indices $a$ and $b$ are always taken in $\{1,2,\ldots, r\}$.
$$\eqalignno{
{\hat {\cal M}}_{a b}
&= {\hat {\cal M}}_{b a}
= B_{a b} + 2\delta_{a b} ( \cosh({x\over t_a}) - 1 )
= 2\cosh({x\over t_a})\Bigl(
\delta_{ab} - {I_{a b} \over 2\cosh({x\over t_a})}\Bigr),
&\cr
&&({\rm B}.10)\cr
{\hat {\cal A}}^{m k}_{a b}&=
{\sinh\bigl(\hbox{min}({m \over t_a}, {k \over t_b})x \bigr)
 \sinh\bigl((\ell-\hbox{max}({m \over t_a}, {k \over t_b}))x \bigr)
\over
\sinh({x\over t_{a b}}) \sinh(\ell x) }, &({\rm B}.11)\cr
{\hat {\cal K}}^{m k}_{a b}&=
{\hat {\cal A}}^{m k}_{a b} {\hat {\cal M}}_{a b},
&({\rm B}.12)\cr
{\hat \Psi}^{m k}_{a b}&= \delta_{a b}\delta_{m k} -
{\hat {\cal K}}^{m k}_{a b},
&({\rm B}.13)\cr
}$$
where $(a,m), (b,k) \in G$ in
(B.11-13) and $B_{a b}$, $I_{a b}$ and $t_{a b}$ are given in
(3.1).
%
%
{}From (B.10-13) we see that
$$
\Psi^{m k}_{a b}(u) = \Psi^{m k}_{a b}(-u)
= \Psi^{k m}_{b a}(u),
\eqno({\rm B}.14)
$$
and $\Psi^{m k}_{a b}(u)$ decays rapidly when
$u \rightarrow \pm \infty$.
The same property holds also for
${\cal A}^{m k}_{a b}(u)$ and ${\cal K}^{m k}_{a b}(u)$.
In particular, we have the asymptotics
$$
{\cal A}^{m k}_{a a}(u) \buildrel{u\rightarrow \pm \infty}
\over \longrightarrow
{\sin{\pi m\over \ell_a}\sin{\pi k\over \ell_a}
\over \ell \sin{\pi \over \ell_a}}
e^{-{\pi \vert u \vert\over \ell}}.
\eqno({\rm B}.15)
$$
Definitions continue.
$$\eqalign{
{\hat {\cal K}}^{m n}_a = {\hat {\cal K}}^{n m}_a &=
\delta_{m n} + {1 \over 2\cosh({x\over t_a})}
\bigl({\bar C}^a_{m n} - 2\delta_{m n}\bigr)\cr
&= \delta_{m n} - {{\bar I}^a_{m n}\over 2\cosh({x\over t_a})}
\quad
1 \le m, n \le \ell_a - 1,\cr}\eqno({\rm B}.16{\rm a})
$$
where ${\bar C}^a$ and ${\bar I}^a$ denote
the Cartan and the incidence matrices of $A_{\ell_a -1}$,
respectively, i.e.,
$${\bar C}^a_{m n} =  2\delta_{m n} - {\bar I}^a_{m n}
= 2\delta_{m n}
- \delta_{m n-1} - \delta_{m n+1},\quad
1 \le m, n \le \ell_a - 1.
\eqno({\rm B}.16{\rm b})
$$
${\hat {\cal K}}^{m n}_a$ above should not be confused with
${\hat {\cal K}}^{m k}_{a b}$ in (B.12).
By the definition we have
$$\eqalignno{
{\hat {\cal K}}_a^{m n} &=
{{\hat {\cal M}}_{m n}\, \hbox{ for }\, X_r = A_{\ell_a - 1}\over
2\cosh({x\over t_a})},&({\rm B}.17{\rm a})\cr
2{\hat {\cal K}}_a^{m n}(0) &= {\bar C}^a_{m n}.
&({\rm B}.17{\rm b})\cr}
$$
The following inversion property is valid:
$$
2\cosh({x\over t_a})\sum_{n=1}^{\ell_a-1}
{\hat {\cal A}}_{a a}^{m n} {\hat {\cal K}}_a^{n k} = \delta_{m k}
\quad \hbox{ for }\,\,1 \le m, k \le \ell_a - 1.
\eqno({\rm B}.18)
$$
We further introduce
$$\eqalignno{
{\hat{\cal J}}^{m k}_{a b}
&= {\hat {\cal M}}_{a b} \sum_{n=1}^{\ell_a - 1}
{\hat {\cal K}}^{m n}_a
{\hat {\cal A}}^{n k}_{a b} &({\rm B}.19{\rm a})\cr
&= {{\hat {\cal M}}_{a b} \over 2\cosh(x/t_a) }
\biggl(
{\sinh(x/t_a) \over \sinh(x/t_{a b}) } \delta_{t_bm,  t_ak} \cr
&+ \sum_{j=1}^{t_b - t_a}
{\sinh(jx/t_b) \over \sinh(x/t_b) }
\bigl(\delta_{t_b(m+1) - t_aj, \, t_ak}
+ \delta_{t_b(m-1) + t_aj, \, t_ak}
\bigr) \biggr),&({\rm B}.19{\rm b})\cr}$$
for $(a,m), (b,k) \in G$.
The sum $\sum_{j=1}^{t_b - t_a}$ in (B.19b)
is to be understood as zero if
$t_a \ge t_b$.
Since the expression (B.19b) does not contain
$\ell$, we extend the definition of ${\cal J}^{m k}_{a b}$
to all the integers $m, k \ge 0$.
${\cal J}^{m k}_{a b}(u)$ is an even function of $u$
but ${\cal J}^{m k}_{a b}(u) \neq {\cal J}^{k m}_{b a}(u)$
in general as opposed to the latter property of (B.14).
Combining (B.18) and (B.19a) we have
$$
2\cosh({x\over t_a})\sum_{m=1}^{\ell_a-1}
{\hat {\cal A}}_{a a}^{j m}{\hat {\cal J}}_{a b}^{m k}
= {\hat {\cal M}}_{a b} {\hat {\cal A}}_{a b}^{j k}
= {\hat {\cal K}}_{a b}^{j k}\quad
\hbox{ for }\,\, (a,j), (b,k) \in G,
\eqno({\rm B}.20)
$$
by noting (B.12).
\par
Special values of these
functions also play an important role and we prepare the
notation
$$\eqalignno{
K^{m k}_{a b} &= {\hat {\cal K}}^{m k}_{a b}(0)
&({\rm B}.21{\rm a})\cr
&= \Bigl(\hbox{min}(t_bm, t_ak) - {m k\over \ell}\Bigr)
(\alpha_a \vert \alpha_b),
&({\rm B}.21{\rm b})\cr}
$$
for $(a,m), (b,k) \in G$.
%
%
We also define
$$\eqalignno{
J^{m k}_{a b} &= {\hat {\cal J}}^{m k}_{a b}(0)
&({\rm B}.22{\rm a})\cr
&= {B_{a b}\over 2}
\bigl({t_{a b} \over t_a}\delta_{t_a k, \, t_b m} +
\sum_{j=1}^{t_b-t_a} j
(\delta_{t_b(m+1)-t_aj, t_a k} + \delta_{t_b(m-1)+t_aj, t_a k})
\bigr),&\cr
&&({\rm B}.22{\rm b})\cr}
$$
for all integers $m, k \ge 0$.
Then one has
$$\eqalignno{
&2J^{k\, m}_{b\, a} = \sum_{n=1}^{\ell_b - 1}
{\bar C}^{b}_{n k} K^{m\, n}_{a \, b}\quad
\hbox{ for }\,\,(a,m), (b,k) \in G,
&({\rm B}.23{\rm a})\cr
&2J^{k\, 0}_{b\, a} = C_{a b} \delta_{k \,\, 0}, \quad
2J^{k\, \ell_a}_{b\, a} = C_{a b} \delta_{k \,\, \ell_b}
\quad \hbox { for }\,\, k \ge 0,&({\rm B}.23{\rm b})\cr
&2J^{\ell_b\, m}_{b\, a} + K^{m \, \ell_b - 1}_{a \, b} =
{m \over \ell_a} C_{a b}\quad \hbox{ for } \,\,
(a,m) \in G. &({\rm B}.23{\rm c})\cr
}$$
Eq.(B.23a) is a direct consequence of (B.19a),
and (B.23b,c) can be
checked for all the possibilities
$(t_a, t_b) = (1,1), (1,2), \ldots, (3,3)$ case by case.

\subsect{B.3. Alternative forms of $Y$-system}
\subskip

Let us formally rewrite the $Y$-system in alternative forms.
Consider the Fourier component of (B.5a) and
take the sum
$\sum_{m=1}^{\ell_a-1}
{\hat {\cal A}}_{a a}^{j m}$ of both sides
by means of (B.18) and (B.20).
After the inverse Fourier transformation
the result reads
$$
\log \bigl(1 + Y^{(a)}_j{}^{-1}\bigr) =
\sum_{(b,k) \in G}
{\cal K}^{j k}_{a b}\ast
\log\bigl(1 + Y^{(b)}_k\bigr)\quad \hbox{ for }\,\,
(a,j) \in G.
\eqno({\rm B}.24)
$$
By (B.13) this can be further rewritten as
$$
\log Y^{(a)}_j =
\sum_{(b,k) \in G}
\Psi^{j k}_{a b}\ast
\log\bigl(1 + Y^{(b)}_k\bigr).
\eqno({\rm B}.25)
$$
Thus we have seen that the level $\ell$ restricted $Y$-system
(B.6) can formally be transformed into three equivalent logarithmic forms
(B.5a), (B.24) and (B.25).
%
%

\subsect{B.4. Constant $Y$-system as $Q$-system}
\subskip

Here we show how  the restricted $Q$-system (3.10)
is recovered from  the restricted
$Y$-system by
dropping the spectral parameter dependence.
This is  in a sense the inverse procedure
 of the Yang-Baxterization
exploited in section 3.

Let $Y\am >0$ be a constant solution of (B.6).
Then, from (B.24) and (B.21a), the resulting
algebraic equation is equivalent to
$$
\log \bigl(1 + Y^{(a)}_j{}^{-1}\bigr) =
\sum_{(b,k) \in G}
K^{j k}_{a b}
\log\bigl(1 + Y^{(b)}_k\bigr).
\eqno({\rm B}.26)
$$
Define
$$
f^{(a)}_m = 1 -
{Q^{(a)}_{m-1}(0)Q^{(a)}_{m+1}(0) \over Q^{(a)}_m(0)^2}\quad
{\rm for }\,\, (a,m) \in G,
\eqno({\rm B}.27{\rm a})
$$
or equivalently,
$$\log(1 - f^{(b)}_n) =
-\sum_{k=1}^{\ell_b-1}{\bar C}^b_{n k} \log Q^{(b)}_k(0),
\eqno({\rm B}.27{\rm b})
$$
due to (3.9b)
and $\forall Q^{(b)}_k(0) > 0$
for $(b,k) \in G$.
{}From (3.10),
we have
$$\eqalign{
\log f^{(a)}_m &= -2 \sum_{(b,k) \in G} J^{k m}_{b a}
\log Q^{(b)}_k(0)\cr
&= - \sum_{(b,k) \in G}\sum_{n=1}^{\ell_b-1}
{\bar C}^b_{n k} K^{m n}_{a b} \log Q^{(b)}_k(0)\cr
&= \sum_{(b,n) \in G} K^{m n}_{a b} \log (1 - f^{(b)}_n),\cr}
\eqno({\rm B}.28)
$$
by using (B.23a) and (B.27).
Comparing this with (B.26) we find a constant solution
$${Y^{(a)}_m \over 1 + Y^{(a)}_m} = f^{(a)}_m,\quad
Y^{(a)}_m =
{Q^{(a)}_m(0)^2 \prod_{(b,k) \in G}
Q^{(b)}_k(0)^{-2J^{k m}_{b a}}\over
Q^{(a)}_{m-1}(0)Q^{(a)}_{m+1}(0)}.
\eqno({\rm B}.29)
$$
%
%
An analogous identification to (B.29)
also holds between the unrestricted $Y$-system and
the unrestricted $Q$-system as eq.(6) of [\rKN].

\subsect{B.5. Simply laced algebra case}
\subskip

Here we shall exclusively consider
the simply laced algebras $X_r = A_r, D_r$ and
$E_{6,7,8}$, where many formulas in sections B.1--B.4 simplify
considerably.
%
%
\par
By virtue of $\forall t_a = 1$, we shall write
${\cal A}_{a b}^{m k}$ (B.11),
${\cal K}_a^{m n}$ (B.16a),
${\bar C}^a$ and ${\bar I}^a$ (B.16b)
just as ${\cal A}^{(\ell)}_{m k}$,
${\cal K}_{m n}$,
${\bar C}$ and ${\bar I}$, respectively.
Then the restricted $Y$-system (B.6)
takes the form
$$
Y^{(a)}_m(u+i)Y^{(a)}_m(u-i) =
{\prod_{b=1}^r\bigl(1 + Y^{(b)}_m(u)\bigr)^{I_{ab}}\over
\prod_{k=1}^{\ell-1}
\bigl(1 + Y^{(a)}_k(u)^{-1}\bigr)^{{\bar I}_{m k}}},
\eqno({\rm B}.30)
$$
where ${\bar I}$ is the incidence matrix
for $A_{\ell - 1}$ as specified just above.
See (3.8) for the corresponding $Q$-system.
Set
$$
{\hat \Phi}^{m k}_{a b} = \delta_{a b}\delta_{m k} -
2\cosh x ({\hat {\cal M}}^{-1})_{a b}{\hat {\cal K}}_{m k}
\quad \hbox{ for }\,\, (a,m), (b,k) \in G,
\eqno({\rm B}.31)
$$
which enjoys the same property as (B.14).
Now we list the useful simplifications.
$$\eqalignno{
{\hat {\cal J}}^{m k}_{a b} &=
{\delta_{m k}{\hat {\cal M}}_{a b}\over 2\cosh x}
= \delta_{m k}\Bigl(\delta_{a b} - {I_{a b}\over 2\cosh x}\Bigr),
&({\rm B}.32{\rm a})\cr
2J^{m k}_{a b} &= \delta_{m k}C_{a b}
= 2\delta_{m k}\delta_{a b} - \delta_{m k} I_{a b},
&({\rm B}.32{\rm b})\cr
K^{m k}_{a b} &= \delta_{a b}\delta_{m k}
- {\hat \Psi}^{m k}_{a b}(0) =
C_{a b}\bigl({\bar C}^{-1}\bigr)_{m k},
&({\rm B}.32{\rm c})\cr
{\hat \Phi}^{m k}_{a b}(0) &= \delta_{a b}\delta_{m k}
- (C^{-1})_{a b}{\bar C}_{m k},
&({\rm B}.32{\rm d})\cr
{\hat {\cal A}}^{(\ell)}_{m k}&=
{\sinh\bigl(\hbox{min}(m, k)x \bigr)
 \sinh\bigl((\ell-\hbox{max}(m, k))x \bigr)
\over
\sinh x \sinh(\ell x) }. &({\rm B}.32{\rm e})\cr
}$$
Due to (B.32b) above, the latter of (B.29) becomes
$$
Y^{(a)}_m ={\prod_{b=1}^r  Q^{(b)}_m(0){}^{I_{a b}}
\over  Q^{(a)}_{m-1}(0)Q^{(a)}_{m+1}(0)}.
\eqno({\rm B}.33)
$$
It is elementary to check that (B.33) is a
 constant solution of (B.30)
by using (3.8) and (3.9b).
Finally we note the property
$$
\check{}\,\bigl({\hat {\cal M}}^{-1}\bigr)_{ab}(u)
\buildrel{u\rightarrow \pm \infty}\over \longrightarrow
{\chi^{(a)}\chi^{(b)}\over 2 \sin{\pi \over g}}
e^{-{\pi \vert u \vert\over g}},
\eqno({\rm B}.34)
$$
where $g$ is the (dual) Coxeter number and
$\chi = (\chi^{(a)})_{1 \le a \le r}$ is the
normalized Perron-Frobenius eigenvector
of the incidence matrix $I_{a b}$, i.e.,
$$\eqalignno{
\sum_{b=1}^r I_{a b}\chi^{(b)} &= (2\cos{\pi \over g})\chi^{(a)}
\quad\hbox{ for } \,\, 1 \le a \le r,&({\rm B}.35{\rm a})\cr
\sum_{a=1}^r \chi^{(a)\,2} &= 1.&({\rm B}.35{\rm b})\cr}
$$
In view of (B.17a) and (B.18), the asymptotics (B.34)
above actually includes
(B.15) as the special case $X_r = A_{\ell_a - 1}$.

\reference

\vfill\eject
\centerline{Table 1}
$$
\epsfxsize=220pt  \epsfysize=315pt \epsfbox{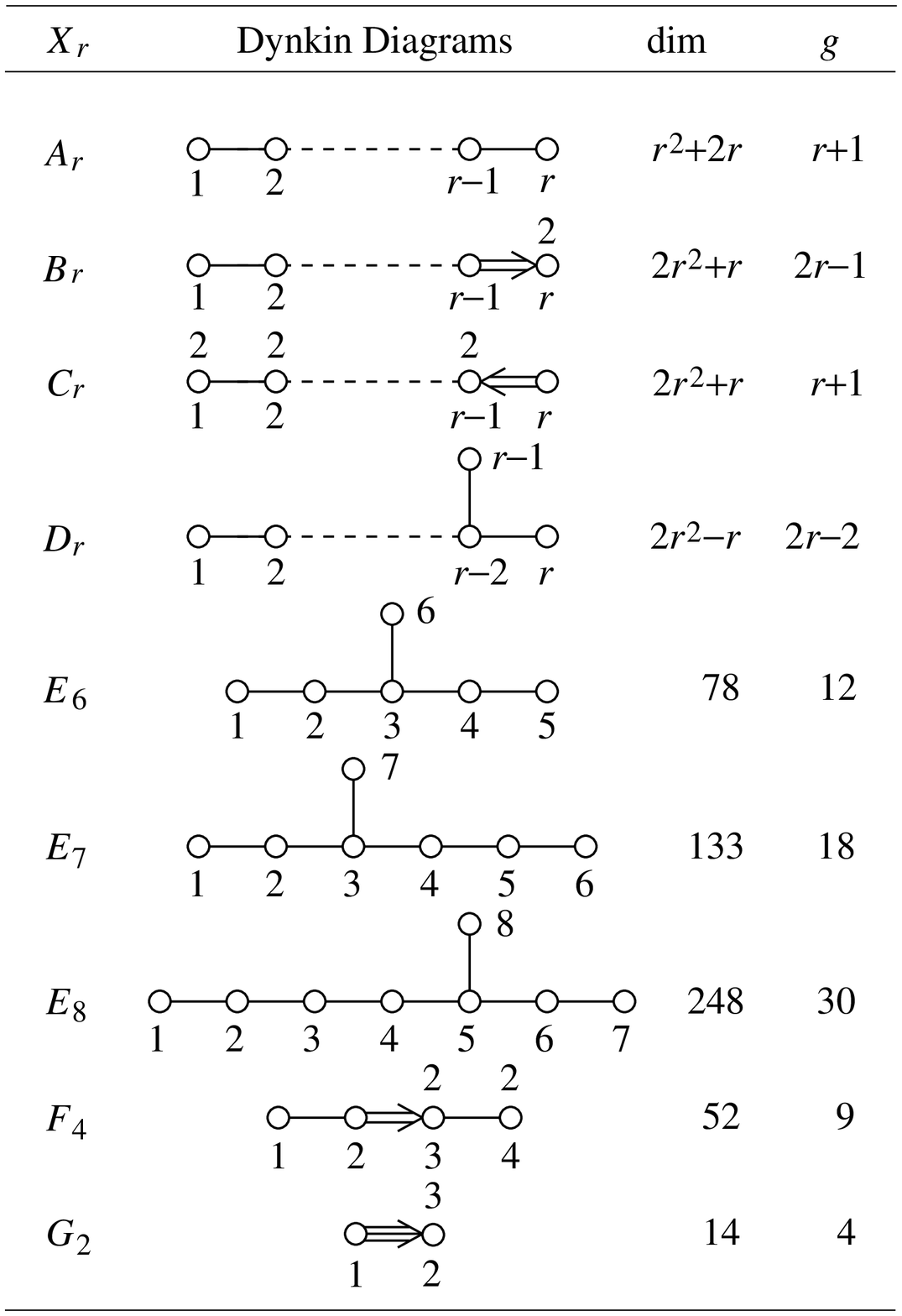}
$$
{The classical simple Lie algebra $X_r$,
the Dynkin diagram, the dimension of $X_r$ and the
dual Coxeter number $g$. In each Dynkin diagram, the
nodes are numerated from $1$ to $r$. The parameter $t_a$
(3.1a) has been given above the node $a$ only when
$t_a \neq 1$.}
%
\vskip6mm

\bye